\documentclass[12pt]{article}
\usepackage{latexsym}
\usepackage{epsfig,amssymb,euscript}
\usepackage{amsmath}
\numberwithin{equation}{section}  %DARIO: added this command
\newsavebox{\ns}
\newsavebox{\dbrane}
\newsavebox{\dbshort}

\def\be{\begin{equation}}
\def\ee{\end{equation}}
\def\bea{\begin{eqnarray}}
\def\eea{\end{eqnarray}}

\newcommand{\nn}{\nonumber}

\def\Dslash{\,\,{\raise.15ex\hbox{/}\mkern-12mu D}}
\def\Dbarslash{\,\,{\raise.15ex\hbox{/}\mkern-12mu {\bar D}}}
\def\delslash{\,\,{\raise.15ex\hbox{/}\mkern-9mu \partial}}
\def\delbarslash{\,\,{\raise.15ex\hbox{/}\mkern-9mu {\bar\partial}}}
\def\pslash{\,\,{\raise.15ex\hbox{/}\mkern-9mu p}}
\def\calDslash{\,\,{\raise.15ex\hbox{/}\mkern-12mu {\cal D}}}

\newcommand\R{\mathbb{R}}
\newcommand\Z{\mathbb{Z}}

\newcommand\C{\mathbb{C}}
\newcommand\T{\mathbb{T}}
\newcommand\TC{\mathbb{T}^s_{\mathbb{C}}}
\newcommand\diff{\mathrm{d}}
\newcommand{\de}{\partial}

\newcommand{\vol}{\mathrm{vol}}
\newcommand{\rdr}{r\partial/\partial r}
\newcommand{\fracrdr}{r\frac{\partial}{\partial r}}
\newcommand{\omegatr}{\omega_T}
\newcommand{\Tr}{\mathrm{Tr}}
\newcommand{\SC}{\mathcal{S}_{\mathcal{C}^*}}
\newcommand{\dd}{\mathrm{d}}

\begin{document}
\begin{titlepage}
\begin{center}
\today
{\small\hfill hep-th/yymmnnn}\\
{\small\hfill CERN-PH-TH/2006-039}\\
{\small\hfill HUTP-06/A0002}\\

\vskip 1.0 cm 
{\Large \bf Sasaki--Einstein Manifolds \\[5mm] \
and Volume Minimisation}\\
\vskip 10mm
{Dario Martelli$^{1}$,~~ James Sparks$^{2,3}$~~and~~ Shing--Tung Yau$^{2}$}\\
\vskip 1.4 cm

1: Department of Physics, CERN Theory Division\\
1211 Geneva 23, Switzerland\\
\vskip 0.5cm
2: Department of Mathematics, Harvard University \\
One Oxford Street, Cambridge, MA 02138, U.S.A.\\
\vskip 0.5cm
3: Jefferson Physical Laboratory, Harvard University \\
Cambridge, MA 02138, U.S.A.\\

\vskip 5mm
{\tt dario.martelli@cern.ch  \ \  sparks@math.harvard.edu }\\
{\tt yau@math.harvard.edu}\\

\end{center}
\vskip 1cm

\begin{abstract}
\noindent
We study a variational problem whose critical point determines
the Reeb vector field for a Sasaki--Einstein manifold. This extends our 
previous work on Sasakian geometry by lifting the condition that the
manifolds are toric. We show that the Einstein--Hilbert action, 
restricted to a space of Sasakian metrics on a link 
$L$ in a Calabi--Yau cone $X$, 
is the volume functional, which in fact is a function on 
the space of Reeb vector fields.
We relate this function 
both to the Duistermaat--Heckman formula and also to a limit of a certain 
equivariant index on $X$ that counts holomorphic functions. 
Both formulae may be evaluated by localisation. This leads to a general 
formula for the 
volume function in terms of topological fixed point data. 
As a result we prove that the volume of a Sasaki--Einstein manifold, 
relative to that of the round sphere, is always an algebraic number. 
In complex dimension $n=3$ these results provide, via AdS/CFT,
the geometric counterpart of $a$--maximisation in four dimensional 
superconformal field theories.
We also show that our variational problem dynamically sets to zero the 
Futaki invariant of the transverse space, the latter being an 
obstruction to the existence of a K\"ahler--Einstein metric. 
\end{abstract}

\end{titlepage}
\pagestyle{plain}
\setcounter{page}{1}
\newcounter{bean}
\baselineskip18pt

\tableofcontents

%%%%%%%%%%%%%%%%%%%%%%%%%%%%%%%%%%%%%%%%%%%%%%%%%%%%%%%%%%%%%%%%%%%%%

\section{Introduction and summary}

\subsection{Background}

The AdS/CFT correspondence \cite{Maldacena} is one of the most important advancements 
in string theory. It provides a detailed correspondence 
between certain
conformal field theories and geometries, and has led to 
remarkable new results on both sides. 
A large class of examples consists of type IIB string theory
on the background AdS$_5\times L$, where $L$ is a Sasaki--Einstein five--manifold and 
the dual theory is a four--dimensional $\mathcal{N}=1$
superconformal field theory \cite{Kehagias,KW,acharya,MP}. This has recently led to 
considerable interest in Sasaki--Einstein geometry.

\subsubsection*{Geometry}

Recall that a \emph{Sasakian} manifold $(L,g_L)$ is a Riemannian 
manifold of dimension $(2n-1)$ whose metric cone 
\be
g_{C(L)} = \diff r^2 + r^2 g_L
\label{metcone}
\ee
is K\"ahler. 
$(L,g_L)$ is \emph{Sasaki--Einstein} if 
the cone (\ref{metcone}) 
is also Ricci--flat. It follows that a Sasaki--Einstein 
manifold is a positively curved Einstein manifold.
The canonical example  
is an odd--dimensional round sphere $S^{2n-1}$; 
the metric cone (\ref{metcone}) 
is then $\C^n$ with its flat metric. 

All Sasakian manifolds have a canonically 
defined Killing vector field $\xi$, 
called the \emph{Reeb} vector field. This vector field will play a central 
role in this paper. To define $\xi$ note that, since the cone is K\"ahler, 
it comes equipped with a covariantly constant complex structure 
tensor $J$. The Reeb vector field is then defined to be
\be
\xi = J\left(r\frac{\partial}{\partial r}\right)~.\ee 
As a vector field on the link\footnote{We prefer this choice of 
terminology to ``base of the cone'', or ``horizon''.} $L=\{r=1\}$ this has norm one, 
and hence its orbits define a \emph{foliation} of $L$. 
Either these orbits all close, or they don't. If they all close, the 
flow generated by $\xi$ induces a $U(1)$ action on $L$ which, since 
the vector field is nowhere--vanishing, is locally free. 
The orbit, or quotient, space is then a K\"ahler orbifold, which is a
manifold if the $U(1)$ action is actually free. These Sasakian metrics 
are referred to as quasi--regular and regular, respectively. 
More generally, the generic orbits of $\xi$ might not close. 
In this case there is no quotient space, and the K\"ahler structure 
exists only as a transverse structure. The closure of the orbits of 
$\xi$ is at least a two--torus, and thus 
these so--called irregular Sasakian metrics admit at least a two--torus 
of isometries. This will also be crucial in what follows. Note that 
Sasakian geometries are then sandwiched between two K\"ahler geometries: 
one of complex dimension $n$ on the cone, and one, which is generally 
only a transverse structure, of dimension $n-1$. For Sasaki--Einstein 
manifolds, the transverse metric is in fact \emph{K\"ahler--Einstein}.

Until very recently, the
only explicit examples of simply--connected Sasaki--Einstein manifolds 
in dimension five (equivalently complex dimension $n=3$) were the round sphere 
$S^5$ and the homogeneous metric on $S^2\times S^3$, known as $T^{1,1}$ in 
the physics literature. These are both regular Sasakian structures, the 
orbit spaces being $\mathbb{C}P^2$ and $\mathbb{C}P^1\times\mathbb{C}P^1$ 
with their K\"ahler--Einstein metrics.
All other Sasaki--Einstein metrics, in dimension five, 
were known only through existence arguments. The remaining regular 
metrics are based on circle bundles over del Pezzo surfaces $dP_k$, 
$3\leq k\leq 8$, equipped with their K\"ahler Einstein metrics  -- 
these are known to exist
through the work of Tian and Yau \cite{tian, tianyau}. On 
the other hand, Boyer and Galicki have produced many examples of 
quasi--regular Sasaki--Einstein metrics using existence results of Koll\'ar 
and collaborators for K\"ahler--Einstein metrics on orbifolds. For 
a review of their work, see \cite{boyerreview}. 

In references \cite{paper1,paper2, paper3} infinite families of 
explicit inhomogeneous 
Sasaki--Einstein metrics 
in all dimensions have been constructed. 
In particular, when $n=3$ there 
is a family of cohomogeneity one five--metrics, denoted 
$Y^{p,q}$ where $q<p$ with $p,q$ 
positive integers \cite{paper2}. 
This has subsequently been generalised to a 
three--parameter
cohomogeneity two family $L^{a,b,c}$ \cite{CLPP,MS2,CLPP2}. 
Provided the integers $a,b,c$ are chosen such that $L^{a,b,c}$ is smooth and 
simply--connected, these manifolds are all diffeomorphic to 
$S^2\times S^3$.
Further generalisations in complex dimension
$n\geq 4$ have appeared in \cite{strasbourg,Chen:2004nq,Lu:2005sn}.
The metrics $Y^{p,q}$ are quasi--regular when $4p^2-3q^2$ is a square. However,
for general $q<p$, they are \emph{irregular}. These were the first examples 
of irregular Sasaki--Einstein manifolds, which in particular 
disproved the conjecture of Cheeger and 
Tian \cite{CT} that irregular Sasaki--Einstein manifolds do not exist. 

\subsubsection*{Field theory}

In general, the field theory duals of Sasaki--Einstein 
five--manifolds may be thought of as arising from a stack of D3--branes
sitting at the apex $r=0$ of the Ricci--flat K\"ahler cone (\ref{metcone}). 
Alternatively, the Calabi--Yau geometry may be thought of as 
arising from the moduli space of 
the Higgs branch of the gauge theory on the D3--branes. 
Through simple 
AdS/CFT arguments, one can show that 
the symmetry generated by the 
Reeb vector field and the volume of the Sasaki--Einstein manifold 
correspond to the R--symmetry and 
the $a$ central charge of the AdS/CFT dual superconformal field theory, 
respectively. 
All $\mathcal{N}=1$ superconformal field theories in 
four dimensions possess a global R--symmetry which is part of the 
superconformal algebra. The $a$ central charge appears as a coefficient in 
the one--point function of the trace of the energy--momentum tensor on 
a general background\footnote{The other coefficient is usually called 
$c$. However, superconformal field theories with a Sasaki--Einstein 
dual have $a=c$.}, and its value may be computed exactly, once  
the R--symmetry is correctly identified.
A general procedure that determines this symmetry is 
 $a$--maximisation \cite{IW}. One defines a function $a_\mathrm{trial}$ on an 
appropriate space of potential (or ``trial'') 
R--symmetries. The local maximum of this 
function  determines the R--symmetry of the theory at its 
superconformal point. Moreover, the critical value of $a_\mathrm{trial}$ is 
precisely the central charge $a$ of the superconformal theory. 
Since $a_\mathrm{trial}$ is 
a cubic function with rational coefficients,
it follows\footnote{Since we make a similar claim in this paper, 
we recall here a proof of this fact: 
suppose we have a vector $v\in\R^s$ which is an isolated 
zero of a set of polynomials in the components of $v$ 
with \emph{rational} coefficients. 
Consider the Galois group $\mathrm{Gal}(\C/\mathbb{Q})$. This group 
fixes the set of polynomials, and thus in particular the Galois orbit of the 
zeros is finite. An \emph{algebraic number} may be defined 
as an element of $\C$ with finite Galois orbit, and thus we see that
the components of the vector $v$ are algebraic numbers. We thank 
Dorian Goldfeld for this argument. Recall also
that the set of algebraic numbers form a field. Thus, in the present 
example, the R--charges of fields being algebraic implies that the 
$a$--central charge, which is a polynomial function of the R--charges 
with rational coefficients, is also an algebraic number.} 
that the R--charges of fields are \emph{algebraic numbers} \cite{IW}.

The AdS/CFT correspondence relates these R--charges to the volume of the dual 
Sasaki--Einstein manifold, as well as the volumes of certain supersymmetric 
three--dimensional submanifolds of $L$. In particular, we have the relation 
\cite{Skenderis, Gubser}
\be
\frac{a_L}{a_{S^5}} = 
\frac{\mathrm{vol}[S^5]}{\mathrm{vol}[L]}~.
\label{central}
\ee
Since the left--hand side is determined by $a$--maximisation, 
we thus learn that the volume of a Sasaki--Einstein five--manifold $\vol[L]$, 
relative to that of the round sphere, is an algebraic number. Moreover, 
in the field theory, this number has been determined by a finite 
dimensional extremal problem. Our aim in \cite{MSY}, of which this 
paper is a continuation, 
was to try to understand, from a purely \emph{geometrical} viewpoint, 
where these statements are coming from.

\subsubsection*{Toric geometries and their duals}

Given a Sasaki--Einstein manifold, it is in 
general a  difficult problem to determine the dual
field theory. However, 
in the case that the local Calabi--Yau singularity is \emph{toric} there exist
techniques that allow one to determine a dual gauge theory, starting from 
the combinatorial data that defines the toric variety. 
Using these methods it has been possible to construct gauge  theory duals
for the infinite family of Sasaki--Einstein manifolds 
$Y^{p,q}$ \cite{toric,quiverpaper}, thus furnishing a 
countably infinite set of 
AdS/CFT duals where both sides of the duality are known explicitly. 
Indeed, remarkable agreement was found between the geometrical 
computation in the case of the $Y^{p,q}$ metrics \cite{paper2,toric}
and the $a$--maximisation calculation \cite{quiverpaper, BBC} 
for the corresponding quiver gauge theories. Thus the relation 
(\ref{central}) was confirmed 
for a non--trivial infinite family of examples. 
Further developments \cite{hanany,Franco:2005rj,Franco:2005sm,BZ,Hanany:2005ss,Franco:2006gc} 
have resulted in the determination of 
families of gauge theories that are 
dual to the wider class of 
toric Sasaki--Einstein manifolds $L^{a,b,c}$ \cite{Franco:2005sm,BFZ} (see also \cite{BK2}).

Of course, given this success, it is natural to try to obtain 
a general understanding of the geometry underlying $a$--maximisation 
and the AdS/CFT correspondence. To this end, in \cite{MSY}
we studied a variational problem on a space of toric Sasakian metrics. 
Let us recall the essential points of \cite{MSY}.
Let $(X,\omega)$ be a toric K\"ahler cone of complex dimension $n$. 
This means that $X$ is an affine toric variety, equipped with a 
conical K\"ahler metric that is invariant under a holomorphic 
action of the $n$--torus $\T^n$. $X$ has an isolated singular 
point at the tip of the cone, the complement of which is 
$X_0=C(L)\cong\R_+\times L$. A conical metric on $X$ which 
is K\"ahler (but in general not Ricci--flat) then gives a Sasakian metric on the link
 $L$. The moment map for the torus action
exhibits $X$ as a Lagrangian $\T^n$ fibration over a
strictly convex rational polyhedral cone\footnote{We make a change 
of notation from our previous paper \cite{MSY}: specifically, we exchange 
the roles of cone and dual cone. This is more in line with algebro--geometric 
terminology, and is more natural in the sense that 
the moment cone $\mathcal{C}^*$ lives in the dual Lie algebra $\mathtt{t}_n^*$ 
of the torus.} $\mathcal{C}^*\subset\mathtt{t}_n^*\cong\R^n$.
This is a subset of $\R^n$ of the form
\be\label{conehead}
\mathcal{C}^* = \{y\in\R^n\mid (y,v_a)\geq 0, \ a=1,\ldots,D\}~.\ee
Thus $\mathcal{C}^*$ is made by intersecting $D$ hyperplanes through the
origin in order to make a convex polyhedral cone.
Here $y\in\R^n$ are coordinates on $\R^n$ and $v_a$ are the inward
pointing normal vectors to the $D$ hyperplanes, or \emph{facets},
that bound the
polyhedral cone.

The condition that $X$ is Calabi--Yau implies
that the vectors $v_a$ may, by an appropriate
$SL(n;\Z)$ transformation of the torus, be all written as $v_a=(1,w_a)$.
In particular, in complex dimension $n=3$ we may therefore represent any
toric Calabi--Yau cone $X$ by a convex lattice polytope in
$\Z^2$, where the vertices are simply the vectors $w_a$.
This is usually called the toric diagram. Note that the cone $\mathcal{C}^*$ may also be 
defined in 
terms of its generating edge vectors $\{u_{\alpha}\}$ giving the 
directions of the lines going through the origin. 
When $n=3$ the projection of 
these lines onto the plane with normal $(1,0,0)$ are 
the external legs of the so--called pq--web appearing in the physics literature.
These are also
weight vectors for the torus action and generalise to the 
non--toric case. 

For a toric K\"ahler cone $(X,\omega)$, one can introduce symplectic coordinates $(y_i,\phi_i)$
where $\phi_i\sim \phi_i+2\pi$ are angular coordinates along the orbits of the torus action,
and $y_i$ are the associated moment map coordinates.  
These may be considered as coordinates on $\mathtt{t}_n^*\cong\R^n$. The symplectic (K\"ahler) form is then 
\be
\omega = \sum_{i=1}^n \dd y_i \wedge \dd \phi_i~.
\ee
In this coordinate system, the metric degrees of freedom are therefore 
entirely encoded in the complex structure tensor $J$ -- see \cite{MSY} 
for further details.

In \cite{MSY} we considered the space of all smooth toric K\"ahler cone 
metrics on such an affine toric variety $X$. The space of metrics naturally factors into the 
space of Reeb vector fields, which live in the interior 
$\mathcal{C}_0$ of the dual done $\mathcal{C}\subset \mathtt{t}_n\cong\R^n$ to $\mathcal{C}^*$, 
and then an infinite dimensional space of transverse K\"ahler metrics. A 
general Reeb vector field may be written 
\be
\xi =\sum_{i=1}^n b_i \frac{\de}{\de \phi_i}
\ee
where $b\in \mathcal{C}_0\subset\R^n$. 
In this toric setting, the remaining degrees of freedom in the metric are 
described by the space of all 
homogeneous degree $1$ functions on $\mathcal{C}^*$ which are smooth 
up to the boundary (together with a convexity requirement). 

The main result of \cite{MSY} is that the 
Einstein--Hilbert action on $L$, restricted to this space of toric Sasakian metrics 
on $L$, reduces to the volume 
function vol$[L]:\mathcal{C}_0\rightarrow\R$, which depends only 
on the Reeb vector field in $\mathcal{C}_0$. Moreover, this is essentially 
just the Euclidean volume of the polytope formed by $\mathcal{C}^*$ and
the hyperplane $2(b,y)=1$. This depends only on $b$ and the toric 
data $\{v_a\}$.
In particular, for $n=3$, we have the formula 
\bea
V(b) \equiv \frac{\mathrm{vol}[L](b)}{\mathrm{vol}[S^5]} = 
\frac{1}{b_1}\sum_{a=1}^D \frac{(v_{a-1},v_a,v_{a+1})}{(b,v_{a-1},v_a)(b,v_a,v_{a+1})}
=\frac{1}{b_1}\sum_{a=1}^D \frac{(v_{a-1},v_a,v_{a+1})}{(b,u_a)(b,u_{a+1})}
\label{toricvolume}
\eea
for the normalised volume of $L$. The symbol $(\cdot,\cdot,\cdot)$ denotes a $3\times 3$  
determinant, while $(\cdot,\cdot)$ is the usual scalar product on $\R^n$ (or dual 
pairing between $\mathtt{t}_n$ and $\mathtt{t}_n^*$, whichever the reader 
prefers). The function $V(b)$ diverges to $+\infty$ at the boundary 
$\partial\mathcal{C}$ of $\mathcal{C}$ -- this is because 
the Reeb vector field develops a fixed point set in 
this limit, as will be explained in section \ref{sectiontwo}.

Once the critical Reeb vector field $b=b_*$ is obtained one can compute the volume
of the Sasaki--Einstein manifold, as well as the volumes of certain toric submanifolds, 
without explicit knowledge 
of the metric\footnote{In \cite{MSY} the issue of existence of this metric was not 
addressed. However, the real Monge--Amp\`ere equation derived in 
\cite{MSY} has recently been shown to always admit a solution 
\cite{Futakinew}, thus solving the existence problem for toric 
Sasaki--Einstein manifolds.}. 
From the explicit form of
the Einstein--Hilbert action, it follows that the ratios of these volumes 
to those of round spheres 
are in general algebraic numbers. 
This method of determining the critical Reeb vector field, and the corresponding volume, 
has been referred 
to as ``$Z$--minimisation'', where $Z$ is just the restriction 
of the function $V(b)$ to the hyperplane $b_1=n$. Indeed, the results 
of \cite{MSY}, for $n=3$, were interpreted as
the geometric ``dual'' 
of $a$--maximisation for the case that the Sasaki--Einstein 
manifolds, and hence the superconformal gauge theories, were toric. 

An analytic proof 
of the equivalence of these two optimisation problems was given in the work of
\cite{BZ}, modulo certain assumptions on the matter content of the field theory.
  The key point is that, following on from results in \cite{Franco:2005sm}, the trial $a$--function
$a_\mathrm{trial}$ for the gauge theory may be defined in closed form
in terms of the toric data, {\it i.e.} the normal vectors $\{v_a\}$, 
independently of the precise details of the gauge theory -- 
for example the form of the 
superpotential. This is {\it a priori} a function of $D-1$ variables, the trial R--charges, 
where $D$ is the number of facets of the cone. The global baryonic symmetries
are $U(1)_B^{D-3}$ \cite{Franco:2005sm}. Once one maximises $a_{\mathrm{trial}}$
over this space, one is left with a function of two variables which geometrically
are the components of the Reeb vector field.
Rather surprisingly, the functions
$a_\mathrm{trial}$ and $1/Z$ are then identically
 equal\footnote{This fact was also observed by two of 
us (D.M. and J.F.S.) in unpublished work.}. This of course explains why maximising
$a$ in the field theory is the same as minimising $Z$ in the geometry. Further work on the relation between $a$--maximisation and $Z$--minimisation 
has appeared in \cite{Tachikawa:2005tq,Barnes:2005bw,LR}.

\subsection{Outline}

The main result of the present work is to extend to 
general Sasaki--Einstein manifolds the toric results obtained in \cite{MSY}. 
This was initially a technical problem -- some of the methods described above 
simply do not extend when $X$ is not toric.
However, in the process of solving this problem, we will also gain 
further insight into the results of \cite{MSY}. 

We begin by fixing a complex manifold $X$, 
which is topologically a real cone over a compact manifold $L$. 
Thus $X_0=\mathbb{R}_+\times L$, where $r>0$ is a 
coordinate on $\R_+$, and $r=0$ is always an isolated singular point of 
$X$, unless $L$ 
is a sphere. For most of the paper it will be irrelevant whether we 
are referring to $X_0$ or the singular cone $X$. This is because 
we shall mainly 
be interested in the Sasakian geometry of the link $L$ -- the 
embedding into $X$ is then purely for convenience, since it is 
generally easier to work with the K\"ahler geometry of the cone.
%, as we may rephrase 
%everything in terms of conical K\"ahler geometry on $X$.
%paper we shall regard $X$ as being defined by $r>0$, thus ignoring 
%the singularity itself. This is because, on the whole, we shall mainly 
%be interested in the Sasakian geometry of the link $L$ -- the 
%embedding into $X$ is then purely for convenience, as we may rephrase 
%everything in terms of conical K\"ahler geometry on $X$. 
%An exception to this is when we 
%come to consider holomorphic functions on $X$, in section \ref{indexsection}. 
Since we are interested in Ricci--flat K\"ahler
metrics, we certainly require the canonical 
bundle of $X_0$ to be trivial. We implement this 
by assuming\footnote{$\Omega$ so defined is far from unique -- one is 
always free to multiply by a nowhere vanishing 
holomorphic function. This is an important difference to the case of 
compact Calabi--Yau manifolds. This degree of freedom can be fixed by 
imposing a ``homogeneous gauge'' for $\Omega$, as we discuss later.} we have a nowhere 
vanishing holomorphic $(n,0)$--form $\Omega$ on $X_0$. By 
definition the singularity 
$X$ is therefore 
\emph{Gorenstein}. 
%At this stage there is 
%no metric on $X$, but we do have a group $\mathrm{Aut}(X)$ of 
%holomorphic transformations of $X$ -- the Lie algebra of this group 
%is identified with the space of holomorphic vector fields on $X$, which is 
%infinite dimensional\footnote{Even for $X=\mathbb{C}^n$, 
%with $n\geq2$, this group is extremely complicated and not well--understood.}.

We also require there to be a 
space of K\"ahler cone metrics on $X$. The space of orbits of 
every homothetic vector field $\rdr$ in this space 
is required to be diffeomorphic to $L$.
The closure of the 
orbits of the corresponding Reeb vector field $\xi=J(\rdr)$  
defines some torus $\T^m\subset\mathrm{Aut}(X)$ since $\xi$ is 
holomorphic. Thus, as for the toric geometries above, we fix 
a (maximal) torus $\T^s\subset\mathrm{Aut}(X)$ and assume that it 
acts isometrically on our space of K\"ahler cone metrics on $X$. 
The Reeb vector fields in our space of metrics 
are all required to lie in the Lie algebra of this torus. Note that there is 
no loss of generality in making these assumptions: the Reeb vector 
field for a Sasaki--Einstein metric defines some torus that acts 
isometrically: by going ``off--shell'' and 
studying a space of \emph{Sasakian} metrics on which this torus 
(or a larger torus containing this) 
also acts isometrically, we shall learn rather a lot 
about Sasaki--Einstein manifolds, realised as critical points 
of the Einstein--Hilbert action on this space of metrics.

The first result is that the Einstein--Hilbert action $\mathcal{S}$ on $L$, 
restricted to the space of Sasakian metrics, is essentially 
just the volume functional $\vol[L]$
of $L$. More precisely, we prove that
\be\label{percy}
\mathcal{S} = 4(n-1)(1+\gamma-n)\vol[L]\ee
where one can show that, for any K\"ahler cone metric, there exists 
a gauge in which $\Omega$ is homogeneous degree $\gamma$ under 
$\rdr$, where $\gamma$ is unique. 
Given 
any homothetic vector field $\rdr$, $c\rdr$ is another homothetic 
vector field for a K\"ahler cone metric on $X$, 
where $c$ is any positive constant\footnote{This is a 
transverse homothety, in the language of Boyer and Galicki 
\cite{eta}.}. Setting to zero 
the variation of 
(\ref{percy}) in this direction gives $\gamma=n$, since 
$\vol[L]$ is homogeneous degree $-n$ under this scaling.
Thus we may think of $\mathcal{S}$ as the volume functional:
\be
\mathcal{S}=4(n-1)\vol[L]\ee
provided we consider only metrics for which $\Omega$ is homogeneous 
degree $n$ under $\rdr$. This condition is the generalisation 
of the constraint $b_1=n$ in the context of toric geometries \cite{MSY}.

The next result is that the 
volume of the link $L$ depends only on the 
Reeb vector field $\xi$, and not on the remaining degrees of 
freedom in the metric. The first and second derivatives of 
this volume function are computed in section \ref{varia}:
\bea\label{devies}
\diff\vol[L](Y)&=& -n\int_L \eta(Y)\diff\mu\nn\\
\diff^2\vol[L](Y,Z)& =&n(n+1)\int_L \eta(Y)\eta(Z)\diff\mu~.\eea
Here $Y,Z$ are holomorphic Killing vector fields in $\mathtt{t}_s$, the
Lie algebra of the torus $\T^s$, and $\eta(Y)$ denotes the contraction of $Y$ 
with
the one--form $\eta$, the latter being dual to the Reeb vector field. In particular, note that 
the second equation shows that $\vol[L]$ is strictly convex -- one 
may use this to argue uniqueness of critical points. 
We shall return to discuss the first equation in detail later. 
Note that, when the background $(L,g_L)$ is Sasaki--Einstein, the right 
hand sides of these 
formulae essentially appeared in \cite{Barnes:2005bw}. In this 
context these formulae arose from Kaluza--Klein reduction on 
AdS$_5\times L$. In particular, we see that the 
first derivative $\diff\vol[L](Y)$ 
is proportional to the coefficient $\tau_{RY}$ of a two--point 
function in the CFT, via the AdS/CFT correspondence. This relates
the geometric problem considered here to $\tau$--minimisation 
\cite{Barnes:2005bm} in the 
field theory.

Since the torus $\T^s$ acts isometrically on each metric, there is again a moment map and a fixed 
convex rational 
polyhedral cone $\mathcal{C}^*\subset\mathtt{t}^*_s$. Any Reeb 
vector field must then lie in the interior of the dual cone $\mathcal{C}$ 
to $\mathcal{C}^*$. The space of Reeb vector fields under which $\Omega$ has
charge $n$ form a convex polytope $\Sigma$ in $\mathcal{C}_0$ -- this is
formed by the hyperplane $b_1=n$ in the toric case \cite{MSY}. 
The boundary of $\mathcal{C}$ is a 
singular limit, since $\xi$ develops a fixed point set there. 
We again write the Reeb vector field as
\be
\xi = \sum_{i=1}^s b_i\frac{\partial}{\partial\phi_i}\ee
where $\partial/\partial\phi_i$ generate the torus action. 
The volume of the link  is then a \emph{function} 
\be\label{volL}
\vol[L]:\mathcal{C}_0\rightarrow\R~.\ee
At this point, the current set--up is not dissimilar to that
in our previous paper \cite{MSY} -- essentially the 
only difference is that the torus $\T^s$ 
no longer has maximal possible dimension $s=n$. The crucial point is 
that, for $s<n$, the volume function (\ref{volL}) is no longer given just 
in terms of the combinatorial data specifying $\mathcal{C}^*$. 
It should be clear, for example by simply restricting the toric case to a 
subtorus, that this data is insufficient to determine the volume 
as a function of $b$.

The key step to making progress in general is to write the
 volume functional of $L$ 
in the form 
\bea\label{DHmeasure}
\mathrm{vol}[L] & = & \frac{1}{2^{n-1} (n-1)!}\int_X 
e^{-r^2/2} \, e^{\omega}~.
\eea
The integrand in (\ref{DHmeasure}) may be interpreted 
as an equivariantly closed form, since $r^2/2$ is precisely 
the Hamiltonian function for the Reeb vector field $\xi$. 
The right hand side of (\ref{DHmeasure}) takes the form 
of the Duistermaat--Heckman formula \cite{DH1,DH2}.
This may then be 
\emph{localised} with respect to the Reeb vector field $\xi$. 
Our general formula is:
\be\label{genvolume}
V(b)\equiv\frac{\vol[L](b)}{\vol[S^{2n-1}]} = \sum_{\{F\}} \frac{1}{d_F}\int_F 
\prod_{\lambda=1}^R \frac{1}{(b,u_{\lambda})^{n_{\lambda}}} 
\left[\sum_{a\geq0} \frac{c_a(\mathcal{E}_{\lambda})}{(b,u_\lambda)^a}\right]^{-1} ~.\ee
Since $\xi$ vanishes only at the tip of the cone $r=0$, 
the right hand side of (\ref{genvolume}) requires one to resolve the singular
cone $X$ -- the left hand side is of course 
independent of the choice of resolution. 
This resolution can always be made, 
and \emph{any} equivariant (orbifold) resolution 
will suffice\footnote{We may in general resolve 
$X$, in an equivariant manner, by blowing up the Fano orbifold $V$ associated 
to any quasi--regular K\"ahler cone structure on $X$, as we shall explain later. 
It is interesting to note that, when constructing the gauge theory that lives 
on D3--branes probing the conical singularity, one also makes such a 
resolution. 
Specifically, an exceptional collection of sheaves  
on $V$ 
may then, in principle, be used to derive the gauge theory
(see \emph{e.g.} \cite{HK, Hanany:2006nm}).}. The first sum in (\ref{genvolume}) 
is over connected components
of the fixed point set of the $\T^s$ action on the resolved space $W$. 
The $u_{\lambda}$ are weights of the $\T^s$ action on the normal bundle to each 
connected component $F$ of fixed point set. These essentially 
enter into defining the moment cone $\mathcal{C}^*$. The 
$c_a(\mathcal{E}_{\lambda})$ are Chern classes of the normal bundle to $F$.
The term $d_F$ denotes the order of an orbifold structure group -- these terms are 
all equal to $1$ when the resolved space $W$ is completely smooth. 
Precise definitions will appear later in section \ref{localsection}. We also note 
that the right hand side 
of (\ref{genvolume}) is homogeneous degree $-n$ in $b$, precisely as in the toric case,
and is manifestly a rational function of $b$ with rational coefficients, 
since the weight vectors $u_\lambda$ and Chern classes $c_a(\mathcal{E}_\lambda)$ 
(and $1/d_F$) are generally \emph{rational}.
From this formula for the volume, which recall is essentially the 
Einstein--Hilbert action, it follows immediately that the 
volume of a Sasaki--Einstein 
manifold, relative to that of the round sphere, 
is an algebraic number. When the complex dimension $n=3$, 
this result is AdS/CFT 
``dual'' to the fact that the central charges of four dimensional 
superconformal field theories are indeed 
algebraic numbers. 

When $X$ is toric, that is the torus action 
is $n$--dimensional, the formula (\ref{genvolume}) simplifies and reduces to a sum over the fixed points 
of the torus action over any toric resolution of the K\"ahler cone:
\be\label{newtoricvolume}
V(b) = \sum_{p_A\in P} \prod_{i=1}^n\frac{1}{(b,u^A_i)}~.
\ee
Here $p_A$ are the vertices of the polytope $P$ of the resolved toric 
variety -- these are the very same vertices that enter into 
the topological vertex in topological string theory. The $u^A_i\in\Z^n$, 
$i=1,\ldots,n$, are the $n$ primitive edge vectors that describe the 
$A-$th vertex.
The right hand side of formula 
(\ref{newtoricvolume}) is of course necessarily independent of the choice of 
resolving polytope $P$,
in order that this formula makes sense.
It is a non--trivial fact that (\ref{newtoricvolume}) is equivalent to the previous toric formula 
for the volume (\ref{toricvolume}). For instance, the number of terms in 
the sum in (\ref{newtoricvolume})
 is given by the Euler number of any crepant resolution -- that is, the number of gauge groups of the dual 
gauge theory; while the number of terms in (\ref{toricvolume}) is $D$ -- that is, the number of facets of
the polyhedral cone. However, it can be shown that (\ref{newtoricvolume}) is 
finite
everywhere in the interior $\mathcal{C}_0$ of the polyhedral cone $\mathcal{C}$, and has simple poles at the 
facets of $\mathcal{C}$, precisely 
as the expression\footnote{This follows from the fact that the Duistermaat--Heckman formula
reduces to the \emph{characteristic function} \cite{Vinberg}
of the cone (see also \cite{oda}), as we will show later.} (\ref{toricvolume}).

These formulae show that the volumes of general Sasakian manifolds, as a 
function of the Reeb vector field, are \emph{topological}. For 
toric geometries, this topological data is captured by the 
normal vectors $v_a$ that define $\mathcal{C}^*$. For non--toric 
geometries, there are additional Chern classes that enter. 
In fact, we will also show that these formulae 
may be recovered from a particular limit of an equivariant 
index on $X$, which roughly counts holomorphic functions 
according to their charges under $\T^s$. Specifically, we define
\be
C(q,X) = \mathrm{Tr}\{q\mid\mathcal{H}^0(X)\}~.\ee
Here $q\in(\C^*)^s$ lives in (a subspace of) the algebraic torus associated to 
$\T^s$, and the notation denotes a trace of the induced action of 
this torus on the space of holomorphic functions on $X$\footnote{We will 
not worry too much about where this trace converges, as we are mainly 
interested in its behaviour near a certain pole.}. 
This equivariant index is clearly a holomorphic 
invariant, and the volume of the corresponding Sasakian link will turn out to appear 
as the coefficient of the leading divergent term of this index in 
a certain expansion:
\be\label{donald}
V(b) = \lim_{t\rightarrow 0} t^n\ C(q=\exp(-tb),X)~.\ee
The character $C(q,X)$ has a pole of order $n$ at $q_i=1$, $i=1,\ldots,n$, 
and this 
limit picks out the leading behaviour near this pole. 
For regular Sasaki--Einstein manifolds this relation, with critical 
$b=b_*$, was noted already in \cite{bergman} -- the essential difference 
here is that we interpret this relation as a \emph{function} of 
$b$, by using the equivariant index rather than just the index.
This result is again perhaps most easily described in the toric setting. In this case, the equivariant 
index counts holomorphic functions on $X$ weighted by their $U(1)^n$ charges.
It is known that these are in one--to--one correspondence with integral points inside the polyhedral 
cone, the $U(1)^n$ charges being precisely the location of the lattice points 
in $\mathcal{S}_{\mathcal{C}^*}=\Z^n\cap\mathcal{C}^*$. In a limit in which 
the lattice spacing tends to zero, the distribution of points gives 
an increasingly better approximation to a volume measure on the cone.
The slightly non--obvious point is that this measure in fact reduces to the measure on the
Sasakian link $L$, giving (\ref{donald}). 

From a physical viewpoint, the equivariant index\footnote{We would 
like to thank S. Benvenuti and A. Hanany for discussions on 
this.} 
is counting BPS mesonic operators
of the dual gauge theory, weighted by their $U(1)^s$ charges. This is because
the set of holomorphic functions on the Calabi--Yau cone correspond to elements
of the chiral ring in the dual guage theory. In 
\cite{Romelsberger:2005eg,maldaindex} some
indices counting BPS operators of superconformal field theories have been
introduced and studied. In contrast to the equivariant index defined here, 
those indices take into account
states with arbitrary spin. On the other hand, the fact that the index considered
here is equivariant means that it is twisted with respect to the global
flavour symmetries of the gauge theory. Moreover, the equivariant index is 
a holomorphic invariant, and may be computed without 
knowledge of the Kaluza--Klein 
spectrum. Rather interestingly, 
our results in section \ref{indexsection}
 may then be interpreted as saying that the
trial central charge of the dual gauge theory emerges as an asymptotic 
coefficient of the generating function of (scalar) BPS operators. It would be
very interesting to study in more detail 
the relation between these results and the work of
\cite{Romelsberger:2005eg,maldaindex}.

Let us now return to the expression for the first derivative of 
$\vol[L]$ in (\ref{devies}). This is zero for a Sasaki--Einstein manifold, 
since Sasaki--Einstein metrics are critical points of the Einstein--Hilbert
action. Moreover, for fixed Reeb vector field $\xi$, this derivative 
is independent of the metric. Thus, $\diff\vol[L]$ is a linear map on a 
space of holomorphic vector fields which is also a holomorphic invariant and 
vanishes identically when $\xi$ is the critical Reeb vector field for a 
Sasaki--Einstein metric. Those readers that are familiar with 
K\"ahler geometry will recognise these as properties of the 
\emph{Futaki invariant} in K\"ahler geometry \cite{Futaki}. Indeed, 
if $\xi$ is quasi--regular, we show that
\be\label{futboy}
\diff\vol[L](Y) = -\frac{\ell}{2}\cdot F[J_V(Y_V)]~.\ee
Here $Y_V$ is the push down of $Y$ to the Fano orbifold $V$, 
$J_V$ is its complex structure tensor, and $\ell$ is the length of 
the circle fibre. We will define the Futaki invariant $F$, and review 
some of its properties, in section \ref{futsection}. Thus 
the dynamical problem of finding the critical Reeb 
vector field can be understood as varying $\xi$ such that the transverse 
K\"ahler orbifold $V$, when $\xi$ is quasi--regular, has zero Futaki 
invariant. 
This is a well--known obstruction to the existence of a K\"ahler--Einstein metric
 on  $V$ \cite{Futaki}. This new interpretation of the Futaki invariant 
places the problem of finding K\"ahler--Einstein metrics on Fanos 
into a more general context. For example, the Futaki invariant of 
the first del Pezzo surface is well--known to be non--zero. 
It therefore cannot admit a K\"ahler--Einstein metric. However, 
the canonical circle bundle over this del Pezzo surface does 
admit a Sasaki--Einstein metric \cite{paper2, toric} -- the K\"ahler--Einstein
metric is only a transverse metric\footnote{In general, we may 
\emph{define} the Futaki invariant for transverse metrics by (\ref{futboy}).}.
From the point of view of our variational problem, there is nothing 
mysterious about this: the vector field that rotates the $S^1$ fibre of 
the circle bundle is simply 
not a critical point of the Einstein--Hilbert action.

This also leads to a result 
concerning the isometry group of Sasaki--Einstein
manifolds. In particular, we will argue\footnote{We will give a rigorous 
proof only for quasi--regular structures.} that the isometry group of a 
Sasaki--Einstein manifold is a
\emph{maximal} compact subgroup $K\supset\T^s$ of the 
holomorphic automorphism group Aut$(X)$ of the K\"ahler cone. 
The Reeb vector field then lies in the centre of the Lie 
algebra of $K$. This gives a rigorous account of the 
expectation that flavour symmetries of the field theory must be realised as isometries 
of the dual geometry, and that the R--symmetry does not mix with non--abelian
flavour symmetries.

Let us conclude this outline with an observation on the types of algebraic
numbers that arise from the volume minimisation problem studied here. 
We have shown that all Sasaki--Einstein manifolds have 
a normalised volume, relative to the round sphere, 
which is an algebraic number, that is the (real) root of a polynomial over the
rationals. Let us say that the \emph{degree}, denoted $\mathrm{deg}(L)$,
of a Sasaki--Einstein manifold $L$ is the degree of this algebraic number. 
Thus, for instance, if  $\mathrm{deg}(L)=2$ the normalised volume is quadratic
irrational. Recall that the \emph{rank} 
of a Sasaki--Einstein manifold is the dimension of the closure of the 
orbits of the Reeb vector field. We write this as $\mathrm{rank}(L)$. 
We then make the following \emph{conjecture}: for a Sasaki--Einstein 
manifold $L$ of dimension
$2n-1$, the degree and rank are related as follows
\be
\mathrm{deg}(L) = (n-1)^{\mathrm{rank}(L)-1}~.
\label{conje}
\ee
For example, all quasi--regular Sasaki--Einstein manifolds have degree one 
since the normalised volume is a rational number. By definition they 
also have rank one. In all the irregular cases that we have examined, 
which by now include a number of infinite families, this relation holds. 
Although we have obtained explicit expressions for 
the volume function $V(b)$, it seems to be a non--trivial fact that 
one obtains 
algebraic numbers of such low degree obeying (\ref{conje}) from extremising this function 
-- {\it a priori}, the degree would seem to be much larger.  It would be
interesting to investigate this further, and prove or disprove the conjecture.

%%%%%%%%%%%%%%%%%%%%%%%%%%%%%%%%%%%%%%%%%%%%%%%%%%%%%%%%%%%%%%%%%%%%%%

\section{Sasakian geometry}
\label{sectiontwo}

In this section we present a formulation of Sasakian geometry in terms 
of the geometry of K\"ahler cones. This way of formulating Sasakian geometry, 
although equivalent to the original description in terms of metric 
contact geometry, turns out to be more natural for describing the problems 
of interest here. 
A review of Sasakian geometry, where further details may be found, is contained 
in \cite{boyerreview}. 

A central fact we use here, that will be useful for later computations, is 
that the radial 
coordinate $r$ determines not only the link $L$ in $X$, but 
also that $r^2$ may be interpreted as the K\"ahler 
potential for the K\"ahler cone \cite{MSY}. A choice of 
Sasakian metric on $L$, for fixed complex structure 
$J$ on $X$, requires a choice of Reeb vector field
$\xi=J(\rdr)$ and a choice of \emph{transverse K\"ahler metric}. 
We also discuss moment maps for torus actions on the cone. 
By a result of \cite{Tomei}, the image of the moment map 
is a convex rational polyhedral cone $\mathcal{C}^*$ in the dual Lie algebra 
$\mathtt{t}_s^*$ of the torus, provided $\xi\in\mathtt{t}_s$. 
Here we identify elements of the Lie algebra with the corresponding 
vector fields on $X$ (or $L$). We show that the space of Reeb vector fields 
in $\mathtt{t}_s$ lies in the interior $\mathcal{C}_0$ of the dual cone to 
$\mathcal{C}^*$. Finally, we discuss the existence of certain 
Killing spinors on Sasakian manifolds and their relation to 
the Sasakian structure. We will make use of some of these 
formulae in later sections.

\subsection{K\"ahler cones}
\label{Kcones}

A Sasakian manifold is a compact Riemannian manifold 
$(L,g_L)$ whose metric cone
$(X,g_X)$ is K\"ahler. Specifically,
\be\label{cone}
g_X = \diff r^2 + r^2 g_L\ee
where $X_0=\{r>0\}$ is diffeomorphic to $\R_+\times L=C(L)$. 
We also typically take $L$ to be simply--connected. 

The K\"ahler condition 
on $(X,g_X)$ means that, by definition, the holonomy group of $(X,g_X)$ 
reduces to a subgroup of $U(n)$, where $n=\dim_{\C}X$. In particular, this 
means that there is a parallel complex structure $J$
\be
\nabla^X J=0~,\ee
where $\nabla^X$ is the Levi--Civita connection of $(X,g_X)$. We refer 
to the vector field 
$\rdr$ as the homothetic vector field. The Reeb vector 
field is defined\footnote{An alternative definition using spinors, 
perhaps more familiar to  physicists, will be given later in subsection
\ref{spinors}.} to be
\be
\xi=J\left(r\frac{\de}{\de r}\right)~.\ee
A straightforward calculation shows that $\rdr$ and $\xi$ are 
both holomorphic. Moreover, $\xi$ is Killing. A proof of these 
statements may be found in Appendix \ref{reebiskilling}.

Provided $L$ is not locally isometric to the round 
sphere\footnote{If $L$ is locally isometric to the round sphere 
then Killing vector fields on the cone $(X,g_X)$ may be constructed 
from solutions to Obata's equation \cite{obata}, which in turn relates to 
conformal Killing vector fields on the link $L$.}, Killing vector fields on the cone $X$ are in one--to--one correspondence with 
Killing vector fields on the link $L$. 
Since $L$ is compact, the group of isometries of $(L,g_L)$ is a compact 
Lie group. Since all 
holomorphic Killing vector fields on $(X,g_X)$ 
arise from Killing vector fields on the link $L$, they therefore 
commute with $\rdr$ and thus also 
commute with 
$\xi=J(\rdr)$. Since $\xi$ is itself Killing, it follows that 
$\xi$ lies in the centre of the Lie algebra of the isometry group.

We now define the 1--form on $X$
\be
\eta = J\left(\frac{\diff r}{r}\right) = \frac{1}{r^2}g_X(\xi,\cdot)~.\ee
This is the contact form of the Sasakian structure when pulled back to the link $L$ 
via the embedding 
\be
i:L\hookrightarrow X
\ee
that embeds $L$ in $X$ at $r=1$. Note that $\eta$ is homogeneous 
degree zero under $\rdr$. 
In terms of the 
$\diff^c$ and $\partial$ operator on $X$ we have
\be
\eta = J\diff \log r = \diff^c \log r = i(\bar{\partial}-\partial)\log r
\label{defeta}\ee
and thus
\be
\diff\eta = 2i{\partial}\bar{\partial}\log r~.\ee
We may now write the metric $g_X$ on $X$ as
\be
g_X = \diff r^2 + r^2\left(\eta\otimes\eta + g_T\right)\ee
where one can show that $g_T$ is a K\"ahler metric on the distribution 
orthogonal to the span of $\rdr$ and $\xi$.
The corresponding transverse K\"ahler form is easily computed to be
\be
\omegatr = \frac{1}{2}\diff\eta~.\ee
The K\"ahler form on $X$ is thus
\be\label{bob}
\omega = \omega_X = \frac{1}{2}\diff(r^2\eta)~.\ee
In particular $\omega$ is exact due to the homothetic symmetry generated 
by $\rdr$. We 
may rewrite (\ref{bob}) as
\be\label{omega}
\omega = \frac{1}{4}\diff\diff^c r^2 = \frac{1}{2}i{\partial}\bar{\partial}r^2~.\ee
The function $r^2$ thus serves a dual purpose: it defines the link $L=X\mid_{r=1}$ 
and is also the \emph{K\"ahler potential} that 
defines the metric.

Note from (\ref{omega}) that any holomorphic vector field $\chi$ which is tangent to 
$L$, $\diff r(\chi)=0$, is automatically Killing. Here we have used the notation 
$\alpha(\chi)$ for the pairing between a 1--form $\alpha$ and vector
 field $\chi$. Conversely, recall that all Killing vector fields on 
 $(X,g_X)$ are tangent to $L$. Thus for any holomorphic Killing vector
  field $Y$ we have\footnote{In this paper we will use extensively 
  the Lie derivative ${\cal L}_Y$ along vector fields $Y$. It is useful to
  recall the standard formula for Lie derivatives acting on a differential form
  $\alpha$: ${\cal L}_Y \alpha= \dd (Y\lrcorner \alpha)+Y \lrcorner \dd \alpha$. Note
  in components we have $(Y\lrcorner \alpha)_{\mu_1\dots \mu_{p-1}}\equiv
   Y^{\mu_p} \alpha_{\mu_p\mu_1\dots \mu_{p-1}}$.}
\be
{\cal L}_{Y}\eta = 0~. 
\label{etapreserved}
\ee

A Riemannian manifold $(X,g_X)$ is a cone if and only if the metric 
takes the form (\ref{cone}). We end this subsection by reformulating this in terms of the 
K\"ahler form $\omega$ on $X$ when $(X,J,g_X)$ is a K\"ahler cone. 
Thus, let $(X_0,J)$ be a complex manifold, with a diffeomorphism onto 
$\R_+\times L$ with $r>0$ a coordinate on $\R_+$. We require that 
$\rdr$ be holomorphic with respect to $J$. We may then simply define
$\omega$ in terms of $r$ by equation (\ref{omega}). One must also 
ensure that the corresponding metric $g=\omega(\cdot,J\cdot)$ is 
positive definite -- typically we shall use this formulation 
only for infinitesimal deformations around some fixed background 
Sasakian metric and thus this will not be an issue. Since $\rdr$ is 
holomorphic, and $\omega $ is clearly homogeneous of degree two 
under $\rdr$,  the metric $g$ 
will be also homogeneous degree two under $\rdr$. 
However, this is 
not sufficient for $g$ to be a cone -- one also requires 
$g(\rdr,Y)=0$ for all vector fields $Y$ tangent to $L$,
$\diff r(Y)=0$. Equivalently, $\omega(\xi,Y)=0$ where we 
define $\xi=J(\rdr)$. It is simple
to check that the necessary condition
\be
\mathcal{L}_{\xi}r = \diff r(\xi)=0\ee
that $\xi$ is tangent to $L$ is also sufficient for $g=\omega(\cdot,J\cdot)$ 
to be a cone. It follows now that 
$\partial/\partial r$ has unit norm and that the metric $g$ 
is a K\"ahler metric which is a cone of the form (\ref{cone}).

\subsection{The Calabi--Yau condition}

So far we have not fixed any Calabi--Yau condition on $(X,J)$. 
We are interested in finding a Ricci--flat K\"ahler metric on $X$, and thus 
we certainly require $c_1(X_0)=0$. We may impose this by  
assuming that there is a nowhere vanishing holomorphic 
section $\Omega$ of $\Lambda^{n,0}X_0$. In particular, $\Omega$ is then closed
\be\label{Omegaclosed}
\diff\Omega=0~.\ee
One can regard the $(n,0)$--form $\Omega$ as defining a reduction of the structure group 
of the tangent bundle of $X_0$ from $GL(2n;\R)$ to $SL(n;\C)$. The corresponding 
almost complex structure is then integrable if (\ref{Omegaclosed}) holds. 
The conical singularity $X$, including the isolated singular point $r=0$, 
is then by definition a \emph{Gorenstein singularity}.

On a compact manifold, such an $\Omega$ is always unique up to a 
constant multiple. However, when $X$ is non--compact, and in particular a K\"ahler cone, $\Omega$ so 
defined is certainly not unique -- one is free to multiply $\Omega$ by 
any nowhere zero holomorphic function on $X$, 
and this will also satisfy (\ref{Omegaclosed}). 
However, for a \emph{Ricci--flat} K\"ahler cone, with homothetic vector field $\rdr$, 
we may always choose\footnote{For the toric geometries studied in \cite{MSY}, 
this condition is equivalent to $b_1=n$, as can be seen by 
writing $\rdr =-\sum_{i=1}^n b_i J (\frac{\de}{\de \phi_i})$ and 
using the explicit form of $\Omega$ given in \cite{MSY}.}  $\Omega$ such that
\be\label{scale}
\mathcal{L}_{\rdr}\Omega=n\Omega~.\ee
In fact, one may construct this $\Omega$ as a bilinear in the covariantly 
constant spinor on $X$, as we recall in section \ref{spinors}. In section
\ref{homogeneous} we show that, for any fixed K\"ahler cone metric -- 
not necessarily Ricci--flat -- 
with homothetic vector field $\rdr$, 
one can always choose a gauge for $\Omega$ 
in which it is homogeneous degree $\gamma$ under $\rdr$, with $\gamma$ a unique 
constant. Then (\ref{scale}) will follow from varying the Einstein--Hilbert
action on the link $L$, as we show in section \ref{EHaction}. However, until 
section \ref{homogeneous}, we fix $(X,\Omega)$
together with a space of 
K\"ahler cone metrics on $X$ such that $\Omega$ satisfies (\ref{scale}) 
for every metric. This $\Omega$ is
then unique up to a constant multiple\footnote{To see this, pick a quasi--regular $\rdr$. Any other 
such holomorphic $(n,0)$--form is $\alpha\Omega$ where $\alpha$ is a 
nowhere zero holomorphic function on $X$. Since $\alpha$ is degree zero under $\rdr$, it descends to a holomorphic 
function on $V$, where $V$ is the space of orbits of $\xi$ on $L$. 
Since $V$ is compact, $\alpha$ is constant.}.
Given such an $\Omega$, for any K\"ahler form $\omega$ on $X$ there exists 
a real function $f$ on $X$ such that
\be\label{MA}
\frac{i^n}{2^n}(-1)^{n(n-1)/2}\Omega\wedge\bar{\Omega} = \exp(f)\frac{1}{n!}\omega^n~.\ee
A Ricci--flat K\"ahler metric with K\"ahler form 
$\omega$ of course has $f$ constant.

\subsection{The Reeb foliation}
\label{reebsection}

The vector field $\xi$ restricts to a unit 
norm Killing vector field on $L$, 
which by an abuse of notation we also denote by $\xi$. Since $\xi$ is 
nowhere--vanishing, 
its orbits define a 
foliation of $L$. There is then a classification of Sasakian 
structures according to the global properties of this foliation:

\begin{itemize}
\item{If all the orbits close, $\xi$ generates a circle action 
on $L$. If, moreover, the action is free the 
Sasakian manifold is said to be \emph{regular}. All the 
orbits have the same length, and $L$ is the total space 
of a principal circle bundle $\pi:L\rightarrow V$ over 
a K\"ahler 
manifold $V$. This inherits a
metric $g_V$ and K\"ahler form $\omega_V$, where $g_V$ is the 
push--down to $V$ of the transverse metric $g_T$.}

\item{More generally, if $\xi$ generates a $U(1)$ action on $L$, 
this action will be \emph{locally free}, but not free. The Sasakian 
manifold is then said to be \emph{quasi--regular}. Suppose that 
$x\in L$ is a point which has some non--trivial isotropy 
subgroup $\Gamma_x\subset U(1)$. Thus $\Gamma_x\cong\mathbb{Z}_m$ 
for some positive integer $m$. The length of the orbit through 
$x$ is then $1/m$ times the length of the generic orbit. The 
orbit space is naturally an orbifold, with $L$ being the 
total space of an orbifold circle bundle $\pi:L\rightarrow V$ 
over a K\"ahler orbifold $V$. Moreover, the point $x$ descends 
to a singular point of the orbifold with local orbifold structure group 
$\mathbb{Z}_m$.}

\item{If the generic orbit of $\xi$ does not close, 
the Sasakian manifold is said to be \emph{irregular}. In this case the 
generic orbits are diffeomorphic to the real line $\R$. Recall 
that the isometry group of $(L,g_L)$ 
is a compact Lie group. The orbits of a Killing vector field define a
one--parameter subgroup, the closure of which will always be an abelian 
subgroup and thus a torus. The dimension of the closure of the 
generic orbit
is called the \emph{rank} of the Sasakian metric, denoted 
$\mathrm{rank}(L,g_L)$. 
Thus irregular Sasakian metrics have $\mathrm{rank}>1$.}
\end{itemize}

A straightforward calculation gives
\be\label{bazza}
\mathrm{Ric}(g_X) = \mathrm{Ric}(g_L)-(2n-2)g_L = \mathrm{Ric}(g_T)-2ng_T\ee
and thus in particular
\be\label{rho}
\rho = \rho_T -2n\omega_T\ee
where $\rho_T$ denotes the transverse Ricci--form. 
We also have\footnote{Note that multiplying $\Omega$ by a nowhere zero 
holomorphic function 
$\alpha$ on $X$ leaves the right hand side of (\ref{Ricciboy}) invariant.}
\be\label{Ricciboy}
\rho = i\partial\bar{\partial}\log\|\Omega\|^2_{g_X}~,\ee
where we have defined $\|\Omega\|^2_{g_X}= \tfrac{1}{n!}\Omega \lrcorner \bar \Omega$. 
The Ricci--potential for the K\"ahler cone is thus $\log\|\Omega\|^2_{g_X}$. Since we assume that $\Omega$ is homogeneous degree $n$ under $\rdr$,
this is homogeneous degree zero {\it i.e.} it is independent of 
$r$, and hence is the pull--back under 
$p^*$ of a global function on $L$. Here
\be
p:X\rightarrow L\ee
projects points $(r,x)\in\R_+\times L$ onto $x\in L$.
Moreover, since $\mathcal{L}_{\xi}\Omega =
ni\Omega$ and $\xi$ is Killing it follows that the Ricci--potential is 
basic with respect to the foliation defined by $\xi$. 
Recall that a $p$--form 
$\alpha$ on $L$ is said to be \emph{basic} with respect to 
the foliation induced by $\xi$ if and only if
\be
\mathcal{L}_{\xi}\alpha = 0,\qquad \qquad \xi\lrcorner\alpha=0~.\ee
Thus $\alpha$ has no component along $g(\xi,\cdot)$ and is 
independent of $\xi$.
It is straightforward to check that the transverse 
K\"ahler form $\omegatr$, and its Ricci--form, are also basic.

Suppose now that $(L,g_L)$ is quasi--regular\footnote{We include the regular 
case when $V$ is a manifold in this terminology.}. Thus the space of 
orbits of $\xi$ is a compact complex orbifold $V$. The transverse 
K\"ahler and Ricci forms push down to 
$\omega_V$, and $\rho_V$ on $V$, respectively.
Thus (\ref{rho}) may be interpreted as an equation on $V$. The 
left hand side is $i\partial\bar{\partial}$ exact on the orbifold $V$, and hence
\be\label{harry}
[\rho_V] - 2n[\omega_V]=0\in H^2(V;\R)~.\ee
In particular this shows that $V$ is Fano, since 
$c_1(V)=[\rho_V/2\pi]$ is positive.

Note that $\eta$ satisfies $\diff\eta=2\pi^*(\omega_V)$. The K\"ahler 
class of $V$ is then proportional 
to the first Chern class of the orbifold circle bundle
$\pi:L\rightarrow V$. By definition, the orbifold is thus \emph{Hodge}. 
We denote the associated orbifold complex line bundle over $V$ by $\mathcal{L}$. 
Note from (\ref{harry}) that, since $[\omega_V]$ is proportional to the anti--canonical class 
$c_1(V)$ of $V$, the orbifold line bundle $\mathcal{L}$ is closely related 
to the canonical bundle $\mathcal{K}\rightarrow V$ over $V$. To see this 
more clearly, let $U\subset V$ denote a smooth open subset of $V$ 
over which $\mathcal{L}$ trivialises. 
We may then introduce a coordinate $\psi$ on the circle fibre of 
$\pi:L\mid_U\rightarrow U$ such that on 
$L\mid_U=\pi^{-1}(U)$ we have
\be
\eta = \diff\psi+\pi^*(\sigma)\ee
where $\sigma$ is a one--form on $U$ with $\diff\sigma=2\omega_V$. 
Note that, although this cannot be extended to a one--form on all of $V$,  
$\eta$ is globally defined on $L$ -- one may cover $V$ by open sets, 
and the $\sigma$ and $\psi$ are related on overlaps by opposite gauge 
transformations.
From equation (\ref{harry}), we see that 
if $\psi\sim\psi+2\pi/n$ 
then $X$ is the total space of the canonical complex cone over $V$ -- 
that is, the associated line bundle $\mathcal{L}$ to $\pi$ is 
$\mathcal{L}=\mathcal{K}$. 
More generally, we may set 
\be\label{minesalargeone}
\psi\sim\psi+\frac{2\pi\beta}{n};\qquad \qquad -c_1(\mathcal{L})=\frac{c_1(V)}{\beta}\in H^2_{\mathrm{orb}}(V;\Z)~.\ee
Then $\mathcal{L}^{\beta}\cong\mathcal{K}$. Here we have introduced 
the integral \emph{orbifold cohomology} $H^*_{\mathrm{orb}}(V;\Z)$ 
of Haefliger, 
such that orbifold line bundles are classified up to isomorphism by
\be
c_1(\mathcal{L})\in H^2_{\mathrm{orb}}(V;\Z)~.\ee
This reduces to the usual integral cohomology when $V$ is a manifold. 
The maximal integer 
$\beta$ in (\ref{minesalargeone}) is called the \emph{Fano index} of $V$. 
If $L$ is simply--connected, then $c_1(\mathcal{L})$ is primitive\footnote{That
is, there is no $\gamma\in H^2_{\mathrm{orb}}(V;\Z)$ and integer $m\in\Z$,
$|m|>1$, such that $m\gamma=c_1(\mathcal{L})$.} in 
$H^2_{\mathrm{orb}}(V;\Z)$ and $\beta$ is then equal to the index of 
$V$. For example, if 
$V=\mathbb{C}P^{n-1}$ then $H^2(V;\Z)\cong\Z$. The line bundle 
$\mathcal{L}$ with $c_1(\mathcal{L})=-1\in H^2(V;\Z)\cong\Z$ gives
$L=S^{2n-1}$, the complex cone being isomorphic to 
$\mathcal{L}=\mathcal{O}(-1)\rightarrow \mathbb{C}P^{n-1}$ with 
the zero section contracted. 
Note that the total space of $\mathcal{L}$ is the 
blow--up of $\mathbb{C}^n$ at the origin. 
On the other hand, the canonical bundle is $\mathcal{K}=\mathcal{O}(-n) 
\rightarrow\mathbb{C}P^{n-1}$ which gives the link $S^{2n-1}/\mathbb{Z}_n$. 
Thus the index is equal to $n$.

From the Kodaira--Bailey embedding theorem 
$(V,g_V)$, with the induced complex structure, is 
necessarily a \emph{normal projective algebraic variety} \cite{boyerreview}.
Let $\mathcal{T}\rightarrow V$ be the orbifold holomorphic line bundle over 
$V$ with 
first Chern class $c_1(V)/\mathrm{Ind}(V)$, with $\mathrm{Ind}(V)$ 
being the index of $V$. 
In particular, $\mathcal{T}$ is ample and has a primitive first 
Chern class. By the Kodaira--Bailey embedding theorem, for $k\in\mathbb{N}$ 
sufficiently large, $\mathcal{T}^k$ defines an embedding of $V$
into $\mathbb{C}P^{N-1}$ via its space of global holomorphic sections. Thus, 
a basis $s_{\alpha}$, $\alpha=1,\ldots,N$, 
of $H^0(V;\mathcal{T}^k)$ may be regarded as homogeneous coordinates 
on $\mathbb{C}P^{N-1}$, with $V\ni p\rightarrow [s_{1}(p),\ldots,s_{N}(p)]$ 
being an embedding. The image is 
a projective algebraic variety, and thus the zero locus of a 
set of homogeneous polynomials $\{f_A=0\}$ in the homogeneous coordinates.
If $H=\mathcal{O}(1)$ denotes the hyperplane 
bundle on $\mathbb{C}P^{N-1}$ then its pull--back to $V$ is of course 
isomorphic to $\mathcal{T}^k$. On the other hand, as described above, 
$\mathcal{L}^*\cong \mathcal{T}^{\mathrm{Ind}(V)/\beta}$ 
for some positive $\beta$, where recall that $\mathcal{T}^{\mathrm{Ind}(V)}
\cong \mathcal{K}^{-1}$ is the anti--canonical bundle. By 
taking the period $\beta=\mathrm{Ind}(V)/k$ in the above it follows that
the corresponding cone $X$ is the affine algebraic variety defined by 
$\{f_A=0\}\subset\C^{N}$. The maximal value of $\beta$, given by 
$\beta=\mathrm{Ind}(V)$, 
is then a $k$--fold cover of this $X$. 
Moreover, by our earlier assumptions, $X$ constitutes an 
isolated Gorenstein singularity.

We note that there is therefore a natural (orbifold) resolution 
of any $(X,J)$ equipped with a quasi--regular K\"ahler cone metric: 
one simply takes the total space 
of the orbifold complex line bundle $\mathcal{L}\rightarrow V$. 
The resulting space $W$ has at worst orbifold singularities, and 
$W\setminus V\cong X_0$ is a biholomorphism. Thus $W$ is birational to $X$. 
One might be able to resolve the cone $X$ completely, 
but the existence of a resolution with at worst orbifold singularities will 
be sufficient for our needs later.

\subsection{Transverse K\"ahler deformations}

In order to specify a Sasakian structure on $L$, one 
clearly needs to give the Reeb vector field $\xi$. By embedding 
$L$ as a link in a fixed non--compact $X$, this is 
equivalent to choosing a homothetic vector field $\rdr$ on $X$. 
Having determined this 
vector field, the remaining freedom in the 
choice of Sasakian metric consists of \emph{transverse K\"ahler deformations}. 

Suppose that we have two K\"ahler potentials $r^2$, 
$\tilde{r}^2$ on $X$ such that 
their respective homothetic vector fields coincide:
\be
\fracrdr = \tilde{r}\frac{\partial}{\partial \tilde{r}}~.\ee
This equation may be read as saying that $\tilde{r}$ is 
a homogeneous degree one function under $\rdr$, and thus
\be
\tilde{r}^2 = r^2\exp\phi\ee
for some homogeneous degree zero function $\phi$. Thus 
$\phi$ is a pull--back of a function, that we also call 
$\phi$, from the link $L$ under $p^*$. 
We compute
\be
\tilde{\eta} = \frac{1}{2}\diff^c\left(\log r^2 + \phi\right) = \eta + \frac{1}{2}\diff^c\phi~.\ee
In order that $\tilde{r}$ defines a metric cone, recall from the end of 
section \ref{Kcones} that we require
\be
\mathcal{L}_{\xi}\tilde{r}=0\ee
which implies that $\phi$ is basic with respect to the 
foliation induced by $\xi$,
$\mathcal{L}_{\xi}\phi=0$.
Introducing local transverse CR coordinates, {\it i.e.} local complex coordinates
on the transverse space, $(z_i,\bar{z}_i)$ on $L$, 
one thus has 
$\phi=\phi(z_i,\bar{z}_i)$. Note also that
\be\label{kahlerdist}
\diff\tilde{\eta} = \diff\eta + i\partial\bar{\partial}\phi\ee
so that the transverse K\"ahler forms $\tilde{\omegatr} = 
(1/2)\diff\tilde{\eta}$, 
$\omegatr = (1/2)\diff\eta$ differ precisely by a transverse K\"ahler deformation.

When the Sasakian structure is quasi--regular, $\phi$ pushes down to a global 
function on the orbifold $V$, 
and transformations of the metric of the form (\ref{kahlerdist}) are
precisely those preserving the K\"ahler class $[\omega_V]\in H^2(V;\R)$.

%%%%%%%%%%%%%%%%%%%%%%%%%%%%%%%%%%%%%%%%%%%%%%%%%%%%%%%%%%%%%%%%%%%%%%%%%%

\subsection{Moment maps}

In this subsection we consider a space of K\"ahler cone metrics on $X$ 
such that each metric has
 isometry group containing a torus $\T^s$. Moreover, the 
flow of the Reeb vector field is assumed to lie in this torus. For each metric, 
there is an associated moment map whose image is a convex 
rational polyhedral cone $\mathcal{C}^*\subset\mathtt{t}_s^*\cong\R^s$. 
Moreover, these cones are all isomorphic.
We show that the space of Reeb vector fields on $L$ is (contained in)
the interior $\mathcal{C}_0\subset\mathtt{t}_s$ of the \emph{dual cone} to
$\mathcal{C}^*$.

Suppose then that $\T^s$ acts holomorphically 
on the cone $X$, preserving a fixed choice of 
K\"ahler form (\ref{omega}) on $X$. Let $\mathtt{t}_s$ denote the Lie algebra 
of $\T^s$. We suppose that
$\xi\in\mathtt{t}_s$ -- the torus action is then said 
to be of \emph{Reeb type} \cite{boyertoric}. Let us 
introduce a basis $\partial/\partial\phi_i$ of 
vector fields generating the torus action, 
with $\phi_i\sim\phi_i+2\pi$. Then we may write
\be
\xi = \sum_{i=1}^s b_i\frac{\partial}{\partial\phi_i}~.\ee 
Since 
\be
\mathcal{L}_{\partial/\partial\phi_i}\omega=0\ee
and $X$ has $b_1(X_0)=0$, it follows that for each $i=1,\ldots,s$ there 
exists a function $y_i$ on $X$ such that
\be
\diff y_i = -\frac{\partial}{\partial\phi_i}\lrcorner\omega~.\ee
In fact, it is simple to verify that
\be\label{mome}
y_i = \frac{1}{2}r^2\eta\left(\frac{\partial}{\partial\phi_i}\right)\ee
is the homogeneous solution. The functions $y_i$ may be considered as coordinates on the 
dual Lie algebra $\mathtt{t}_s^*$. This is often referred to as 
the \emph{moment map}
\be
\mu:X\rightarrow \mathtt{t}_s^*\ee
where for $Y\in\mathtt{t}_s$ we have
\be\label{magic}
(Y,\mu)=\frac{1}{2}r^2\eta(Y)~.\ee
Under 
these conditions, \cite{Tomei} proved that
the image of $X$ under $\mu$ is a \emph{convex 
rational polyhedral cone} $\mathcal{C}^*\subset\mathtt{t}_s^*$. 
This is a convex polyhedral cone whose generators are all 
vectors whose components are rational numbers. 

The image of the link $L=X\mid_{r=1}$ is given by
\be\label{hp}
2(b,y)=1\ee
as follows by setting $Y=\xi$ in (\ref{magic}). The hyperplane
(\ref{hp}) intersects the cone $\mathcal{C}^*$ to form 
a compact polytope 
\be
\Delta(b)=\mathcal{C}^*\cap\{2(b,y)\leq 1\}\ee
if and only if $b$ lies in the interior of the 
\emph{dual cone} to $\mathcal{C}^*$, which 
we denote $\mathcal{C}\subset\mathtt{t}_s$. Note that
this analysis is essentially the same as that appearing 
in the toric context in our previous paper \cite{MSY}. 
The only difference is that the torus no longer has 
maximal possible rank $s=n$, and thus the cone need not be 
toric. However, 
the Euclidean volume of $\Delta(b)$ is no longer the 
volume of the Sasakian metric on $L$.

The cone $\mathcal{C}$ is a convex rational polyhedral cone
by Farkas' Theorem. Geometrically, the limit in which 
the Reeb vector field $\xi$ approaches the boundary $\partial \mathcal{C}$ of 
this cone is precisely the limit in which $\xi$ develops 
a non--trivial fixed point set on $X$. Recall that
$\xi$ has square norm $r^2$ on $X$ and thus in particular 
is nowhere vanishing on $X_0=\{r>0\}$. Thus the boundary of 
the cone $\mathcal{C}$ is a singular limit of the space
of Sasakian metrics on $L$. To see this, 
let $\mathcal{F}_{\alpha}$ denote the facets of $\mathcal{C}\subset\mathtt{t}_s$, and 
let the associated primitive inward pointing normal vectors be $u_{\alpha}\in\mathtt{t}_s^*$. 
The $u_{\alpha}$ are precisely the generating rays of the dual cone 
$\mathcal{C}^*$ to $\mathcal{C}$. Thus we may exhibit $\mathcal{C}^*\subset
\mathtt{t}^*_s\cong\R^s$ as
\be
\mathcal{C}^*=\left\{\sum_{\alpha} t_{\alpha} u_{\alpha}\in\mathtt{t}_s^*\mid 
t_{\alpha}\geq 0\right\}~.\ee
If $\xi\in\mathcal{F}_{\alpha}$, for some $\alpha$, then $(\xi,u_{\alpha})=0$. 
We may reinterpret this equation in terms of the moment cone 
$\mathcal{C}^*$. Let $\mathcal{R}_{\alpha}$ denote the 1--dimensional 
face, or ray, of $\mathcal{C}^*$ generated by the vector $u_{\alpha}$. 
The inverse image $X_{\alpha}=\mu^{-1}(\mathcal{R}_{\alpha})$ is a $\T^s$--invariant 
conic symplectic subspace of $X$ \cite{Tomei} and  
is a vanishing set for the vector 
field $\xi\in\mathcal{F}_{\alpha}\subset\mathtt{t}_s$. 
%To see this, let 
%$\partial/\partial\psi_j$, $j=1,\ldots,s-1$, be any basis for 
%the vector subspace of $\mathtt{t}_s$ containing the facet
%$\mathcal{F}_{\alpha}\subset\mathtt{t}_s$. We complete this to 
%a basis of $\mathtt{t}_s$ by choosing an additional vector 
%$\partial/\partial\psi_0$. We denote the components of 
%$\xi$ in this basis by $c_j$, $j=0,\ldots,s-1$. Thus
%\be
%\xi=c_0\frac{\partial}{\partial\psi_0}+\sum_{j=1}^{s-1}c_j\frac{\partial}{\partial\psi_j} 
%= c_0\frac{\partial}{\partial\psi_0}+c^{\prime}~.\ee
%The vector field $c^{\prime}$ is in the stabiliser group of 
%$X_{\alpha}$ and thus vanishes on $X_{\alpha}$. Thus on $X_{\alpha}$
%\be
%0=(\xi,u_{\alpha})=c_0\left(\frac{\partial}{\partial\psi_0},u_{\alpha}\right)~.\ee
%Since $(\partial/\partial\psi_0,u_{\alpha})$ cannot be zero\footnote{If it 
%were, $\{\partial/\partial\psi_j,j=0,\ldots,s-1\}$ would not 
%form a basis for $\mathtt{t}_s$.} we conclude that $c_0=0$ and 
%thus $\xi=0$ identically on $X_{\alpha}$.

It may help to give a simple example. Thus, let $X=\C^n$. Taking the flat metric 
gives
\be
\omega = \sum_{i=1}^n \frac{1}{2}\diff(\rho_i^2)\wedge\diff\phi_i\ee
where $(\rho_i,\phi_i)$ are polar coordinates on the $i$th complex plane 
of $\C^n$. We have 
$\mathcal{C}^*=(\mathbb{R}_+)^n$, with coordinates $y_i=\rho_i^2/2\geq 0$, 
which happens to be isomorphic to its dual cone, $\mathcal{C}=(\R_+)^n$. 
This orthant is bounded by $n$ hyperplanes, with primitive inward--pointing 
normals $u_i=e_i$, where $e_i$ is the $i$th standard orthonormal basis 
vector: $e_i^j=\delta_i^j$. The $\{u_i\}$ indeed also generate the 
moment cone $\mathcal{C}^*$. The inverse image under the moment map 
of the ray $\mathcal{R}_i$ generated by $u_i$ is the subspace 
\be
X_i = \{z_j=0\mid j\neq i\}\cong \C \subset \C^n~.\ee
Any vector field $\xi$ of the form
\be
\xi = \sum_{j\neq i} c_j\frac{\partial}{\partial\phi_j}\ee
clearly vanishes on $X_i$. 

%%%%%%%%%%%%%%%%%%%%%%%%%%%%%%%%%%%%%%%%%%%%%%%%%%%%%%%%%%%%%%%%%%%%%%%%%%%%%

\subsection{Killing spinors and the $(n,0)$--form}
\label{spinors}

We now discuss the existence of 
certain Killing spinors on Sasakian
manifolds, and their relation to the differential forms defining the Sasakian 
structure which
we have already introduced. The use of spinors often provides a quick and
 elegant method for obtaining various results, as we shall see in 
section \ref{futsection}. In
fact, these methods are perhaps more familiar to physicists. 

Recall that all K\"ahler  
manifolds admit a gauge covariantly constant spinor. More precisely, 
the spinor in question is in fact a section of a specific spin$^c$ bundle 
that is intrinsically defined on any K\"ahler manifold. We will not give 
a complete account of this in the following, but simply make a note
of the results that we will need in this paper, especially in section
\ref{futsection}. 
For further details, the reader might consult a number of standard 
references \cite{besse, lawson}. The spinors on a Sasakian manifold 
are induced from those on the K\"ahler cone, again in a rather standard 
way. This is treated, for Sasaki--Einstein manifolds, in the paper of 
B\"ar \cite{bar}. The extension to Sasakian manifolds is straightforward.

Let $(X,J,g)$ be a K\"ahler manifold. The bundle of complex spinors 
$\mathcal{S}$ does not necessarily exist globally on $X$, the 
canonical example 
being $\mathbb{C}P^2$. However, the spin$^c$ bundle
\be\label{spinc}
\mathcal{V} = \mathcal{S}\otimes \mathcal{K}_X^{-1/2}\ee
always exists. Here $\mathcal{K}_X$ denotes the canonical bundle 
of $X$, which is the (complex line) bundle $\Lambda^{n,0}(X)$ of forms of Hodge 
type $(n,0)$ with respect to $J$. The idea in (\ref{spinc}) is that, 
although neither bundle may exist separately due to $w_2(X)\in H^2(X;\Z_2)$ 
being non--zero, the obstructing cocycle cancels out in the tensor product and 
$\mathcal{V}$ exists as a genuine complex vector bundle. The metric on 
$X$ induces the usual spin connection on $\mathcal{S}$, and the 
canonical bundle $\mathcal{K}_X$ inherits a connection one--form with curvature 
$-\rho$, where $\rho$ is the Ricci--form on $X$. Thus the bundle $\mathcal{V}$ 
has defined on it a standard connection form. 

A key result is that, as a complex vector bundle, 
\be\label{iso}
\mathcal{V}\cong \Lambda^{0,*}(X)~.\ee
In fact, since $X$ is even dimensional, the spin$^c$ bundle 
$\mathcal{V}$ decomposes into spinors of positive and negative 
chirality, $\mathcal{V}=\mathcal{V}^+\oplus\mathcal{V}^-$ and
\be
\mathcal{V}^+ \cong \Lambda^{0,\mathrm{even}}(X), \qquad\qquad 
\mathcal{V}^- \cong \Lambda^{0,\mathrm{odd}}(X)~.\ee
The connection on $\mathcal{V}$ referred to above is then 
equal to the standard metric--induced 
connection on $\Lambda^{0,*}(X)$. In particular, there is 
always a covariantly constant section of $\Lambda^{0,*}(X)$ -- it 
is just the constant function on $X$. Via the isomorphism\footnote{This 
is essentially the same twisting of spinors that occurs on 
the worldvolumes of D--branes wrapping calibrated submanifolds.} 
$(\ref{iso})$ this gets interpreted as a gauge covariantly constant 
spinor $\Psi$ on $X$, the gauge connection being the standard metric--induced 
one  on $\mathcal{K}_X^{-1/2}$. 

We conclude then that there is always a spinor field\footnote{By an abuse of 
terminology we 
refer to sections of $\mathcal{V}$ as ``spinors''.} $\Psi$ on $X$ satisfying 
(in a local coordinate patch over which $\mathcal{V}$ trivialises)
\be\label{kahlerspinor}
\nabla^X_Y \Psi - \frac{i}{2} A(Y)\Psi = 0\ee
where $Y$ is any vector field on $X$ and $\nabla^X$ denotes the spin 
connection on $(X,g)$. 
The connection one--form $A/2$ on $\mathcal{K}_X^{-1/2}$ satisfies, as mentioned above, 
\be
\diff A = \rho~.\ee 

For our Calabi--Yau cone\footnote{Strictly, one should write $X_0$ 
in most of what follows.} $(X,\Omega)$, $\mathcal{K}_X$ is of course topologically 
trivial by definition. Thus, topologically, $\Psi$ is a genuine spinor field. 
However, unless the metric is Ricci--flat, the connection form $A$ will 
be non--zero. The reduction of $\Psi$ to a spinor on 
the link $L=X\mid_{r=1}$ is again rather standard \cite{bar}. 
Either spinor bundle $\mathcal{V}^+\mid_{r=1}$ or $\mathcal{V}^-\mid_{r=1}$ is isomorphic 
to the spinor (or rather spin$^{c}$) bundle on $L$. By writing 
out the spin connection on the cone in terms of that on the link $L$, 
one easily shows that there is a spinor $\theta=\Psi\mid_{r=1}$ on $L$ 
satisfying
\be\label{killing}
\nabla^L_Y\theta  -\frac{i}{2}Y\cdot\theta - 
\frac{i}{2}A (Y)\theta = 0 ~.\ee
Here $Y\cdot\theta$ denotes Clifford multiplication:
$Y\cdot\theta=Y^{\mu}\gamma_{\mu}\theta$,
and $\gamma_{\mu}$ generate the Clifford algebra $\mathrm{Cliff}(2n-1,0)$. 
Thus
$\{\gamma_{\mu},\gamma_{\nu}\}=2g_{L \ \mu\nu}$.
It is now simple to check 
from (\ref{killing}) that
$\theta$  has constant norm, and  we normalise it so that
$\bar{\theta}\theta = 1$.
Then the contact one--form $\eta$ on $L$ is given by the bilinear
\be
\eta = \bar{\theta}\gamma_{(1)}\theta~.\ee
Note that one may define the Reeb vector field as the dual of the contact
one--form $\eta$. Thus, in components, $\xi^\mu=g_L^{\mu\nu}\eta_\nu$. It is then
straightforward to check, using (\ref{killing}), that
$\nabla^L_{(\mu}\eta_{\nu)}=0$, so that $\xi$ is indeed a Killing vector field.
One also easily verifies, again using (\ref{killing}), that
\be
\diff\eta = -2i\bar{\theta}\gamma_{(2)}\theta = 2\omega_T~.\ee
We may define an $(n,0)$--form $K$ on $X$ 
as a bilinear in the spinor $\Psi$, namely
\be
K = \bar\Psi^c\gamma_{(n)}\Psi~.
\ee
It is important to note that this is \emph{different} from the 
holomorphic $(n,0)$--form $\Omega$ on $X$ we introduced
earlier. The two are of course necessarily proportional, 
and in fact are related by
\be\label{OmegaK}
\Omega = \exp{({f}/{2})} K~.
\ee
Here $f$ is the same function as that appearing in (\ref{MA}). Indeed,
we can write the Ricci--form on $X$ as 
$\rho = i{\partial}\bar{\partial}f$, 
so that we may take
\be
A = \frac{1}{2}\diff^c f~.\ee
Equivalently, we have
\be
f=\log\|\Omega\|^2_{g_X}~.\ee
From (\ref{kahlerspinor}) we have, as usual on a K\"ahler manifold,
\be
\diff K = iA\wedge K = \frac{i}{2}\diff^c f\wedge K~.\ee
Then
\be
\diff\Omega = \frac{1}{2}(\diff f + i\diff^c f)\wedge \Omega = \partial f\wedge \Omega 
=0~.\ee
In an orthonormal frame $e_{\mu}$, $\mu=1,\ldots,2n$, for $(X,g)$ note that
\bea
\omega & = & e_1\wedge e_2+\cdots+e_{2n-1}\wedge e_{2n}\nn \\
K & = & (e_1+ie_2)\wedge\cdots\wedge (e_{2n-1}+ie_{2n})\eea
so that
\be
\frac{i^n}{2^n}(-1)^{n(n-1)/2}K\wedge\bar{K} = \frac{1}{n!}\omega^n~.\ee
The relation (\ref{OmegaK}) is then consistent with the 
normalisation in (\ref{MA}).

Finally, we introduce the space of Reeb vector fields under which 
$\Omega$ has charge $n$. Thus, we define
\be
\Sigma=\left\{\xi\in\mathcal{C}_0\subset\mathtt{t}_s\mid \mathcal{L}_{\xi}\Omega=in
\Omega\right\}~.\ee 
Clearly, if $\xi^{\prime}\in\Sigma$ is fixed, any 
other $\xi\in\Sigma$ is given by $\xi=\xi^{\prime}+Y$ where $\Omega$ is 
uncharged with respect to $Y$:
\be\label{affineX}
\mathcal{L}_Y\Omega=0~.\ee
The space of all $Y$ satisfying (\ref{affineX}) forms a 
vector subspace of $\mathtt{t}_s$. Moreover, this subspace has 
codimension one, so that the corresponding plane through 
$\xi^{\prime}$ forms a finite polytope
with $\mathcal{C}$. Thus 
$\Sigma$ is an $(r-1)$--dimensional polytope. In the toric 
language of \cite{MSY}, this is just the intersection of the 
plane $b_1=n$ with Reeb polytope $\mathcal{C}$.

\subsection{The homogeneous gauge for $\Omega$}
\label{homogeneous}

Suppose that $X_0=\R_+\times L$ is a complex manifold admitting a 
nowhere vanishing holomorphic section $\Omega^{\prime}$ of $\Lambda^{n,0}(X_0)$. 
Recall that, in contrast to the case of compact $X$, 
$\Omega^{\prime}$ is unique only up to multiplying by a 
nowhere vanishing holomorphic function. 
Suppose moreover that $g_X$ is a quasi--regular 
K\"ahler cone metric on $X$, with homothetic vector field $\rdr$. 
The aim in this section is to prove that there always exists a 
``gauge'' in which the holomorphic $(n,0)$--form is homogeneous of constant 
degree $\gamma\in\R$ under $\rdr$, where $\gamma$ is unique. We shall use 
this result in section \ref{EHaction} to argue that $\gamma=n$ arises 
by varying the Einstein--Hilbert action of the link. 

Since $\rdr$ is a holomorphic vector field
\be
\mathcal{L}_{\rdr}\Omega^{\prime} = \kappa\Omega^{\prime}\ee
where $\kappa$ is a holomorphic function. 
We must then find a nowhere zero holomorphic 
function $\alpha$ and a constant $\gamma$ such that
\be\label{equationoftheday}
\mathcal{L}_{\rdr}\log \alpha = \gamma-\kappa\ee
since then $\Omega\equiv\alpha\Omega^{\prime}$ is homogeneous degree $\gamma$. 
Let us expand
\be
\gamma-\kappa = \sum_{k\geq 0} a_k\ee
where $a_k$ are holomorphic functions of weight $k$ under the 
$U(1)$ action generated by $\xi=J(\rdr)$. Roughly, we are 
doing a Taylor expansion in the fibre of $\mathcal{L}^*\rightarrow V$, 
where $V$ is the K\"ahler orbifold base defined by the quasi--regular 
Reeb vector field $\xi$. Here $\mathcal{L}$ is the associated 
complex line orbifold bundle to $U(1)\hookrightarrow L\rightarrow V$. 
The $a_k$, in their
$V$ dependence, may be considered as sections of $(\mathcal{L}^*)^k
\rightarrow V$. If $\de/\de\nu$, $\nu\sim\nu+2\pi$, rotates 
the fibre of $L$ with weight \emph{one}, then we have
\be
\frac{\de}{\de\nu} = h\xi\ee
for some positive constant $h$. Thus
\be
\mathcal{L}_{\xi}a_k = i\frac{k}{h}a_k~.\ee
The equation (\ref{equationoftheday}) is then straightforward to solve:
\be\label{alphasolution}
\log\alpha = \sum_{k\geq1} \frac{h}{k}a_k + \log \delta~.\ee
Here we have used the fact that each $a_k$ is holomorphic. 
Moreover, $\delta$ is holomorphic of homogeneous degree $a_0$ where
\be
a_0 = \gamma - \kappa_0\ee
and the constant $\kappa_0$ is the degree zero part of $\kappa$. 
In order that $\alpha$ be nowhere vanishing, we now require 
$a_0=0$. This is because $\delta$ is homogeneous of fixed degree, and 
thus corresponds to a section of $(\mathcal{L}^*)^m\rightarrow V$, 
where $a_0=m/h$. However, since $(\mathcal{L}^*)^m$ is a non--trivial 
bundle for $m\neq 0$, any section must vanish somewhere, unless $m=0$. 
Thus $a_0=0$, which in fact fixes $\gamma$ uniquely because of 
the latter argument. Finally, the resulting expression (\ref{alphasolution}) for $\alpha$ 
is clearly nowhere vanishing, holomorphic, and satisfies (\ref{equationoftheday}). This completes the proof.

%%%%%%%%%%%%%%%%%%%%%%%%%%%%%%%%%%%%%%%%%%%%%%%%%%%%%%%%%%%%%%%%%%%%%%%%

\section{The variational problem}
\label{varia}

In this section we show that the Einstein--Hilbert action on $L$, 
restricted to the space of Sasakian metrics on $L$, is 
essentially the \emph{volume functional} of $L$. Moreover, 
the volume depends only on the choice of Reeb vector field, 
and not on the remaining degrees of freedom in the metric. 
We give general formulae for the first and second variations, 
in particular showing that the volume of $L$ 
is a strictly convex function. The derivations of these formulae, 
which are straightforward but rather technical,
are relegated to Appendix \ref{variations}. The first variation will be related 
to the Futaki invariant for quasi--regular Sasakian metrics in 
section \ref{futsection}. From the second variation formula it follows 
that there is a unique critical point of the Einstein--Hilbert 
action in a given Reeb cone $\Sigma\subset\mathcal{C}_0\subset\mathtt{t}_s$.

\subsection{The Einstein--Hilbert action}
\label{EHaction}

As is well--known, a metric $g_L$ on $L$ satisfying the Einstein equation
\be
\mathrm{Ric}(g_L) = (2n-2)g_L\ee
is a critical point of the Einstein--Hilbert action
\be
\mathcal{S}:\mathrm{Met}(L)\rightarrow\R\ee
given by
\be
\mathcal{S}[g_L] = \int_L \left[s(g_L)+2(n-1)(3-2n)\right]\diff\mu\ee
where $s(g_L)$ is the scalar curvature of $g_L$, and 
$\diff\mu$ is the associated Riemannian measure. 
We would like to restrict $\mathcal{S}$ to a space of 
\emph{Sasakian} metrics on $L$. Recall that we require 
$X_0=\R_+\times L$ to be Calabi--Yau, meaning that there is a nowhere vanishing 
holomorphic $(n,0)$--form $\Omega$. At this stage, we are not assuming that
$\Omega$ obeys 
any additional property. The Ricci--form on $X$ is then
$\rho = i\partial\bar{\partial}f$ where
\be
f = \log \|\Omega\|^2_{g_X}~.\ee
Note that of course 
the Ricci--form is independent of multiplying $\Omega$ by any 
nowhere zero holomorphic function on $X$. The scalar curvature of $(X,g_X)$ is 
\be
s(g_X)=\mathrm{Tr}\left(g_X^{-1}\mathrm{Ric}(g_X)\right) 
=-\Delta_X f \ee
where $\Delta_X$ is the Laplacian on $(X,g_X)$. Using the relation
\be
s(g_X) = \frac{1}{r^2}\left[s(g_L)-2(n-1)(2n-1)\right]\ee
one easily sees that
\be
\mathcal{S}[g_L] = 2(n-1)\left(R + 2\vol[L]\right)\ee
where $R$ is defined by 
\be
R = \int_{r\leq 1} s(g_X)\ \frac{\omega^n}{n!}=-\int_{r\leq 1} 
\Delta_X f\ \frac{\omega^n}{n!}~.\ee
Note that this is independent of the gauge choice for $\Omega$, {\it i.e.} 
it is independent of the choice of nowhere zero holomorphic multiple. 
However, in order to relate $R$ to an expression on the link, it is useful to 
impose a homogeneity property on $\Omega$. This can always be done, as 
we showed in section \ref{homogeneous}. Strictly speaking, 
we only proved this for \emph{quasi--regular} Sasakian metrics. 
However, since the rationals are dense in the reals and $\mathcal{S}$ 
is continuous, this will in fact be sufficient. Thus, let $\Omega$ be 
homogeneous degree $\gamma$ and $\rdr$ be quasi--regular.  
It follows that $f$ satisfies
\be
\mathcal{L}_{\rdr} f = 2(\gamma-n)~.\ee
Recall that on a cone 
\be
\Delta_X = \frac{1}{r^2}\Delta_{L}-\frac{1}{r^{2n-1}}\frac{\partial}{\partial r}
\left(r^{2n-1}\frac{\partial}{\partial r}\right)\ee
where $\Delta_{L}$ is the Laplacian on $(L,g_L)$. We therefore have
\be
R = 2(\gamma-n)\vol[L]-\int_{r\leq 1}\Delta_{L} f\ r^{2n-3}\diff r\wedge \diff\mu~.\ee
The integrand in the second term is now homogeneous under $\rdr$, so we may 
perform the $r$ integration trivially. Using Stokes' theorem 
on $L$ we conclude that this term is identically zero. Thus we find that
\be
\mathcal{S}[g_L] = 4(n-1)(1+\gamma-n)\vol[L]~.
\label{Zvol}\ee
Thus the Einstein--Hilbert action on $L$ is related simply to the 
volume functional on $L$. Given any homothetic vector field $\rdr$, 
$c\rdr$ is always\footnote{That is, the space of homothetic vector fields 
is itself a cone.} another homothetic vector field for a K\"ahler cone 
metric on $X$, where $c$ is a positive constant. Since $\vol[L]$ is 
homogeneous degree $-n$ under this scaling, one may immediately extremise 
(\ref{Zvol}) in this direction, obtaining 
\be
\gamma=n~.\ee
Note that this is precisely analogous to the argument in our previous 
paper \cite{MSY}, which sets $b_1=n$, in the notation there. 
Thus, provided we restrict to K\"ahler cone metrics for which $\gamma=n$, 
{\it i.e.} there is a nowhere vanishing
holomorphic $(n,0)$--form of homogeneous degree $n$, 
the Einstein--Hilbert action is just the volume functional for the 
link:
\be
\mathcal{S}[g_L]=4(n-1)\vol[L]~.\ee
This reduces our problem to studying the volume of the link in 
the remainder of the paper.

We next show that $\mathcal{S}$ 
is independent of the choice of transverse K\"ahler metric. Hence 
$\mathcal{S}$ is a function on the space of 
Reeb vector fields, or equivalently, of homothetic vector fields. Thus, consider
\be\vol[L]:\mathrm{Sas}(L)\rightarrow \R~.\ee
We may write the volume as
\be
\vol[L]=\int_L\diff\mu = 2n\ \vol[X_1] = 2n\int_{r\leq 1}\frac{\omega^n}{n!}\ee
where we define $X_1=X\mid_{r\leq 1}$. Here we have simply written 
the measure on $X$ in polar coordinates. Note that, since we are regarding 
$r^2$ as the K\"ahler potential, changing the metric also changes the definition of 
$X_1$. Let us now fix a background with K\"ahler potential $r^2$ and set
\be
r^2(t)=r^2\exp(t\phi)\ee
where $t$ is a (small) parameter and $\phi$ is a basic function on $L=X\mid_{r=1}$. Thus
\be
\frac{\diff\omega}{\diff t}(t=0) = \frac{1}{4}\diff\diff^c (r^2\phi)~.\ee
To first order in $t$, the hypersurface $r(t)=1$ is given by
\be
r=1-\frac{1}{2}t\phi~.\ee
We now write
\be
\vol[L] = \int_{r\leq 1} \diff(r^{2n})\wedge\diff\mu~.\ee
The first order variation in $\vol[L](t)$ from that of $t=0$ contains 
a contribution from the domain of integration, as well as from 
the integrand. The former is slightly more subtle. 
Consider the expression
\be\label{domainchange}
\int_{r\leq 1-\tfrac{1}{2}t\phi} \diff(r^{2n})\wedge \diff\mu\ee
which is $\vol[L]$ together with the first order variation 
due to the change of integration domain. By performing the $r$ integration 
pointwise over the link $L$ we obtain, to first order in $t$,
\be
\vol[L] - t\ n\int_L \phi \diff\mu~.\label{domainchange2}\ee
The total derivative of $\vol[L]$ at $t=0$ is thus
\be\label{joe}
\frac{\diff\vol[L]}{\diff t}(t=0) = 
-n\int_L \phi\diff\mu + \frac{n}{2}\int_{r\leq 1} 
\diff\diff^c(r^2\phi)\wedge\frac{\omega^{n-1}}{(n-1)!}\ee
where the second term arises by varying the Liouville 
measure $\omega^n/n!$. We may now apply Stokes' theorem to the second 
term on the right hand side of (\ref{joe}). Using 
\be
i^*\omega = \omegatr\ee
together with the equation $\diff^c r^2 = 2r^2\eta$ (see eq. (\ref{defeta})), 
we obtain
\be
\frac{\diff\vol[L]}{\diff t}(t=0) = -n\int_L\phi\diff\mu 
+ n\int_L\phi \eta\wedge \frac{\omegatr^{n-1}}{(n-1)!}~.\ee
Notice that the term involving $\diff^c\phi$ does not 
contribute, since $\xi$ contracted into the integrand is 
identically zero. Indeed, we have
\be
\xi\lrcorner \diff^c\phi = \mathcal{L}_{\rdr}\phi = 0\ee
and $\omegatr=(1/2)\diff\eta$ is basic. Noting that
\be
\diff\mu = \eta\wedge\frac{\omegatr^{n-1}}{(n-1)!}\ee
we have thus shown that
\be
\frac{\diff\vol[L]}{\diff t}(t=0)=0\ee
identically for all transverse K\"ahler deformations\footnote{Note 
that we didn't use the equation $\mathcal{L}_{\xi}\phi=0$ 
anywhere. Any deformation $\phi$ of the metric not satisfying this equation 
will preserve the homothetic scaling of the metric on $X$, but 
it will no longer be a cone.}. It follows 
that $\vol[L]$ may be interpreted as a \emph{function} 
\be
\vol[L]:\mathcal{C}_0\rightarrow\R~.\ee
Our task in the remainder of this paper is to understand the 
properties of this function.

\subsection{Varying the Reeb vector field}

In the previous subsection we saw that 
$\vol[L]$ may be regarded as a function on the space 
of Reeb vector fields, since the volume is 
independent of transverse K\"ahler deformations. 
We now fix a 
maximal torus $\T^s\subset\mathrm{Aut}(X)$, acting by 
isometries on each metric in a space of Sasakian metrics, and consider the 
properties of the functional $\vol[L]$ on this space 
as we vary the Reeb vector field. In particular, in the remainder
of this section we give formulae for the first and second variations. 

We would like to differentiate the function $\vol[L]$. Thus, 
we fix an arbitrary background K\"ahler cone metric, 
with K\"ahler potential $r^2$, and linearise 
the deformation equations around this. We set
\bea
\xi(t) & = & \xi+tY\\
r^2(t) & = & r^2(1+ t\phi)\eea
where $t$ is a (small) parameter and $Y\in\mathtt{t}_s$ is holomorphic 
and Killing. 
{\it A priori}, $\phi$ is any 
function on $X$. Working to first order in $t$, the calculation goes 
much as in the last subsection. The details may be found 
in Appendix \ref{variations}. We obtain
\be\label{firstder}
\diff\vol[L](Y)  = -n\int_L \eta(Y)\diff\mu~.\ee
This is our general result for the first derivative of $\vol[L]$. 
As one can see, it manifestly depends only on 
the direction $Y$ in which we deform the Reeb vector field, 
and not on the function $\phi$, in accord with the previous section.

Note that  
the integrand in (\ref{firstder}) is twice the 
Hamiltonian function for $Y$. Indeed, in this case, 
$\diff r(Y)=0$ and hence $\mathcal{L}_Y\omega=0$. 
Since $X$ necessarily 
has $b_1(X_0)=0$, there is therefore a function $y_Y$ such that
\be
\diff y_Y = -\frac{1}{4}Y\lrcorner \diff\diff^c r^2~.\ee
Using 
\be
Y \lrcorner \omega = -\frac{1}{2} \left[ \eta (Y) \diff r^2 - r^2 Y \lrcorner \diff \eta \right]
\ee
and (\ref{etapreserved}), one finds that the homogeneous solution to this equation is
\be\label{moment}
y_Y = \tfrac{1}{2}r^2\eta(Y)~.\ee
The Hamiltonian $y_Y$ is then homogeneous degree two under $\rdr$. Substituting this 
into (\ref{firstder}) we recover the toric formula (3.18) in \cite{MSY}, 
obtained in a completely different way using convex polytopes.

In order to compute the second variation, we note that, since 
$Y$ is Killing, it commutes with $\rdr$: $[Y,\rdr]=0$. 
As 
a result, $\eta(Y)$ is independent of $r$. 
Using this property, we may write
\be\label{firstd}
\diff\vol[L](Y) = -n(n+1)\int_{r\leq1}(\diff^c r^2)(Y)\frac{\omega^n}{n!}\ee
where recall that $\diff^c r^2=2r^2\eta$. We now differentiate again to 
obtain
\be\label{secondder}
\diff^2\vol[L](Y,Z)  = n(n+1)\int_L \eta(Y)\eta(Z)\diff\mu~.\ee
This is a general form for the second variation of the volume 
of a Sasakian manifold. The derivation, which is a little lengthy, 
is contained in Appendix \ref{variations}. Note that (\ref{secondder}) 
is manifestly positive definite, and hence the volume is a 
strictly convex function. Again, for toric geometries, this reduces to a 
formula in \cite{MSY}.

\subsection{Uniqueness of critical points}

Using this last result, we may prove uniqueness of a critical 
point rather simply. We regard the volume as a function
\be
\vol[L]:\mathcal{C}_0\rightarrow\R~.\ee
The Reeb vector field for a Sasaki--Einstein metric 
is a critical point of $\mathrm{vol}[L]$, restricted to the 
subspace $\Sigma$ for which the holomorphic 
$(n,0)$--form has charge $n$. This defines a 
compact convex polytope $\Sigma\subset\mathtt{t}_s$, and 
we may hence regard the Einstein--Hilbert action as a 
function\be
\mathcal{S}:\Sigma\rightarrow\R~.\ee
Since we have just shown that $\mathcal{S}$ is 
strictly convex on this space, and $\Sigma$ is 
itself convex, standard convexity arguments 
show that $\mathcal{S}$ has a unique critical point.
Thus, assuming a Sasaki--Einstein metric exists in our space 
of Sasakian metrics on $L$, 
its Reeb vector field is unique in $\Sigma\subset\mathtt{t}_s$.

%%%%%%%%%%%%%%%%%%%%%%%%%%%%%%%%%%%%%%%%%%%%%%%%%%%%%%%%%%%%%%%%%%%%%%%%%%%%%%%%%%%%%%%%%

\section{The Futaki invariant}
\label{futsection}

In this section, we consider a fixed background Sasakian metric which
is \emph{quasi--regular}. We 
show that the first derivative 
$\diff\vol[L]$, as a linear function on the Lie algebra $\mathtt{t}_s$, 
is closely related 
to the Futaki invariant of $V$. This is a well--known \cite{Futaki}
obstruction to the existence of a K\"ahler--Einstein metric on $V$. 
Using this relation, 
together with Matsushima's theorem \cite{Matsushima}, we show that the 
group of holomorphic isometries of 
a quasi--regular Sasaki--Einstein metric on $L$ is a maximal 
compact subgroup of $\mathrm{Aut}(X)$. We conjecture 
this to be true also for the more generic irregular case.

\subsection{Brief review of the Futaki invariant}

Let $(V,J_V,g_V)$ be a K\"ahler orbifold\footnote{One usually 
works with manifolds. Passing to the larger orbifold category 
involves no essential differences.} with K\"ahler form $\omega_V$ 
and corresponding Ricci--form $\rho_V$ such that 
\be
[\rho_V]=\lambda[\omega_V]\in H^2(V;\R)\ee
where $\lambda$ is a real positive constant. The value of $\lambda$ is
irrelevant since one can always rescale the metric, leaving $\rho_V$ invariant\footnote{
Nevertheless, the value $\lambda=2n$ is that relevant for K\"ahler cones 
in complex dimension $n$.}. By the global $i{\partial}\bar{\partial}$--lemma, 
there is a globally defined smooth function $f=f_{g_V}$ such that
\be
\rho_V - \lambda\omega_V = i\partial\bar{\partial}f\ee
where, throughout this section, the $\bar{\partial}$ operator is 
that defined on $V$. Note that this is the same $f$ as that appearing 
in section \ref{sectiontwo}. 
%Indeed,
%\be
%\rho = i\partial\bar{\partial}f\ee
%where $\rho$ is the Ricci--form of the complex K\"ahler cone over $V$.
We now define a linear map 
\be
F:\mathrm{aut}_{\R}(V)\rightarrow \R\ee
from the Lie algebra of real holomorphic vector fields $\mathrm{aut}_{\R}(V)$ 
on $V$ to the real numbers by assigning to each holomorphic 
vector field $\zeta$ on $V$ the number
\be\label{futaki}
F[\zeta]=\int_V \left(\mathcal{L}_{\zeta} f\right)\ \frac{\omega_V^{n-1}}{(n-1)!}~.\ee
It is more natural, and more standard, to define $F$ on the space of 
\emph{complex} holomorphic vector fields:
\be
F_{\C}:\mathrm{aut}(V)\rightarrow\C\ee
by simply complexifying (\ref{futaki}).

The functional $F$, introduced by Futaki in 1983 \cite{Futaki}, 
has the following rather striking properties:
\begin{itemize}

\item $F$ is independent of the choice of K\"ahler metric representing 
$[\omega_V]$. That is, it is invariant under K\"ahler deformations. In this sense
it is a topological invariant.

\item $F_{\C}$ is a Lie algebra homomorphism.

\item If $V$ admits a K\"ahler--Einstein metric, so that for 
some $g_V$ the function $f$ is constant, the Futaki invariant 
$F$ clearly vanishes identically.

\end{itemize}

Because of the second item, Calabi named $F_{\C}$ the \emph{Futaki character} 
\cite{Calabi} -- it is a character because $F_{\C}$ is a homomorphism 
onto the complex numbers.
Let $G=\mathrm{Aut}(V)$ denote the group of holomorphic automorphisms of 
$V$, and $\mathtt{g}$ its Lie algebra. Thus $F_{\C}:\mathtt{g}\rightarrow\C$. 
Mabuchi \cite{Mabuchi2} proved that the 
nilpotent radical of $\mathtt{g}$ lies in $\ker F_{\C}$. Thus if
\be
\mathtt{g} = \mathtt{h}\oplus \mathrm{Lie}(\mathrm{Rad}(G))
\ee
denotes the Levi decomposition of $\mathtt{g}$, it follows
that $F_{\C}$ is completely determined by its restriction to the 
maximal reductive algebra $\mathtt{h}$. Since $F_{\C}$ is 
a Lie algebra character of $\mathtt{h}$, it vanishes on 
the derived algebra $[\mathtt{h},\mathtt{h}]$, and is therefore determined 
by its restriction to the Lie algebra of the centre of 
$G=\mathrm{Aut}(V)$. The upshot then is that $F_{\C}$ is non--zero only 
on the centre of $\mathtt{g}$.

\subsection{Relation to the volume}

Note that our derivative $\diff\vol[L]:\mathtt{t}_s\rightarrow \mathbb{R}$ 
is a linear map which is also independent of 
transverse K\"ahler deformations. This follows since $\vol[L]$ itself has this property for 
all K\"ahler cones. Moreover, 
if $L$ admits a Sasaki--Einstein metric 
with Reeb vector field $\xi$ then $\diff\vol[L](Y)=0$  
for all vector fields $Y\in\mathtt{t}_s$ with ${\cal L}_Y\Omega=0$.
This is true simply because Sasaki--Einstein metrics are critical points 
of the Einstein--Hilbert action, which is equal to $\vol[L]$ on the 
subspace $\Sigma$. Thus, given 
the exposition in the previous subsection, it is not 
surprising that $\diff\vol[L]$ is related to Futaki's invariant. 
We now investigate this in more detail.

We regard
\be\label{lin}
\diff\vol[L](Y)=-n\int_L\eta(Y)\diff\mu\ee
as a linear map 
\be
\diff\vol[L]:\mathtt{t}_s\rightarrow \R\ee 
with $Y\in\mathtt{t}_s$. Since all such $Y$ commute
with $\rdr$ and $\xi$, $Y$ descends to a holomorphic vector field 
$Y_V=\pi_*Y$ on $V$, where recall that $\pi:L\rightarrow V$ is the 
orbifold circle fibration. In fact, when interpreting 
$\vol[L]$ as the Einstein--Hilbert action, we should consider 
only those $Y\in\mathtt{t}_s$ such that $\mathcal{L}_Y\Omega=0$. 
These form a linear subspace in $\mathtt{t}_s$.

In order to relate $\diff\vol[L]$ to the Futaki invariant, we shall 
use the existence of a certain spinor field $\theta$ on the Sasakian link $L$, 
as discussed 
in subsection \ref{spinors}.
We shall also need the Lie derivative, acting\footnote{Recall that 
the differential forms act on spinors via Clifford multiplication; 
that is for a $p$--form 
$A$, we have $A\cdot\theta\equiv \tfrac{1}{p!}A_{\mu_1\dots \mu_p}\gamma^{\mu_1 \dots
\mu_p}\theta$.} 
on spinor fields, along  a Killing vector field\footnote{We denote the 
one--form dual to $Y$ by $Y^{\flat}=g_L(Y,\cdot)$.} Y:
\be\label{lieboy}
\mathcal{L}_Y\theta = \nabla_Y\theta +\frac{1}{4}\dd Y^\flat \cdot\theta~.\ee
Suppose now that $Y$ is Killing and satisfies
\be
\mathcal{L}_Y\theta = i\alpha\theta~.\ee
Of course, we are interested in those $Y$ with $\alpha=0$ since 
then $\mathcal{L}_Y\Omega=0$ also, as follows by writing the holomorphic
$(n,0)$--form $\Omega$ as a bilinear in the spinor field. The 
equivalence of these two conditions is not quite obvious -- 
a detailed argument proving this is given in Appendix \ref{moremore}.

We now compute
\bea\label{halfway}
\frac{1}{2}\int_L \eta(Y) \diff\mu  & = & \int_L \left(-i\bar{\theta}\nabla_Y\theta -\frac{1}{2} A(Y)\bar{\theta}\theta\right)\diff\mu \nn\\
& = & \int_L \bar{\theta}\left(\alpha\theta + \frac{i}{4}
\dd Y^\flat\cdot\theta-\frac{1}{2}A(Y)\theta\right)\diff\mu\nn \\
& = & \alpha\vol[L] -\frac{1}{8}\int_L (\diff\eta,\diff Y^{\flat})\diff\mu
-\frac{1}{2}\int_L A(Y)\diff\mu~,\eea
where recall that the spinor is normalised so that $\bar\theta \theta =1$.
Here we have denoted the pointwise inner product between two 
two--forms $A, B$ as
\be
(A,B) \equiv \frac{1}{2}A_{\mu\nu}B^{\mu\nu}~.\ee
Next, we obtain
\bea
-\frac{1}{8}\int_L (\diff\eta,\diff Y^{\flat})\diff\mu & = & 
-\frac{1}{8}\int_L (\diff^*\diff\eta,Y^{\flat})\diff\mu\nn \\
& = & -\frac{1}{8}\int_L (\Delta_L\eta,Y^{\flat})\diff\mu\nn\\
& = & - \frac{n-1}{2}\int_L \eta(Y)\diff\mu~.
\label{nicetrick}\eea
Here we have used the fact that $\diff^*\eta=0$ as $\eta$ is a Killing 
one--form. Moreover, we have
\be
\Delta_L\eta = 2\mathrm{Ric}(g_L)(\eta) = 4(n-1)\eta+2\mathrm{Ric}(g_X)(\eta) = 4(n-1)\eta~.\ee
The first equality is true for any Killing one--form. In the second we have 
used equation (\ref{bazza}) to relate the Ricci curvature of the link 
$L$ to that of the cone $X$. Note that from the same equation 
it also follows that  $\xi$ contracted into $\mathrm{Ric}(g_X)$ is zero, 
which gives the 
last equality. Finally, substituting (\ref{nicetrick}) into
(\ref{halfway}) gives us
\be
\diff\vol[L](Y) = -n\int_L \eta(Y) \diff\mu = 
-2\alpha\vol[L]+\int_L A(Y)\diff\mu~.\ee
Clearly, to relate to the Futaki invariant (\ref{futaki}), we must now 
integrate over the circle fibre of $\pi:L\rightarrow V$. Indeed, recall that
\be
\rho = \rho_V -2n\omega_V = \frac{1}{2}\diff\diff^c f\ee
as an equation on the Fano orbifold $V$. We thus have
\be
A = \frac{1}{2}\pi^*(\diff^c f)~.\ee
Since $f$ is basic by definition, and $Y$ commutes with $\xi$, it follows that 
$A(Y)$ is also a basic function, {\it i.e.} it is independent of 
$\xi$. Thus we may trivially integrate over the circle fibre. Denote the 
length of this by 
\be
\ell = \frac{2\pi\beta}{n}~.\ee
Then
\bea
\diff\vol[L](Y) & = & -2\alpha\vol[L] +\frac{\ell}{2}\int_V \diff^c f(Y_V) 
\ \frac{\omega_V^{n-1}}{(n-1)!}\nn \\
& = & -2\alpha\vol[L] -\frac{\ell}{2}\int_V \left(\mathcal{L}_{J_V(Y_V)}f\right)
\ \frac{\omega_V^{n-1}}{(n-1)!}~.\eea
Thus we have shown that
\bea\label{relly}
\diff\vol[L](Y)  =  -2\alpha\vol[L] -\frac{\ell}{2}\cdot F[J_V(Y_V)]~.\eea
Of course, for $Y$ preserving $\Omega$ we have $\alpha=0$ and we are 
done. Note also that, when $Y=\xi$ the spinor has charge 
$\alpha=n/2$, as follows from a simple calculation. 
In this case $Y_V=0$ and (\ref{relly}) shows 
that $\vol[L]$ is homogeneous degree $-n$ under deformations along the
Reeb vector field. Later we will see this rather directly from our explicit 
formula for the volume.

Thus, as expected, the derivative of the volume is directly related 
to the Futaki invariant of $V$. The 
dynamical problem of finding a critical point of 
the Sasakian volume $V$ may be interpreted
as choosing the Reeb vector field in such a way so that 
the corresponding K\"ahler orbifold $V$ has zero Futaki invariant. 
This is certainly a necessary condition for existence 
of a K\"ahler--Einstein metric on $V$, which clearly is 
also necessary in order for a Sasaki--Einstein metric 
to exist on $L$. Of course, this interpretation 
requires us to stick with quasi--regular stuctures, 
which is unnatural. Nevertheless, it gives a very interesting new 
interpretation of the Futaki invariant.

\subsection{Isometries of Sasaki--Einstein manifolds}
\label{mtype}

Using the result of the last subsection, together with known properties of the 
Futaki invariant described above, we may now deduce 
some additional properties of Sasaki--Einstein manifolds. In 
particular, we first 
show that the critical Reeb vector field is in the centre 
of a maximal compact subgroup $K\subset \mathrm{Aut}(X)$, where 
$\T^s\subset K$. Then
we use this fact to argue that the group of (holomorphic) isometries of a Sasaki--Einstein
metric on $L$ is a maximal compact subgroup of Aut$(X)$.

Let $K\subset \mathrm{Aut}(X)$ 
be a maximal compact subgroup of the holomorphic automorphism group of 
$X$, containing the maximal torus $\T^s\subset K$. 
Let $\mathtt{z}\subset \mathtt{k}$ denote the centre 
of $\mathtt{k}$, and write
\be
\mathtt{t}_s = \mathtt{z}\oplus\mathtt{t}^{\prime}~.\ee
Pick a basis for $\mathtt{z}$ so that we may identify 
$\mathtt{z}\cong\R^m$. We may consider the space of 
Reeb vector fields in this subspace -- by the same reasoning 
as in section \ref{sectiontwo}, these form a cone $\mathcal{C}_0^{(m)}$ where 
we now keep track of the dimension of the cone. There 
will be a unique critical point of the Einstein--Hilbert action 
on this space. Let us suppose that $b_*^{(m)}\in\mathbb{Q}^m$, so that 
this critical point is quasi--regular.
Clearly we have considered only 
a subspace $\mathtt{z}\subset\mathtt{t}_s$, or $\R^m\subset\R^s$, and the minimum we have found 
in $\R^m$ might not be a minimum on the larger space.
However, using the relation between $\diff\vol[L]$ and the 
Futaki invariant we may in fact argue that the critical 
point $b_*^{(m)}\in\R^m$ is necessarily a critical point of the minimisation 
problem on $\R^s$. 

To see this, note that $\mathtt{t}^{\prime}\cong\R^{s-m}$ descends to a subalgebra 
$\mathtt{t}^{\prime}\subset\mathrm{aut}_{\R}(V)$. Since $b=(b_*^{(m)},0)\in 
\R^m\oplus\R^{s-m}$ 
is in the centre of $\mathtt{k}$ by construction, 
in fact the whole of $\mathtt{k}$ descends to a subalgebra of $\mathrm{aut}_{\R}(V)$. 
Recall now that 
$F_{\C}:\mathrm{aut}(V)\rightarrow\C$ vanishes on 
the complexification of $\mathtt{t}^{\prime}$. This is because 
$F_{\C}$ is non--zero only on the centre of $\mathrm{aut}(V)$. 
Thus the derivative of $\vol[L]$ in the directions $\mathtt{t}^{\prime}$ 
is \emph{automatically zero}. Since the critical point is unique, 
this proves that $b_*=(b_*^{(m)},0)\in \R^m\oplus\R^{s-m}$ is the critical 
point also for the larger extremal problem on $\R^s$. This
 argument may of course be 
made for any $K\supset \T^s$, but in particular applies to a maximal 
such $K$. Hence 
we learn that the critical Reeb vector field, for a Sasaki--Einstein 
metric, necessarily lies in the centre\footnote{Note this 
statement is different from the statement 
that the Reeb vector field for any Sasakian 
metric lies in the centre of the isometry group: 
the group of automorphisms of $(L,g_L)$ {\it i.e.} the isometry group, depends on the choice 
of Reeb vector field, whereas $K$ depends only on $(X,J)$.} of the Lie 
algebra of a maximal compact 
subgroup of $\mathrm{Aut}(X)$. 

Using this last fact, we may now prove that, for 
fixed $\T^s\subset K$, the isometry group of a Sasaki--Einstein metric on $L$ with 
Reeb vector field in $\mathtt{t}_s$ is a 
maximal compact subgroup of $\mathrm{Aut}(X)$. 
To see this, note that since the critical Reeb vector field
$\xi_*\subset\mathtt{z}$ lies in the centre of $\mathtt{k}$, 
the whole of $K$, modulo the $U(1)$ generated by 
the Reeb vector field, descends to a compact subgroup of the complex 
automorphisms $G=\mathrm{Aut}(V)$ of the orbifold $V$. The latter is 
K\"ahler--Einstein, and by Matsushima's theorem\footnote{Recall that Matsushima's
theorem states that on a K\"ahler--Einstein manifold $V$ (or, more generally, orbifold) the 
Lie algebra of holomorphic vector fields aut$(V)$ is the complexification of the Lie algebra
generated by Killing vector fields.} \cite{Matsushima}, 
we learn that $K$ in fact acts \emph{isometrically} on $V$. 
Thus $K$ acts isometrically on $L$, and 
we are done.

Of course, this is also likely to be true for 
irregular Sasaki--Einstein metrics. Thus we make a more general conjecture
\begin{itemize}
\item The group of holomorphic isometries of 
a Sasaki--Einstein metric on $L$ is a maximal compact 
subgroup of $\mathrm{Aut}(X)$.\end{itemize}

%%%%%%%%%%%%%%%%%%%%%%%%%%%%%%%%%%%%%%%%%%%%%%%%%%%%%%%%%%%%%%%%%%%%%%%%%%%

\section{A localisation formula for the volume}
\label{localsection}

In this section we explain that the volume of a 
Sasakian manifold may be interpreted in terms of the
Duistermaat--Heckman formula. 
The essential point is that the K\"ahler potential 
$r^2/2$ may also be interpreted as the Hamiltonian 
function for the Reeb vector field. By taking any 
equivariant orbifold resolution of the cone $X$, 
we obtain an explicit formula for the volume, as a 
function of the Reeb vector $\xi\in\mathcal{C}_0$, in 
terms of topological fixed point data. If we write
$\xi\in\mathcal{C}_0\subset\mathtt{t}_s$ as 
\be
\xi = \sum_{i=1}^s b_i\frac{\partial}{\partial\phi_i}
\label{pippo}
\ee
then as a result the volume $\vol[L]:\mathcal{C}_0\rightarrow\R$, relative 
to the volume of the round sphere, is a 
rational function of $b\in\R^s$ with rational coefficients. 
The Reeb vector field $b_*$ for a Sasaki--Einstein
metric is a critical point of $\vol[L]$ on the convex 
subspace $\Sigma$, formed by intersecting $\mathcal{C}_0$ with the 
rational hyperplane of Reeb vectors under which 
$\Omega$ has charge $n$. It follows that
$b_*$ is an algebraic vector -- that is, a
vector whose components are all algebraic numbers. 
It thus also follows that 
the volume $\vol[L](b_*)$ of a Sasaki--Einstein metric,
relative to the round sphere, is an algebraic number.

\subsection{The volume and the Duistermaat--Heckman formula}

There is an alternative way of writing the 
volume of $(L,g_L)$. In the previous subsections
we used the fact that
\be
\vol[L] = \int_L \diff\mu = 2n\int_{r\leq1}\frac{\omega^n}{n!}~.\ee
This follows simply by writing out the measure on the 
cone $X$ in polar
 coordinates and cutting off the $r$ integral at $r=1$. However, 
we may also write
\bea
\mathrm{vol}[L] & = & \frac{1}{2^{n-1} (n-1)!}\int_X 
e^{-r^2/2} \,\frac{\omega^n}{n !}~.
\label{newvolume}
\eea
This is now an integral over the whole cone. 
Note that the term in the exponent is the 
K\"ahler potential, which here acts as a convergence factor.

So far, the function $r^2$ has played a dual role: it determines 
the link $L=X\mid_{r=1}$, and is also the K\"ahler potential. 
However, the function $r^2/2$ is also the Hamiltonian function 
for the Reeb vector field. To see this, recall from 
(\ref{moment}) that the Hamiltonian function associated to the 
vector field $Y$ is $y_Y=\tfrac{1}{2}r^2\eta(Y)$. Setting 
$Y=\xi$, we thus see that $r^2/2$ is precisely the Hamiltonian 
associated to the Reeb vector field. Thus we may suggestively write 
the volume (\ref{newvolume}) as 
\be\label{volDH}
\vol[L] \ = \ \frac{1}{2^{n-1} (n-1)!}\int_X e^{-H}\ e^{\omega}\ee
where $H=r^2/2$ is the Hamiltonian. 
The integrand on the right hand side of (\ref{volDH}) is precisely that 
appearing in the
Duistermaat--Heckman formula \cite{DH1,DH2} for a (non--compact) 
symplectic manifold $X$ with symplectic form $\omega$. $H$ is the Hamiltonian 
function for a Hamiltonian 
vector field. The Duistermaat--Heckman theorem expresses 
this integral as an integral of local data over the 
fixed point set of the vector field. 
Of course, for a K\"ahler cone, the action generated by the flow of the 
Reeb vector field is locally free on $r>0$, since 
$\xi$ has square norm $r^2$. The fixed point contribution is 
therefore, formally, entirely from the isolated singular point $r=0$. 
Thus in order 
to apply the theorem, one must first resolve the singularity. 
Taking a limit in which the resolved space approaches the 
cone, we will obtain a well--defined expression for the volume 
in terms of purely topological fixed point data. The volume is of course 
independent of the choice of resolution.

\subsection{The Duistermaat--Heckman Theorem}

In this subsection we give a review of the Duistermaat--Heckman
theorem \cite{DH1,DH2} for compact K\"ahler manifolds. 
The non--compact case of interest will 
follow straightforwardly from this, as we shall explain.
Since the proof of the Duistermaat--Heckman theorem \cite{DH2} is 
entirely differentio--geometric, the result is also valid for
orbifolds, with a simple modification that we describe.

Let $W$ be a compact K\"ahler manifold with K\"ahler form $\omega$, 
$\dim_{\C}W=n$, and 
let $\T^s\subset\mathrm{Aut}(W)$ act on $W$ in a Hamiltonian fashion. Let 
$\xi\in\mathtt{t}_s$ with Hamiltonian $H$. Thus
\be
\diff H = -\xi\lrcorner\omega~.\ee
The flow on $W$ generated by $\xi$ 
will have some fixed point set, which is also the zero set of the vector field $\xi$. 
In general, the fixed point set $\{F\}$ will consist of a number of disconnected 
components $F$ of different dimensions; each component is a 
K\"ahler submanifold of $W$. Let $f:F\hookrightarrow W$ 
denote the embedding, so that $f^*\omega$ is a 
K\"ahler form on $F$. From now on, we focus on 
a particular connected component $F$, of complex codimension 
$k$.

The normal bundle $\mathcal{E}$ of $F$ in $W$ is a rank $k$ complex vector bundle 
over $F$. The flow generated by $\xi$ induces a linear 
action on $\mathcal{E}$. Let $u_1,\ldots,u_R\in\Z^s\subset \mathtt{t}_s^*$ be 
the set of distinct weights of this linear action on $\mathcal{E}$. 
This splits $\mathcal{E}$ into a direct sum of complex 
vector bundles
\be\label{banana}
\mathcal{E} = \bigoplus_{\lambda=1}^{R} \mathcal{E}_{\lambda}~.\ee
Here $\mathcal{E}_{\lambda}$ is a complex vector bundle over $F$ 
of  rank $n_{\lambda}$, and hence
\be
\sum_{\lambda=1}^R n_{\lambda} = k~.\ee
Thus, for example, if $\mathcal{E}$ splits into a sum of 
complex line bundles then each $n_{\lambda}=1$ and 
$R=k$. We denote the linear action of $\xi$ on 
$\mathcal{E}$ by $L\xi$. This acts on the 
$\lambda$th factor in (\ref{banana})
by multiplication by $i(b,u_{\lambda})$, where recall that $b \in \R^s$ are
the components of the vector field $\xi$ as in (\ref{pippo}). 
 With respect to the decomposition (\ref{banana}), we thus have
\be
L\xi = i\ \mathrm{diag}\left(1_{n_1}(b,u_1),1_{n_2}(b,u_2),\ldots,1_{n_R}(b,u_R)\right)\ee
where $1_{n_{\lambda}}$ denotes the unit $n_{\lambda}\times n_{\lambda}$ matrix. Hence the 
determinant of this transformation is
\be
\det \left(\frac{L\xi}{i}\right)=\prod_{\lambda=1}^R (b,u_{\lambda})^{n_{\lambda}}~.\ee
Note this is homogeneous degree $k$ in $b$.

Finally, choose any $\T^s$--invariant 
Hermitian connection on $\mathcal{E}$, with curvature $\Omega_\mathcal{E}$. 
The Duistermaat--Heckman theorem then states that
\be\label{DHtheorem}
\int_W e^{-H}\ \frac{\omega^n}{n!} = 
\sum_{\{F\}} \int_F \frac{e^{-f^*H}\ e^{f^*\omega}}{\det 
\left(\frac{L\xi-\Omega_\mathcal{E}}{2\pi i}\right)}~.\ee
The sum is over each connected component $F$ of 
the fixed point set. The determinant is a $k\times k$ 
determinant and should be expanded formally into a 
differential form of mixed degree. Moreover, the 
inverse is understood to mean one should expand this 
formally in a Taylor series. This notation is standard in 
index theory. We will now examine the right hand side of (\ref{DHtheorem}) 
in more detail.

If we let $\Omega_{\lambda}$ be the curvature of $\mathcal{E}_{\lambda}$, we may write
\bea
\det\left(\frac{L\xi-\Omega_\mathcal{E}}{2\pi i}\right)  =  
\frac{1}{(2\pi)^k}\det\left(\frac{L\xi}{i}\right) \prod_{\lambda=1}^R
\det (1-(L\xi)^{-1}\Omega_{\lambda})~.\eea
Fix one of the vector bundles $\mathcal{E}_{\lambda}$. Then
\bea\label{cherny}
\det(1-(L\xi)^{-1}\Omega_{\lambda}) = 
\det\left(1+w\frac{i\Omega_{\lambda}}{2\pi}\right)\eea
where 
\be
w=\frac{2\pi}{(b,u_{\lambda})}~.\ee
The right hand side of (\ref{cherny}) is precisely the 
\emph{Chern polynomial} of $\mathcal{E}_{\lambda}$. As a 
cohomology relation, we have
\bea
\det\left(1+w\frac{i\Omega_{\lambda}}{2\pi}\right) 
= \sum_{a\geq0} c_a(\mathcal{E}_{\lambda})w^a\in H^*(F;\R)~.\eea
Recall that each $\mathcal{E}_{\lambda}$ has Chern classes 
\be
c_a(\mathcal{E}_{\lambda})\in H^{2a}(F;\R)\ee
where $0\leq a \leq n_{\lambda}$ and $c_0=1$. Thus we may write
\bea
\left[\det\left(\frac{L\xi-\Omega_\mathcal{E}}{2\pi i}\right)\right]^{-1}   
&  =  & (2\pi)^k\ \left[\det\left(\frac{L\xi}{i}\right)\right]^{-1} \sum_{ 
a\geq 0} \beta_a(b)\nn \\ 
& = & \frac{(2\pi)^k}{\prod_{\lambda=1}^R (b,u_{\lambda})^{n_{\lambda}}} 
\sum_{a\geq 0} \beta_a(b)~.\eea
The $\beta_a(b)$ are closed differential forms on $F$ of
degree $2a$, with $\beta_0(b)=1$. The cohomology class
of $\beta_a(b)$ in $H^{2a}(F;\R)$ is then a 
polynomial in the Chern classes of the $\mathcal{E}_{\lambda}$. 
Moreover, the coefficients are \emph{rational} functions 
of $b$, of degree $-a$.

Since $f^*H$ is constant on each connected component of $F$, it 
follows that, for each connected component, the integrand on the right hand side of (\ref{DHtheorem}) 
is a polynomial in the Chern classes of $\mathcal{E}_{\lambda}$ and 
the pulled--back K\"ahler form $f^*\omega$ on $F$. The coefficients are 
memomorphic\footnote{We may analytically 
continue the right hand side to $b\in\C^s$.} in $b$, and analytic 
provided $(b,u_{\lambda})\neq 0$ for all $\lambda=1,\ldots,R$. 
Of course, the left hand side of (\ref{DHtheorem}) is certainly 
analytic. 

As a special case of this result, suppose that $F$ is an isolated fixed point. 
Thus $k=n$, and $\mathcal{E}$ is a trivial bundle. We may then 
write the $n$, \emph{possibly indistinct}, weights 
as $u_{\lambda}$, $\lambda=1,\ldots,n$. Neither the 
Chern classes nor the Liouville measure $\exp(f^*\omega)$ 
contribute non--trivially, and we are left with 
\be\label{isolated}
(2\pi)^n\ e^{-f^*H}\prod_{\lambda=1}^n \frac{1}{(b,u_{\lambda})}~.\ee
This is the general formula for the contribution of an isolated 
fixed point to the Duistermaat--Heckman formula (\ref{DHtheorem}).

The proof of the 
Duistermaat--Heckman theorem \cite{DH2} is entirely differentio geometric, and 
thus the proof also goes through easily for non--compact manifolds, and orbifolds. 
The proof goes roughly as follows. The integrand on the 
left hand side of the Duistermaat--Heckman formula (\ref{DHtheorem})
is exact on $W$ minus the set of fixed points $\{F\}$. One then applies 
Stokes' theorem to obtain a sum of integrals over boundaries around 
each connected component $F$. This boundary is diffeomorphic, via the 
exponential map, 
to the total space of the normal sphere bundle to 
$F$ in $W$, of radius $\epsilon$. One should eventually take the limit 
$\epsilon\rightarrow 0$. 
By introducing an Hermitian connection on the normal bundle $\mathcal{E}$ 
with curvature $\Omega_\mathcal{E}$, one can 
perform the integral over the normal sphere explicitly, resulting in 
the formula (\ref{DHtheorem}). Thus we may extend this proof straightforwardly 
to our non--compact case, and to orbifolds, as follows:
\begin{itemize}
\item For non--compact manifolds, provided the fixed point sets are in the interior, 
and that the measure tends to zero at infinity (so that the boundary at 
infinity makes no contribution in Stokes' theorem), the formula (\ref{DHtheorem}) is 
still valid.

\item For orbifolds, we must modify the formula (\ref{DHtheorem}) slightly. 
The normal space to a \emph{generic} point in a connected component $F$ is not a sphere, 
but rather a quotient $S^{2k-1}/\Gamma$, where $\Gamma$ is a finite 
group of order $d$. Thus, when we integrate over this normal space, 
we pick up a factor $1/d$. For each connected component $F$ we 
denote this integer, called the order of $F$, by $d_F$. 
The general formula is then almost identical
\be\label{DHorbifold}
\int_W e^{-H}\ \frac{\omega^n}{n!} = 
\sum_{\{F\}} \int_F \frac{1}{d_F}\ \frac{e^{-f^*H}\ e^{f^*\omega}}{\det 
\left(\frac{L\xi-\Omega_\mathcal{E}}{2\pi i}\right)}~.\ee
In the orbifold case, 
the Chern classes, defined in terms of a curvature form $\Omega_\mathcal{E}$ on 
the vector orbibundle $\mathcal{E}$, are in general \emph{rational}, {\it i.e.} 
images under the natural map
\be
H^*(F;\mathbb{Q})\rightarrow H^*(F;\R)~.\ee
\end{itemize}
These slight generalisations will be crucial for the 
application to Sasakian geometry, to which we now turn.

%%%%%%%%%%%%%%%%%%%%%%%%%%%%%%%%%%%%%%%%%%%%%%%%%%%%%%%%%%%%%%%%

\subsection{Application to Sasakian geometry}
\label{application}

Let $(X,\omega)$ be a K\"ahler cone, with K\"ahler potential 
$r^2$. The volume of the link $L=X\mid_{r=1}$ may be written 
as in (\ref{volDH}). In order to apply the Duistermaat--Heckman 
theorem, we must first resolve the cone $X$. In fact in order 
to compute the volume $\vol[L](b)$ as a function of $b$, we 
pick, topologically, a fixed (orbifold) resolution $W$ of $X$. Thus 
we have a map
\be
\Pi: W\rightarrow X\ee
and an exceptional set $E$ such that $W\setminus E\cong X_0$ is 
a biholomorphism. Moreover, the map $\Pi$ should be equivariant 
with respect to the action of $\T^s$. Thus all the fixed points of 
$\T^s$ on $W$ necessarily lie on the exceptional set.
There is a natural way to do this: we simply choose a (any) quasi--regular 
Reeb vector field $\xi_0$ and 
blow up the K\"ahler orbifold $V_0$ to obtain the total space $W$ of the bundle 
$\mathcal{L}\rightarrow V_0$. Note that then $W\setminus V_0\cong X_0$, 
and that $\T^s$ acts on $W$. Thus this resolution is 
obviously equivariant. 

We then assume\footnote{We assume there is no obstruction to doing this. 
In any case, we shall also prove the localisation formula (\ref{normvol}) 
in the next section using a different relation to an equivariant index 
on the cone $X$. This doesn't assume the existence of any metric on $W$.} that, for 
every Reeb vector field $\xi\in\mathcal{C}_0$, 
there is a 1--parameter family of $\T^s$--invariant 
K\"ahler metrics $g(T)$, $0<T<\delta$ for some $\delta>0$, 
on $W$ such that $g(T)$ smoothly approaches a K\"ahler cone 
metric with Reeb vector field $\xi$ as $T\rightarrow 0$. We may then apply the Duistermaat--Heckman 
theorem (\ref{DHorbifold}) to the K\"ahler metric $g(T)$ on the orbifold $W$.
In the limit that $T\rightarrow 0$, the exceptional set collapses 
to zero volume and we recover the cone $X$. Since all fixed points 
of $\T^s$ lie on the exceptional set, the pull--back of the Hamiltonian 
$f^*H$ tends to zero in this limit, as $H\rightarrow r^2/2$
which is zero at $r=0$. Moreover, the pull--back of the 
K\"ahler form $f^*\omega$ is also zero in this limit. Hence 
the exponential terms on the right--hand side of 
the Duistermaat--Heckman formula are equal to $1$ in 
the conical limit $T\rightarrow 0$. This leaves us with the formula
\be
\vol[L](b) = \frac{1}{2^{n-1}(n-1)!}\sum_{\{F\}}\int_F \frac{1}{d_F}\ \frac{(2\pi)^k}
{\prod_{\lambda=1}^R (b,u_{\lambda})^{n_{\lambda}}} 
\sum_{a\geq 0} \beta_a(b)~.\ee
This formula is valid for $b$ a generic element of $\mathtt{t}_s$. Then 
the vanishing set of the Reeb vector field is the fixed point set of $\T^s$. 
For certain special values of $b$ the set of fixed points changes. For example, 
when $W$ is obtained by taking a quasi--regular Reeb vector field $\xi_0$ 
and blowing up $V_0$, the fixed point set of $\xi_0$ is the whole of $V_0$.
Note, however, that $\vol[L](b)$ is still a smooth function of $b$.

The integral over $F$ picks out the term in the sum of 
degree  $a=(n-k)$. Recall that the $a$th term is 
homogeneous in $b$ of degree $-a$. In particular, we 
may extract the factor $(2\pi)^{n-k}$ and write
\be
\vol[L](b) = \frac{2\pi^n}{(n-1)!}\sum_{\{F\}} \beta(b)\prod_{\lambda=1}^R \frac{1}{(b,u_{\lambda})^{n_{\lambda}}} ~.\ee
Here $\beta(b)$ is a sum of Chern numbers of the normal bundle $\mathcal{E}$ of $F$ in $W$, 
with coefficients which are homogeneous degree $-(n-k)$ in $b$. Specifically,
\bea\label{betabadger}
\beta(b) = \int_F\frac{1}{d_F} \ \prod_{\lambda=1}^R \left[\sum_{a\geq0} 
\frac{c_a(\mathcal{E}_{\lambda})}{(b,u_{\lambda})^a}\right]^{-1}~.\eea
The Chern polynomials in (\ref{betabadger}) should be expanded in a Taylor series, 
and the integral over $F$ picks out the differential form of degree $2(n-k)$.
Recall that Chern numbers are defined as integrals over $F$ of wedge products of 
Chern classes. Thus when $W$ is a manifold the coefficients in $\beta(b)$ are integers. For the 
more general case of orbifolds, the Chern numbers are rational numbers.

Noting that 
\be
\vol[S^{2n-1}]=\frac{2\pi^n}{(n-1)!}\ee
is precisely the volume of the round $(2n-1)$--sphere, we thus have
\be\label{normvol}
V(b)\equiv\frac{\vol[L](b)}{\vol[S^{2n-1}]} = \sum_{\{F\}} \frac{1}{d_F}\int_F 
\prod_{\lambda=1}^R \frac{1}{(b,u_{\lambda})^{n_{\lambda}}}\left[ 
\sum_{a\geq0} \frac{c_a(\mathcal{E}_{\lambda})}{(b,u_{\lambda})^a}\right]^{-1} ~.\ee
Here we have defined the normalised volume $V(b)$. The right hand side 
of (\ref{normvol}) is homogeneous degree $-n$ in $b$, and is 
manifestly a rational function of $b$ with \emph{rational} coefficients, 
since the weights and Chern numbers are generally rational numbers.
This is our general formula for the volume of a Sasakian 
metric on $L$ with Reeb vector field $b$. 
Using this result, we may now prove:
\begin{itemize} \item The volume of a Sasaki--Einstein 
manifold, relative to that of the round sphere, 
is an algebraic number. 
\end{itemize}
This follows trivially, since the Reeb vector $b_*$ for a 
Sasaki--Einstein metric is a critical point of (\ref{normvol}) on the 
subspace of vector fields under which the holomorphic $(n,0)$--form 
$\Omega$ has charge $n$. 
Thus $b_*$ is an isolated zero of 
a system of algebraic equations with rational 
coefficients, and hence 
is algebraic. The normalised volume $V(b_*)$ is thus also 
an algebraic number.

We conclude this section by relating the  formula (\ref{normvol}) to the volume 
of quasi--regular Sasakian metrics. 
%At this point we may relate the formula (\ref{normvol}) to the 
%volume of the Sasakian metric in a rather different manner. 
Let $\xi$ be the Reeb vector field for a quasi--regular
Sasakian structure, and choose the resolution $W$
above so that $V$ is the K\"ahler orbifold of the Sasakian structure. 
$\xi$ generates an action of $U(1)$ on $W$ which is 
locally free outside the zero section, and fixes the 
zero section $V$. Thus in this case $\mathcal{E}=\mathcal{L}$ and 
the formula (\ref{normvol}) simplifies considerably. 
The weight $u=1$, the codimension $k=1$, $d=1$, and hence (\ref{normvol}) gives
\bea
V(b) & = &
\frac{1}{b}\int_V\left[1+b^{-1}c_1(\mathcal{L})\right]^{-1}\nn \\
& = & \frac{1}{b^n}\int_V c_1(\mathcal{L}^*)^{n-1}\eea
where $c_1(\mathcal{L})=-c_1(\mathcal{L}^*)$. Recall from section
\ref{sectiontwo} that 
$\mathcal{L}\rightarrow V$ is always given by some 
root of the canonical bundle over $V$. The first Chern class of $\mathcal{L}^*$ is then 
\be\label{class}
c_1(\mathcal{L}^*)=\frac{c_1(V)}{\beta}\in H^2_{\mathrm{orb}}(V;\Z)\ee
where $c_1=c_1(V)\in H^2_{\mathrm{orb}}(V;\Z)$ is the first Chern class of the 
complex orbifold $V$. The total space $L$ 
of the associated circle bundle to $\mathcal{L}$ will be simply connected 
if and only if $\beta$ is the \emph{maximal} positive integer such 
that (\ref{class}) is an integer class, which recall is 
called the index of $V$. The general volume formula 
(\ref{normvol}) gives
\be\label{voly}
V(b) = \frac{1}{b^n\beta^{n-1}}\int_V c_1^{n-1}~.\ee

We now compute the volume $V(b)$ directly. 
Recall that, in general, the volume of a \emph{quasi--regular} Sasakian 
manifold is given by the formula
\be\label{qrvol}
\vol[L] = \frac{2\pi^n\beta}{n^n(n-1)!}\int_V c_1^{n-1}~.\ee
To see this, recall from (\ref{harry}) that the K\"ahler class of $V$ is $[\omega_V]=[\rho_V/2n]$, 
where in the latter equation we have
used the fact that the holomorphic $(n,0)$--form $\Omega$ is homogeneous
degree $n$ under $\rdr$. We also have $c_1=[\rho_V/2\pi]$. Thus the volume 
of $V$ is given by 
\be
\mathrm{vol}[V] = \int_V \frac{\omega_V^{n-1}}{(n-1)!} = \frac{\pi^{n-1}}
{n^{n-1}(n-1)!}\int_V c_1^{n-1}~.\ee
The length of the circle fibre is $2\pi\beta/n$, and  
hence (\ref{qrvol}) follows. For example, when $V=\C P^{n-1}$ 
and $\beta=n$ this leads to the formula
\be
\vol[S^{2n-1}] = \frac{2\pi^n}{(n-1)!}\ee
for the volume of the round sphere. Thus from (\ref{qrvol}) we have
\be\label{barry}
V(b)= \frac{\beta}{n^n}\int_V c_1^{n-1}\ee
which is precisely formula (\ref{voly}) with 
\be\label{brel}
b=n/\beta~.\ee This equation for $b$ must be imposed to compare (\ref{barry}) 
to (\ref{voly}), since in the former equation we have assumed that $\Omega$ has 
charge $n$ under the Reeb vector field. More precisely, in (\ref{voly}) we have written
\be\label{frog}
\xi = b\frac{\partial}{\partial\nu}\ee
where $\nu\sim\nu+2\pi$, and $\partial/\partial\nu$ rotates the fibre of the 
line bundle $\mathcal{L}$ with weight \emph{one}. Thus
\be
\nu = \frac{n\psi}{\beta}\ee
where $\psi$ was defined in section \ref{reebsection}. Recalling that $\Omega$ has 
charge $n$ under $\partial/\partial\psi$, we may thus also impose this 
on the vector field $\xi$ in (\ref{frog}):
\be
in\Omega = \mathcal{L}_{\xi}\Omega = b\mathcal{L}_{\partial/\partial\nu}\Omega = 
b \frac{\beta}{n}\mathcal{L}_{\partial/\partial\psi}\Omega = ib\beta\Omega\ee
from which (\ref{brel}) follows. 

%Since any Sasakian structure may be approximated arbitrarily 
%closely by a quasi--regular structure\footnote{{\it i.e.} 
%the rationals are dense in the reals.}, this constitutes a 
%different proof of the relation between the volume of Sasakian manifolds 
%and the localisation formula (\ref{normvol}).

\subsection{Sasakian 5--manifolds and an example}

The most physically interesting dimension is $n=3$. Then the AdS/CFT 
correspondence conjectures that string theory on $AdS_5\times L$ is 
dual to an $\mathcal{N}=1$ superconformal field theory in 
four dimensions. In this section we therefore specialise the formula 
(\ref{normvol}) to complex dimension $n=3$.

Suppose that $X_0\cong\R_+\times L$ is a Calabi--Yau 3--fold, and 
suppose we have a space of K\"ahler cone metrics on $X$ which are 
invariant under a $\T^2$ action.
We choose this case since a $\T^3$ action would mean that 
$L$ were toric, which we treat in section \ref{toricsection}. 

Resolving the cone $X$ to $W$, as before, the fixed point sets on $V_0$ will 
consist of isolated fixed points $p\in V_0$ and curves $C\subset V_0$. 
We have already treated the isolated fixed points in (\ref{isolated}).
Let the normal bundle of $C$ in $V_0$ be $\mathcal{M}$. The total 
normal bundle of $C$ in $W$ is then $\mathcal{E}=h^*\mathcal{L}\oplus\mathcal{M}$
where $h:C\hookrightarrow V_0$ is the inclusion. Thus the normal bundle $\mathcal{E}$ 
to $C$ in $W$ splits into a sum of two line bundles. We denote the 
weights as $u_{\lambda}\in\Z^2$, $\lambda=1,2$ for $\mathcal{L}$ 
and $\mathcal{M}$, respectively. We then get the 
following general formula for the volume, in terms of topological 
fixed point data:
\bea\label{fiveformula}
V(b) = \sum_{\{p\}} \frac{1}{d_p} \ \prod_{\mu=1}^3 \frac{1}{(b,u_{\mu})} - 
\sum_{\{C\}} \frac{1}{d_C}\ \left[\prod_{\lambda=1}^2 \frac{1}{(b,u_{\lambda})}\right] \int_C 
\left[\frac{c_1(\mathcal{L})}{(b,u_1)}+\frac{c_1(\mathcal{M})}{(b,u_2)}\right]~.\eea
Here $u_{\mu}\in\Z^2$, $\mu=1,2,3$ are the weights on the tangent 
space at $p$. Note that $V(b)$ is clearly homogeneous degree $-3$ in $b$, as it should be.

{\bf Example}: As an example of formula (\ref{fiveformula}), let us calculate the volume of 
Sasakian metrics on the complex cone over the first 
del Pezzo surface. Of course, this is toric, so one can use 
the toric methods developed in \cite{MSY}. However, the point here 
is that we will rederive the result using \emph{non--toric} 
methods. Specifically, we'll use the formula (\ref{fiveformula}).

We think of the del Pezzo as the first Hirzebruch surface, $\mathbb{F}_1$. 
This is a $\mathbb{C} P^1$ bundle over $\mathbb{C} P^1$, which may be 
realised as the projectivisation
\be
\mathbb{F}_1 = \mathbb{P}(\mathcal{O}(0)\oplus\mathcal{O}(-1))\rightarrow \mathbb{C} P^1~.\ee
We take $W$ to be the total space of the canonical bundle $\mathcal{K}\rightarrow \mathbb{F}_1$, and 
the $\T^2$ to act by rotating the fibre of $\mathcal{K}$ and the fibre  
$\mathbb{C} P^1$ of $\mathbb{F}_1$. The fixed point set of this
$\T^2$ action consists of two curves on $V_0=\mathbb{F}_1$, which are the north and south 
pole sections of $\mathbb{F}_1$. We denote these by $H$ and $E$, which 
are two copies of $\mathbb{C} P^1$. In fact, $[H]$ is the hyperplane 
class on $dP_1$, and $[E]$ is the exceptional divisor. The normal 
bundles over $H$ and $E$ are
\bea\label{normalboys}
H  & : & \mathcal{E}=\mathcal{O}(-3)\oplus\mathcal{O}(1)\nn \\
E  & : & \mathcal{E}=\mathcal{O}(-1)\oplus\mathcal{O}(-1)~.\eea
Note that the Chern numbers sum to $-2$ in each case,
as is necessary to cancel the Chern number $+2$ of 
$\mathbb{C} P^1$. 

Write $\T^2=U(1)_1\times U(1)_2$. Let $U(1)_1$ rotate the fibre of 
$\mathcal{L}=\mathcal{K}$ with weight one, and let $U(1)_2$ rotate the fibre 
$\mathbb{C} P^1$ of $\mathbb{F}_1$ with weight one, where we take
the canonical lifting of this action to the canonical bundle $\mathcal{K}$. 
We write
\be
\xi = \sum_{i=1}^2b_i\frac{\partial}{\partial\phi_i}\ee
We need to compute the weights $u_{\lambda}\in\Z^2\subset\mathtt{t}_2^*$ of the $\T^2$ action 
on the normal bundles (\ref{normalboys}). Here $\lambda=1,2$ denote the two 
line bundles in the splitting (\ref{normalboys}).
We compute the weight of each $U(1)\subset\T^2$ 
in turn. 
$U(1)_1$ has weights $[1,0]$ with respect to the splitting (\ref{normalboys}) 
for both $H$ and $E$ since this 
simply rotates the fibre of $\mathcal{K}$ with weight one. As for $U(1)_2$, 
note that $\mathcal{K}$ restricted to any point $x$ on the 
base $\mathbb{C} P^1$ of $\mathbb{F}_1$ is 
a copy of $T^*\mathbb{C} P^1 = \mathcal{O}(-2)\rightarrow\mathbb{C} P^1$.
By definition, $U(1)_2$ rotates this fibre $\mathbb{C} P^1$ with weight one, and thus fixes 
its north and south poles. Thus as we vary $x$ on the base $\mathbb{C}P^1$, 
we sweep out $H$ and $E$, respectively.
The weights of $U(1)_2$ on the tangent space to $x$ in $T^*\mathbb{C}P^1$ are 
thus $[-1,1]$ and $[1,-1]$, 
respectively -- the opposite signs appear because we have the cotangent bundle, 
rather than the tangent bundle. 
These 
also give the weights for $U(1)_2$ acting on $H$ and $E$, with respect to 
the decomposition (\ref{normalboys}), respectively. 

Thus, to summarise, the weights $u_{\lambda}\in\Z^2\subset\mathtt{t}_2^*$ are
\bea
H & : & u_1=(1,-1), \qquad u_2=(0,1)\nn\\
E & : & u_1=(1,1), \qquad u_2=(0,-1)~.\eea
The formula (\ref{fiveformula}) thus gives
\bea\label{dp1simple}
V(b) & = & \frac{1}{(b_1-b_2)b_2}\left(\frac{3}{b_1-b_2}+\frac{-1}{b_2}\right)
+ \frac{1}{(b_1+b_2)(-b_2)}\left(\frac{1}{b_1+b_2}+\frac{1}{-b_2}\right)\nn \\
& = & \frac{8b_1+4b_2}{(b_1^2-b_2^2)^2}~.\eea
One can verify that, on setting $b_1=3$, corresponding to the 
holomorphic $(3,0)$--form $\Omega$ having charge $3$, the remaining function of 
$b_2$ has a critical point, inside the Reeb cone, at $b_2=-4+\sqrt{13}$.
The volume at the critical point is then
\be
V_{*} = \frac{43+13\sqrt{13}}{324}\ee
which is indeed the correct result \cite{paper2, toric}. 

The reason that we get the correct result here is that 
any circle that rotates the \emph{base} $\mathbb{C} P^1$ 
of $\mathbb{F}_1$ is not in the 
centre of the Lie algebra of the compact group 
$K=U(1)^2\times SU(2)$ acting on the cone $X$. From the results 
%earlier work 
on the Futaki invariant in section \ref{futsection}, it follows that
$b_3=0$ necessarily at any critical point. Indeed, 
the Lie algebra $\mathtt{k}=
\mathtt{t}_2\oplus \mathtt{su}(2)$. The reason 
that we do not need to extremise over 
$\mathtt{t}_3$ is that $\mathtt{k}=\mathtt{z}\oplus\mathtt{t}^{\prime}$ 
where $\mathtt{t}^{\prime}\subset\mathtt{su}(2)$. Thus the derivative 
of $\vol[L]$ is automatically zero in the direction $\mathtt{t}^{\prime}$, provided 
$\xi\in\mathtt{z}=\mathtt{t}_2$. The isometry algebra of the 
metric is then, according to our conjecture, 
the maximal $\mathtt{k}=\mathtt{t}_2\oplus\mathtt{su}(2)$, which 
indeed it is \cite{paper2,toric}. 
The formula (\ref{dp1simple}) may indeed be recovered from 
the toric result on setting $b_3=0$, as we show in section \ref{toricsection}.

%%%%%%%%%%%%%%%%%%%%%%%%%%%%%%%%%%%%%%%%%%%%%%%%%%%%%%%%%%%%%%%%%%%%%%%%%

\section{The index--character}
\label{indexsection}

In this section we show that the volume of a Sasakian 
manifold, as a function of $b$, is also related to a limit of the 
equivariant index of the Cauchy--Riemann operator $\bar{\partial}$
on the cone $X$. This equivariant index essentially counts 
holomorphic functions on $X$ according to their 
charges under the $\T^s$ action.
The key to this relation is the Lefschetz fixed point theorem 
for the $\bar{\partial}$ operator. In fact, by taking a 
limit of this formula, we will recover our general 
formula (\ref{normvol}) for the volume in terms of fixed point data.

\subsection{The character}

Recall that $\bar{\partial}$ is the Cauchy--Riemann operator on 
$X$. We may consider the elliptic complex
\be
0\longrightarrow \Omega^{0,0}(X)\stackrel{\bar{\partial}}{\longrightarrow}
\Omega^{0,1}(X)\stackrel{\bar{\partial}}{\longrightarrow}\cdots\stackrel{\bar{\partial}}{\longrightarrow}
\Omega^{0,n}(X)\longrightarrow 0\ee
on $X$. 
Here $\Omega^{0,p}(X)$ denotes the differential forms of 
Hodge type $(0,p)$ with 
respect to the complex structure of $X$. 
We denote the 
cohomology groups of this sequence as 
$\mathcal{H}^p(X)\cong H^{0,p}(X;\C)$. 
In fact, the groups $\mathcal{H}^p(X)$, for $p>0$, 
are all zero. This follows since $X$ may be realised as the 
total space of a negative complex line bundle $\mathcal{L}$ over a compact Fano 
orbifold $V$. The property is then inherited from the Fano $V$. 
On the other hand, $\mathcal{H}^0(X)$ is 
clearly infinite dimensional, in contrast to the compact case. 

The action of $\T^s$ on $X$, since it is holomorphic,
commutes with $\bar{\partial}$. Hence there is an induced action of $\T^s$  
on the cohomology groups of $\bar{\partial}$. 
The equivariant index, or holomorphic Lefschetz number, 
for an element $q\in\T^s$, is defined to be
\be\label{lef}
L(q,\bar{\partial},X) = \sum_{p=0}^n (-1)^p \
\Tr \{q \mid \mathcal{H}^p(X)\}~.\ee
Here the notation $\Tr \{q\mid \mathcal{H}^p(X)\}$ means one should take 
the trace of the induced action of $q$ on $\mathcal{H}^p(X)$.
The index itself, given by setting $q=1$, is clearly infinite:
the action of $q$ is trivial and the trace simply 
counts holomorphic functions on $X$. 
However, the equivariant index is well--defined, provided the 
trace converges. In fact, we may analytically continue (\ref{lef})
to $q\in\TC$. 
The singular behaviour at $q=1$ will then show up as a 
pole. Note we have not imposed any type of boundary conditions 
on $\bar{\partial}$. We shall henceforth write the 
equivariant index as 
\be
C(q,X)=L(q,\bar{\partial},X)=\Tr \{q\mid \mathcal{H}^0(X)\}\ee
and refer to it simply as the character.

\subsection{Relation to the ordinary index}

Suppose that we have a \emph{regular} Sasakian structure on 
$L$, and consider the corresponding circle action on $X$. 
Holomorphic functions on the cone $X$ of charge $k$ 
under this circle action may be identified with holomorphic sections of the 
bundle
\be
(\mathcal{L}^{*})^k\rightarrow V\ee
where recall that $\mathcal{L}^*$ is an ample line bundle over $V$ 
whose dual is the associated complex line bundle to the projection 
$\pi:L\rightarrow V$. Canonically, we may take $\mathcal{L}=\mathcal{K}$, 
the canonical bundle over $V$.

The trace of $q\in\C^*$ on the space of holomorphic functions of 
charge $k$ is given by
\be
\Tr \{q\mid \mathcal{H}^0(X)_k\} = q^k \dim H^0\left(V;(\mathcal{L}^*)^k\right)~.\ee
The right hand side is given by the Riemann--Roch theorem. 
Indeed, we have
\be
\dim H^0\left(V;(\mathcal{L}^*)^k\right) = \chi\left(V,(\mathcal{L}^*)^k\right) 
= \int_V e^{-kc_1(\mathcal{L})}\cdot\mathrm{Todd}(V)~.\ee
In the first equality we have used the fact that 
$\dim H^i\left(V;(\mathcal{L}^*)^k\right) = 0$ for $i>0$ since $V$ is Fano 
and hence $\mathcal{L}^*$ is ample. The second equality is 
the Riemann--Roch theorem\footnote{This does \emph{not} generalise as 
straightforwardly to orbifolds as one might have hoped. We shall make 
some comments on the (equivariant) Riemann--Roch theorem for orbifolds 
at the end of this section.}. The Todd class is a certain polynomial in 
the Chern classes of $V$, whose precise form we won't need. 

It follows then that the character, for a regular $U(1)$ action, is simply 
given by
\be\label{regch}
C(q,X) = \sum_{k\geq 0} q^k \int_V e^{-kc_1(\mathcal{L})}\cdot\mathrm{Todd}(V)~.\ee
More generally, we can interpret the character $C(q,X)$ in terms of the 
equivariant index theorem for $\T^{s-1}$ on $V$. However, 
it is easier to keep things defined on the cone. 

The relation between 
$C(q,X)$ defined in (\ref{regch}) and the volume of 
a regular Sasakian (--Einstein) manifold has been noted before in 
\cite{bergman}. The key in this section is to extend this to 
the equivariant case. Then the relation of the volume to 
the equivariant index becomes a function of the Reeb 
vector field.

\subsection{Localisation and relation to the volume}

In the above 
discussion we have been slightly cavalier in defining 
$C(q,X)$ as a trace over holomorphic functions on $X$, 
since $X$ is \emph{singular}. Recall from 
section \ref{sectiontwo} that $X$ is an affine algebraic 
variety with an isolated Gorenstein singularity at $r=0$, that is defined 
by polynomial equations $\{f_1=0,\dots,f_S=0\}\subset \C^{N}$.
The space of holomorphic 
functions on $X$ that we want is then given by elements of the 
coordinate ring of $X$, 
\be
\C[X]=\mathbb{C}[z_1,\ldots,z_{N}]/\langle f_1,\dots,f_S\rangle~,
\label{ring}\ee
where
$\C[z_1,\dots,z_{N}]$ is simply the polynomial ring on $\C^{N}$ and 
$\langle f_1,\dots,f_S\rangle$ is the ideal generated by the functions $\{f_A, A=1,\ldots,S\}$. 
%recall that the polynomial equations $\{f_i=0\}$ define $M$. 
Also as discussed in section \ref{sectiontwo}, we may always 
resolve $X$ to a space $W$ with at worst orbifold 
singularities: for example, one can take any quasi--regular Sasakian 
structure and take $W$ to be the total space of $\mathcal{L}\rightarrow {V}_0$. It is important that the resolution is equivariant 
with respect to the torus action. Thus, in general, we have a 
$\T^s$--equivariant birational map
\be
f:W\rightarrow X\ee
which maps some exceptional set $E\subset W$ to the singular point 
$p=\{r=0\}\in X$. In particular, $f:W\setminus E\rightarrow X_0=X\setminus \{p\}$ 
is a biholomorphism. The fixed point set of $\T^s$ on $W$ is then 
necessarily supported on $E$. The character is conveniently computed 
on $W$, as in the previous section, and is independent of the choice of 
resolution. In fact this set--up is identical 
to that in section \ref{localsection} -- the resulting formula is 
rather similar to (\ref{DHtheorem}), although for orbifolds 
the equivariant index theorem is rather more involved than for 
manifolds\footnote{The additional technicalities for orbifolds drop out on taking 
the limit to obtain the volume.}. 

We claim that the volume $V(b)$ is given in terms of the 
character $C(q,X)$ by the simple formula
\be\label{limitformula}
V(b) = \lim_{t\rightarrow 0} \ t^n \ C(\exp(-tb),X)~.\ee
Recall that $C(q,X)$ is singular at $q=1$. 
By defining $q_i=\exp(-tb_i)$, and sending $t\rightarrow 0$, 
we are essentially picking out the leading singular behaviour. 
As we shall explain, the leading term in $t$ is always a pole of 
order $n$. 

Let $W$ be a completely smooth resolution of the cone $X$. For example, 
if $X$ admits any \emph{regular} Sasakian structure, as in the 
previous section we make take $W$ to be the
total space of $\mathcal{L}\rightarrow V_0$ with $V_0$ a \emph{manifold}. 
With definitions as in the last section, the 
equivariant index theorem in \cite{AS} gives
\be\label{leffe}
C(q,W) = \sum_{\{F\}} \int_F \frac{\mathrm{Todd}(F)}{\prod_{\lambda=1}^R 
\prod_j \left(1-q^{u_{\lambda}}e^{-x_j}\right)}~.\ee
Here the $x_j$ are the basic characters\footnote{We suppress 
the $\lambda$--dependence for simplicity of notation.} for the 
bundle $\mathcal{E}_{\lambda}\rightarrow F$. These are defined via 
the splitting principle. This says that, for practical 
calculations, we may assume that  
${\mathcal{E}}_{\lambda}$ splits as a direct sum of complex line 
bundles
\be
{\mathcal{E}}_{\lambda} = \bigoplus_{j=1}^{n_{\lambda}} \mathcal{L}_j~.\ee
Then 
\be
x_j=c_1(\mathcal{L}_j)\in H^2(F;\Z)~.\ee For a justification of this, 
the reader might consult reference \cite{BT}. 
The Chern classes 
of $\mathcal{E}_{\lambda}$ may then be written straightforwardly 
in terms of the basic characters. For example, 
\be\label{chernclass}
c(\mathcal{E}_{\lambda})=\prod_j(1+x_j)\ee
so that $c_1=\sum_j x_j$. The Chern character is given by $\mathrm{ch}(\mathcal{E}_{\lambda}) 
=\sum_j \exp(x_j)$.

To illustrate this, we note that one defines the Todd class as 
\be\label{Todd}
\mathrm{Todd} = \prod_a \frac{x_a}{1-\exp(-x_a)} = 1+\frac{1}{2}c_1 
+ \frac{1}{12}(c_1^2+c_2)+\ldots~.\ee
Here $x_a$ are the basic characters for the complex tangent 
bundle of $F$. We have also expanded the Todd class in terms of Chern classes 
of $F$. It will 
turn out that the Todd class does not contribute to the volume 
formula in the limit (\ref{limitformula}). The denominator in (\ref{leffe}),
and (\ref{Todd}), are again understood 
to be expanded in formal a Taylor expansion. 

Before proceeding to the limit (\ref{limitformula}), we note that 
one can recover (\ref{regch}) from (\ref{leffe}) rather straighforwardly. 
The fixed point set of the free $U(1)$ action is the zero section, or exceptional divisor, $V$. 
The normal bundle is then $\mathcal{L}$, and the weight $u=1$. Hence (\ref{leffe}) gives
\bea
C(q,X) = C(q,W) & = & \int_V \frac{\mathrm{Todd}(V)}{1-q e^{-c_1(\mathcal{L})}} 
\nn \\
& = & \int_V \sum_{k\geq0} q^k e^{-kc_1(\mathcal{L})}\cdot\mathrm{Todd}(V)~.\eea

We now turn to proving (\ref{limitformula}). Set $q=\exp(-tb)$ with $t$ a (small) real positive number. 
The denominator in (\ref{leffe}) is then, to leading order in 
$t$, given by 
\bea\label{half}
\prod_{\lambda=1}^R\prod_{j=1}^{n_{\lambda}}\left[t(b,u_{\lambda})+x_j\right]
 = t^{k}\ \left[\prod_{\lambda=1}^R (b,u_{\lambda})^{n_{\lambda}} \right]
 \prod_{\lambda=1}^R\prod_{j=1}^{n_{\lambda}}\left[1+z x_j\right]\eea
where
\be
z = \frac{1}{t(b,u_{\lambda})}~.\ee
The higher order terms in $x_j$ will not contribute 
at leading order in $t$, once one integrates over the fixed point set, 
which is why they do not 
appear in (\ref{half}). Recalling the definition of 
the Chern polynomial, we thus see that, to leading order in $t$, 
(\ref{leffe}) is given by 
\bea
C(e^{-tb},W) & \sim & \sum_{\{F\}} t^{-k}\prod_{\lambda=1}^R \frac{1}{(b,u_{\lambda})^{n_{\lambda}}}
\int_F \frac{\mathrm{Todd}(F)}{\prod_{\lambda=1}^R
\det\left(1+zi\frac{\Omega_\lambda}{2\pi}\right)}\nn \\
& = & \sum_{\{F\}} t^{-k}\prod_{\lambda=1}^R \frac{1}{(b,u_{\lambda})^{n_{\lambda}}}
\int_F \frac{\mathrm{Todd}(F)}{\prod_{\lambda=1}^R \sum_{a\geq0}
c_a(\mathcal{E}_{\lambda})z^a}~.\eea
Again, it is simple to verify that the Todd class (\ref{Todd})
contributes only the constant term $1$ to leading order in $t$. 
Moreover, the integral over $F$ picks out the differential form of 
degree $2(n-k)$, the coefficient of which is homogeneous degree 
$-(n-k)$ in $t$. Thus we have
\be
C(e^{-tb},W) \sim  t^{-n}\sum_{\{F\}} \prod_{\lambda=1}^R \frac{1}{(b,u_{\lambda})^{n_{\lambda}}}
\beta(b)\ee
where 
\be
\beta(b)=\int_F \prod_{\lambda=1}^R \left[\sum_{a\geq0} \frac{c_a(\mathcal{E}_{\lambda})}{(b,u_{\lambda})^a}\right]^{-1}~.\ee
Thus we have shown that the expression
\be
\lim_{t\rightarrow0}\ t^n\ C(e^{-tb},W) \ee
gives precisely the earlier volume formula (\ref{normvol}). We have recovered it 
here by taking a limit of the equivariant index theorem for 
the $\bar{\partial}$ operator. One can argue that, for a \emph{fixed} 
\emph{quasi--regular}
Sasakian metric, this limit of the index gives the volume -- such 
an argument was given in \cite{bergman} and is similar to that at the end 
of section \ref{application}. Since the rationals are dense in the reals, this proves 
(\ref{limitformula}) in general, as a \emph{function} of the Reeb vector field.

We finish this subsection with some comments on the extension 
of this result to orbifold resolutions of $X$ -- that is, 
resolutions with at worst orbifold singularities. Unfortunately, the 
Lefschetz formula (\ref{leffe}) is \emph{not} true for 
orbifolds. Recall that, for the Duistermaat--Heckman 
theorem, the only essential difference was the 
order $d$ of the fixed point set which enters into the 
formula. For the character $C(q,W)$ the difference is more 
substantial. The order again appears, but the integral over 
a connected component of the fixed point set $F$ is replaced 
by an integral over the associated orbifold $\hat{F}$ 
to $F$. Moreover, there are additional terms in the integrand. 
For a complete account, we refer the reader to the 
(fairly recent) original paper \cite{Vergne}. 
We shall not enter into the details of the general 
equivariant index theorem for orbifolds, since we 
do not need it. Instead, we simply note that the 
orbifold version of Duistermaat--Heckman may be 
recovered from an expression for $C(q,W)$ using the 
general results of \cite{Vergne} in much the same way 
as the smooth manifold case treated here.

%%%%%%%%%%%%%%%%%%%%%%%%%%%%%%%%%%%%%%%%%%%%%%%%%%%%%%%%%%%%%%%%%%%%%%%%%%%%%%%%%%%%%%%%%%%%%%%%%%%%%%%%

\section{Toric Sasakian manifolds}
\label{toricsection}

In this section we turn our attention to toric Sasakian manifolds. In this case the K\"ahler cone $X$
is an affine toric variety.  The equivariant index, which is 
a character on the 
space of holomorphic functions, may be computed as a sum over 
integral points inside the polyhedral cone $\mathcal{C}^*$. The toric setting also allows us to obtain 
a ``hands--on'' derivation of the volume function from the index--character. For the purpose of
being self--contained, we begin by recalling the well--known correspondence between the combinatorial
data of an affine toric variety and the set of holomorphic functions defined on it.

\subsection{Affine toric varieties}

When $X$ is toric, that is the torus has maximal possible rank 
$s=n$, it is specified by a convex rational polyhedral cone 
$\mathcal{C}^*\subset\R^n$. 
Let $\SC = \mathcal{C}^*\cap \Z^n$.  As is well--known, 
$\SC\subset\Z^n$ is 
an abelian semi--group, which by Gordan's lemma is finitely generated. 
This means that there are a finite number of generators $m_1,\dots,m_{N} \in \SC$, 
such that  every element of $\SC$ is of the form
\be
a_1 m_1 +\cdots + a_{N} m_{N}~, \qquad a_A \in \mathbb{N}~.
\ee 
To $\SC$ there is an associated semi--group algebra, denoted $\C[\SC]$,
given by the characters $w^m: (\C^*)^{n}\to \C^*$ defined as
\bea
w^m = \prod_{i=1}^n w_i^{m^i}
\eea
with multiplication rule
\be
w^m \cdot w^{m'} = w^{m+m'}~.
\ee
Notice that $\C[\SC]$ is generated by the elements $\{w^{m_A}| m_A ~\mathrm{generate}~ \SC \}$.
In algebraic geometry, the toric variety $X_{\mathcal{C}^*}$ associated to a strictly convex rational
polyhedral cone $\mathcal{C}^*$ is  defined 
as the maximal spectrum\footnote{The maximal spectrum of an algebra $A$
is defined to be the set $\mathrm{Spec}_\mathrm{m}\ A =\{\mathrm{maximal~ideals~in~}A \}$ equipped
with the Zariski topology. An ideal $I$ in $A$ is said to be maximal if $I\neq A$ and the only
proper ideal in $A$ containing $I$ is $I$ itself.}
\bea
X_{\mathcal{C}^*} = \mathrm{Spec}_\mathrm{m}\ \C[\SC]
\eea
of the semi--group algebra $\C[\mathcal{C}^* \cap \Z^n ]$. In general, one can show 
that 
there exist suitable binomial\footnote{A binomial is a difference of two monomials. Then
the affine toric variety is defined by equations of the type `monomial equals monomial'.}
functions $f_A \subset \C^{N}$, where $N$ is the number 
of generators of $\SC$, such that 
\be
\C[\SC] = \C [z_1,\dots,z_{N}]/ \langle f_1,\dots,f_S\rangle~.
\label{chiralring}
\ee
Then, more concretely,
\be
X_{\mathcal{C}^*} = \{ f_1=0,\dots,f_S=0 \}\subset\C^{N}
\ee
presents $X_{\mathcal{C}^*}$ as an affine variety, with ring of holomorphic functions given precisely 
by (\ref{chiralring}).
%$\C[\SC]=\C[z_1,\dots,z_{N}]/<f_1,\dots,f_s>$. 
%where $\C[\SC]$ denotes the polynomial ring generated by 
%$w^m$, where $m\in\SC$ and $w=(w_1,\ldots,w_n)$ are formal 
%variables. This means that

To exemplify this, let us describe briefly the conifold singularity. A set of generators of 
$\SC$ is given, in this case, by the four outward primitive  
edge vectors that generate the polyhedral cone
$\mathcal{C}^*$, which we will present shortly, see (\ref{edgesconi}). Denoting $w=(x,y,z)$, 
the corresponding generators of $\C[\SC]$ are given by  
\be
Y = y ~,\quad W=xz^{-1}~,\quad X=xy^{-1}~,\quad Z=z 
\ee
respectively. It then follows, as is well--known, that the conifold singularity can be represented 
as the single equation 
\be
f=XY-ZW=0 \subset \C^4
\ee
and the coordinate ring is simply $\C[X,Y,Z,W]/\langle XY-ZW \rangle$. 
The vanishing of $XY-ZW$ follows from the
relation $m_1+m_3=m_2+m_4$ between the generators of $\mathcal C^*\cap \Z^3$, and 
the ideal
$\langle XY-ZW\rangle$ is determined by the integer linear relation among these generators. 
While this
is a rather trivial example, it is important to note that in general to construct the monomial 
ideal one has to include all the generators of $\SC$ (otherwise the resulting variety is not normal)
and these are generally many more than the generating edge vectors of $\mathcal{C}^*$. For instance, for
the complex cone over the first del Pezzo surface, whose link is the Sasaki--Einstein manifold
$Y^{2,1}$, there are 9 generators of $\SC$, so that $N=9$, while there are 20 relations
among them, so that $S=20$. As a result this is not a complete intersection.
Some discussion illustrating these points in the physics
literature can be found in \cite{Berenstein:2005xa,Pinansky:2005ex}.

\subsection{Relation of the character to the volume}

As we have explained, when $X$ is toric, by construction 
a basis of holomorphic functions on $X$ is given by the $w^m$
above.
Thus counting holomorphic functions on $X$ according to their 
charges under $\T^n$ is equivalent  to 
counting the elements of the semi--group
$\SC$. 
The character\footnote{In 
fact, the character is also very closely related to the 
\emph{Ehrhart polynomial} when $L$ admits a regular Sasakian 
structure, with Fano $V$. In this case $V$ is also toric and is thus 
associated to a convex lattice polytope $\Delta$ in 
$\R^{n-1}$. The Ehrhart polynomial $E(\Delta,k)$ is defined to 
be the number of lattice points inside the dilated polytope 
$k\Delta$. One can then show that $E(\Delta,k)$ is a polynomial 
in $k$ of degree $n-1$. The coefficient of the leading term 
is precisely the volume of the polytope $\Delta$. This is 
analogous to the relation we discussed in section \ref{indexsection}. 
For a nice account of this, 
and the relation to toric geometry, see David Cox's notes \cite{cox}.}  
 $C(q,X)$ is thus given by
\be\label{character}
C(q,X) = \sum_{m\in\SC} q^m~.\ee
We are again tacitly assuming here 
that the series defining $C(q,X)$ converges.

As we proved in the last section, in general the normalised 
volume $V(b)$ is related to the character $C(q,X)$ by the simple 
formula
\be\label{doughnut}
V(b) = \lim_{t\rightarrow 0} \ t^n\ C(e^{-tb},X)~.\ee
We again remind the reader that the notation $q=e^{-tb}$ is 
shorthand for defining the components
\be
q_i = e^{-tb_i}~.\ee
In this section we shall prove this relation more directly, using 
the formula (\ref{character}). The limit $t\rightarrow 0$ may be 
understood as a Riemann integral, with the limit giving the 
volume formula (\ref{newvolume}).

We first discuss a toy example -- 
the generalisation will be straightforward. Consider the following limit
\bea
\lim_{t\to 0 } \frac{t}{1-e^{-tb}} = \frac{1}{b}~.
\eea
Now, let us expand the fraction in a Taylor series. The radius of convergence of
this series is precisely 1, so that for $b>0$, we require 
that $t$ also be positive. We will be particularly interested in isolating the
singular behaviour as $t\to 0$. We claim that one can deduce the above 
limit via
\bea
\lim_{t\to 0} \sum_{m=0}^\infty t e^{-tmb} =
\int_0^\infty e^{-yb} \ \dd y = \frac{1}{b}~.
\eea
The integral arises simply from the definition of the Riemann integral. In
particular, we subdivide the interval $[0,\infty]$ into intervals of length 
$t$, and sum the contributions of the function $e^{-yb}$ evaluated at the 
end--points of each interval $y_m=mt$. The limit $t\to 0$ is then precisely a
definition of the Riemann integral. 

This easily generalises, and we obtain  
\bea\label{nocoffee}
\lim_{t\to 0} \ t^n \ C(e^{-tb},X) = \lim_{t\to 0} \sum_{m\in \SC} t^n \
e^{-t (b,m)} = \int_{\mathcal{C}^*} e^{-(b,y)} \ \dd y_1\dots \dd y_n~.
\eea
The term appearing on the right hand side is 
called the \emph{characteristic function} of the cone,
and was introduced in \cite{Vinberg}. Of course, the interpretation of 
this function as the volume of a toric Sasakian manifold is new. Specialising 
(\ref{newvolume}) to the toric case, we can relate this to
the volume of the Reeb polytope. Thus, from (\ref{newvolume}), we have
\bea
 \int_{\mathcal{C}^*} e^{-(b,y)} \ \dd y_1\dots \dd y_n = n! 2^n \, \vol (\Delta(b)) =
 \frac{(n-1)!}{2 \pi^n} \,\mathrm{vol}[L](b)~.
 \label{charfuntoric}
\eea
Putting (\ref{charfuntoric}) together with (\ref{nocoffee}), 
we have thus shown that the volume of $L$ 
follows from a simple limit of the
index--character 
\bea
\lim_{t\to 0} \ t^n \ C(e^{-tb},X) = \frac{(n-1)!}{2 \pi^n} \,\mathrm{vol}[L](b)
\label{charvolume}
\eea
in a very direct manner. Note that here we did not use any 
fixed point theorems. 
One can verify (\ref{charvolume}) directly in some simple cases. 
Consider, for instance, $X=\C^3$, with the
canonical basis $v_i=e_i$ for the toric data. We then have 
\bea
\lim_{t\to 0} \frac{t^3}{(1-e^{-tb_1})(1-e^{-tb_2})(1-e^{-tb_3})}&  = &
 \int_{{\cal C}^*} e^{-(b,y)} \ \dd y_1 \dots \dd y_n \nn \\
& = & \frac{1}{b_1b_2b_3} = \ 48\
\mathrm{vol}(\Delta(b))
\eea
where the last equality is computed using the formulae in \cite{MSY}.

Finally, we can also give an independent proof of 
(\ref{charfuntoric}) by induction, 
which uses the particular
structure of polyhedral cones. First, we note that (\ref{charfuntoric}) 
can be proved by direct calculation for $n=2$: without loss of
generality, we can take the primitive normals to the cone to be $(v_1,v_2)$ and
$(0,-1)$ respectively. Then the evaluation of the integral yields
\bea
\int_{{\cal C}^*} e^{-(b,y)} \ \dd y_1 \dd y_2 & = & \int_0^\infty \dd y_1 \int_0^{-
\frac{v_1}{v_2}y_1} \dd y_2\, e^{-(b_1 y_1 +b_2 y_2)} \nn \\
& = & \frac{v_1}{b_1(v_1 b_2- v_2b_1)} = \ 8\  \mathrm{vol}(\Delta(b))
\eea 
where the last equality follows by calculating the area of the 
triangular region $\Delta(b)$.

A result in \cite{Lasserre} shows that the integral of an 
exponential of a linear function on a polyhedral
cone (more generally on a polytope) can be reduced to integrals 
over its facets ${\cal C}^*_a$,
 namely
\bea
b \int_{\mathcal {C}^*} e^{-(b,y)} \ \dd y_1 \dots \dd y_n = \sum_{a=1}^d \frac{v_a}{|v_a|}
\int_{\mathcal{C}^*_a} e^{-(b,y)} \ \dd \sigma
\label{lass}
\eea
for any $b\in \R^n$. Now we proceed by induction, where the hypothesis is that 
(\ref{charfuntoric}) holds at the $(n-1)$--th step. Using the 
first component of 
(\ref{lass}), one obtains
\bea
\int_{\mathcal {C}^*} e^{-(b,y)} \ \dd y_1 \dots \dd y_n = \frac{2^{n-1} (n-1)!}{b_1} 
\sum_{a=1}^d \frac{1}{|v_a|}\mathrm{vol}(\mathcal{F}_a)~,
\eea
which upon using the relation (2.91) of \cite{MSY} gives
\bea
2^n n! \mathrm{vol}(\Delta(b))
\eea
concluding the proof.

Notice that this expression for the volume allows one to 
compute derivatives in $b$ straightforwardly:
\bea
\frac{\de^k}{\de b_{i_1} \cdots \de b_{i_k}} V(b) = (-1)^k 
\int_{{\cal C}^*} y_{i_1}\dots y_{i_k} e^{-(b,y)} \ \dd y_1 \dots \dd y_n
\eea
thus generalising in a natural way the formulae in \cite{MSY}. In particular, 
convexity of $V(b)$ is now immediate from this form of the volume.

\subsection{Localisation formula}

In the case of toric cones $X$, the general fixed point formula 
(\ref{leffe}) for the character has a very simple presentation. 
Recall that every toric $X$ may be completely resolved by 
intersecting the cone $\mathcal{C}^*$ with enough hyperplanes 
in generic position. Specifically, the primitive normal vectors 
$v_a\in\Z^n$ that define the cone $\mathcal{C}^*$ may, by an 
$SL(n;\Z)$ transformtion, be put in the form $v_a=(1,w_a)$ 
where each $w_a\in\Z^{n-1}$. The convex hull of $\{w_a\}$ in 
$\R^{n-1}$ defines a convex lattice polytope. Each interior 
point in this polytope defines a normal vector to a
hyperplane in $\R^n$. If all such hyperplanes are included, 
in generic position\footnote{The positions are the Fayet--Iliopoulos 
parameters, in the language of gauged linear sigma models.}, it is well--known that the corresponding 
toric manifold
is in fact completely smooth.

Let $W=X_P$ be the resolved toric Calabi--Yau manifold\footnote{
The construction described above ensures that the resolution is 
Calabi--Yau. However, more generally 
there is no need to impose this condition in order to compute the character. 
For example, one can compute the character for the canonical action 
of $\T^2$ on $\C^2$ by blowing up origin of the latter to give 
$\mathcal{O}(-1)\rightarrow\mathbb{C}P^1$. This is not a Calabi--Yau 
manifold. Similarly, the character for the conifold may be computed 
by resolving to $\mathcal{O}(-1,-1)\rightarrow \mathbb{C}P^1\times\mathbb{C}P^1$.} corresponding to 
the resulting non--compact polytope $P$. Thus 
$P\subset\R^n$ is the image of $X_P$ under the moment map for the 
$\T^n$ action.
The vertices of $P$ are precisely the images of the fixed points under the 
$\T^n$ action on $X_P$. Denote these as $p_A$. Since each $p_A$ 
corresponds to a smooth point, it follows 
that there are $n$ primitive edge vectors $u_i^A\in\Z^n\subset\mathtt{t}_n^*$, $i=1,\ldots,n$, meeting 
at $p_A$, which moreover span $\Z^n$ over $\Z$. In particular, 
this ensures that a small neighbourhood of $p_A$ is equivariantly 
biholomorphic to $\C^n$. The action of $q\in (\C^*)^n$ on 
complex coordinates $(z_1,\ldots,z_n)$ in this neighbourhood 
is given by 
\be
q:(z_1,\ldots,z_n)\rightarrow 
(q^{u_1^A}z_1,\ldots,q^{u_n^A}z_n)~.\ee
We may then define the character:
\be
C(q,\C^n;\{u_i^A\})= \prod_{i=1}^n \frac{1}{(1-q^{u_i^A})}~.\ee
The fixed point theorem for the equivariant index of $\bar{\partial}$ on 
$X_P$ is 
now very simple to state. It says that
\be\label{toricchar}
C(q,X_P)= \sum_{p_A\in P} C(q,\C^n;\{u_i^A\})\ee
where the $\{u_i^A\}$ are the outward--pointing primitive edge 
vectors at each vertex $p_A\in P$.

As explained earlier, $C(q,X)=C(q,X_P)$. 
Thus we may in fact choose \emph{any} toric resolution $X_P$. 
One can prove invariance directly in dimension $n=3$ as follows. 
For dimension $n=3$, any toric resolution of the cone $X$ may be 
reached from any other by a sequence of local toric flop transitions. 
Since each flop is a local modification of the formula (\ref{toricchar}), 
one needs to only focus on the relevant vertices $p$ near the flop at 
each step. 
One can show that the formula for the conifold is itself 
invariant under the flop transition, as we discuss in the next 
subsection. This proves rather directly that the fixed point 
formula is invariant under toric flops, in complex dimension $n=3$.

It is now simple to take the limit (\ref{doughnut}), giving
\be\label{fpsumtoric}
V(b) = \sum_{p_A\in P} \prod_{i=1}^n \frac{1}{(b,u_i^A)}\ee
where again the $u_i^A$ are the outward--pointing primitive 
normals at the vertex $p_A$. Clearly, this is a
special case of our general result (\ref{normvol}).
The number of fixed points in the sum (\ref{fpsumtoric}) 
is given by the Euler number $\chi(X_P)$ of the resolution 
$X_P$. One can deduce 
this simply from the Lefschetz fixed point formula: one 
applies the equivariant index theorem to the de Rham complex. 
This expresses the Euler number of the resolved space $X_P$ as 
a sum of the Euler numbers of the fixed point sets. The 
Euler number of each fixed point contributes $1$ to the 
total Euler number, and the result follows.

\subsection{Examples}

In this subsection, we compute explicitly the character $C(q,X)$ 
in a number of examples, and verify that we correctly reproduce 
the volume $V(b)$ of the Sasakian metric as a limit. In particular, 
we consider three smooth resolutions of simple toric Gorenstein 
singularities. In order to demonstrate that in general we only 
need to consider orbifold resolutions, we recover the Sasakian volume 
$V(b)$ for the $Y^{p,q}$ singularities by applying our more 
general orbifold localisation formula to a partial resolution of
the singularity obtained by blowing up a Fano.

\subsubsection{The conifold}

We take the toric data $w_1=(0,0)$, $w_2=(0,1)$, $w_3=(1,1)$, $w_4=(1,0)$. 
\begin{figure}[!th]
\vspace{5mm}
\begin{center}
\epsfig{file=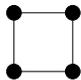,width=1cm,height=1cm}\\
\end{center}
\caption{Toric diagram for the conifold.} \label{con} \vspace{5mm}
\end{figure}
The outward primitive edge vectors for the polyhedral cone are then 
easily determined to be
\be
(0,1,0), \quad (1,0,-1), \quad (1,-1,0), \quad (0,0,1)~.
\label{edgesconi}\ee
Each vector has zero dot products with precisely two of the $v_a=(1,w_a)$, 
and positive dot products with the remaining two. We must now 
choose a resolution of the cone. There are two choices, 
related by the flop transition. There are two vertices in each 
case. 

{\bf First resolution}: We choose the following resolution:
\bea
p_1 & : & u_1^{(1)}=(0,1,0),\ u_2^{(1)} = (0,0,1), \ u_3^{(1)} = 
(1,-1,-1)\nn \\
p_2 & : & u_1^{(2)}=(1,-1,0),\ u_2^{(2)} = (1,0,-1), \ u_3^{(2)} = 
(-1,1,1)~.\eea

\begin{figure}[!th]
\vspace{5mm}
\begin{center}
\epsfig{file=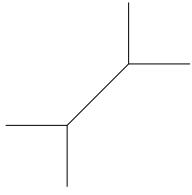,width=2.5cm,height=2.5cm}\\
\end{center}
\caption{A small resolution of the conifold.} \label{webcon} \vspace{5mm}
\end{figure}

The fixed point formula thus gives
\bea\label{firstres}
C(q,X) & = & \sum_{p_{\alpha}} \prod_{i=1}^3 \frac{1}{1-q^{u_i^{(\alpha)}}} \nn\\
& = & \frac{1}{(1-q_2)(1-q_3)(1-q_1q_2^{-1}q_3^{-1})} + 
\frac{1}{(1-q_1q_2^{-1})(1-q_1q_3^{-1})(1-q_1^{-1}q_2q_3)}\nn \\
& = & \frac{1-q_1}{(1-q_2)(1-q_3)(1-q_1q_2^{-1})(1-q_1q_3^{-1})}~.\eea
This is the general result for the character\footnote{Similar computations in related
contexts have appeared before in \cite{Nekrasov:2004vw,Grassi:2005jz,Grassi:2006wh}.}. 
We may now 
set $q=\exp(-tb)$ and take the limit
\be
V(b) = \lim_{t\rightarrow 0} \ t^3 \ C(e^{-tb},X) = 
\frac{b_1}{b_2b_3(b_1-b_2)(b_1-b_3)}~.\ee
This indeed correctly reproduces the result of \cite{MSY} for the volume.

{\bf Second resolution}: The other small resolution of the conifold 
has fixed points
\bea
p_1 & : & u_1^{(1)}=(1,-1,0),\ u_2^{(1)} = (0,0,1), \ u_3^{(1)} = 
(0,1,-1)\nn \\
p_2 & : & u_1^{(2)}=(0,1,0),\ u_2^{(2)} = (1,0,-1), \ u_3^{(2)} = 
(0,-1,1)~.\eea

\begin{figure}[!th]
\vspace{5mm}
\begin{center}
\epsfig{file=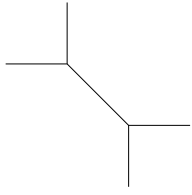,width=2.5cm,height=2.5cm}\\
\end{center}
\caption{The other small resolution of the conifold.} \label{webcon2} \vspace{5mm}
\end{figure}

The fixed point formula thus gives
\bea
C(q,X) & = & \sum_{p_{\alpha}} \prod_{i=1}^3 \frac{1}{1-q^{u_i^{(\alpha)}}} \nn\\
& = & \frac{1}{(1-q_1q_2^{-1})(1-q_3)(1-q_2q_3^{-1})} + 
\frac{1}{(1-q_2)(1-q_1q_3^{-1})(1-q_2^{-1}q_3)}\nn \\
& = & \frac{1-q_1}{(1-q_2)(1-q_3)(1-q_1q_2^{-1})(1-q_1q_3^{-1})}~.\eea
Of course, as expected, this is the same as (\ref{firstres}).

\subsubsection{The first del Pezzo surface}

Recall that this singularity is the lowest
member of the $Y^{p,q}$ family of toric singularities \cite{toric}.
Here we take the toric data $w_1=(-1,-1)$, $w_2=(-1,0)$, $w_3=(0,1)$, $w_4=(1,0)$. 
\begin{figure}[!th]
\vspace{5mm}
\begin{center}
\epsfig{file=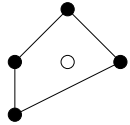,width=1.6cm,height=1.6cm}\\
\end{center}
\caption{Toric diagram for the complex cone over $dP_1$.} \label{dp1} \vspace{5mm}
\end{figure}
The outward primitive edge vectors for the polyhedral cone are then 
easily determined to be
\be
(1,1,0), \quad (1,1,-1), \quad (1,-1,-1), \quad (1,-1,2)~.\ee
We resolve the cone by simply blowing up the del Pezzo surface, 
corresponding to including the interior point $w=(0,0)$. This 
leads to four vertices, with edges:
\bea
p_1 & : & u_1^{(1)}=(1,1,0),\ u_2^{(1)} = (0,0,-1), \ u_3^{(1)} = 
(0,-1,1)\nn \\
p_2 & : & u_1^{(2)}=(1,1,-1),\ u_2^{(2)} = (0,0,1), \ u_3^{(2)} = 
(0,-1,0)\nn \\
p_3 & : & u_1^{(3)}=(1,-1,-1),\ u_2^{(3)} = (0,0,1), \ u_3^{(3)} = 
(0,1,0)\nn \\
p_4 & : & u_1^{(4)}=(1,-1,2),\ u_2^{(4)} = (0,0,-1), \ u_3^{(4)} = 
(0,1,-1)~.\eea

\begin{figure}[!th]
\vspace{5mm}
\begin{center}
\epsfig{file=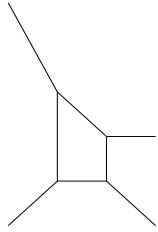,width=2cm,height=3cm}\\
\end{center}
\caption{Canonical bundle over $dP_1$.} \label{webdp1} \vspace{5mm}
\end{figure}

The fixed point formula gives, after some algebra:
\be
C(q,X) = \frac{N(q)}{(1-q_1q_2)(1-q_1q_2q_3^{-1})(1-q_1q_2^{-1}q_3^{-1})(1-q_1q_2^{-1}q_3^2)}\ee
where the numerator is given by
\bea
N(q) & = & 1+q_1+q_1q_3+q_1q_3^{-1}+q_1q_2^{-1}+q_1q_2^{-1}q_3 \nn\\
& & - q_1^2(1+q_1+q_2+q_3+q_3^{-1}+q_2q_3^{-1})~.\eea
Either by taking a limit of this expression, or else using 
(\ref{fpsumtoric}) directly, one obtains 
\be
V(b) = \lim_{t\rightarrow 0} \ t^3\ C(e^{-tb},X) = 
\frac{2(4b_1+2b_2-b_3)}{(b_1+b_2)(b_1-b_2+2b_3)(b_1-b_2-b_3)(b_1+b_2-b_3)}~.\ee
This is indeed correct, although we have chosen a different basis 
from that of \cite{MSY}. Note also that, setting $b_3=0$, we recover 
the formula (\ref{dp1simple}) derived earlier without using 
toric geometry.

\subsubsection{The second del Pezzo surface}

We take the toric data $w_1=(-1,-1)$, $w_2=(-1,0)$, $w_3=(0,1)$, $w_4=(1,0)$, 
$w_5=(0,-1)$. 

\begin{figure}[!th]
\vspace{5mm}
\begin{center}
\epsfig{file=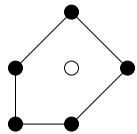,width=1.7cm,height=1.7cm}\\
\end{center}
\caption{Toric diagram for the complex cone over $dP_2$.} \label{dP2} 
%\vspace{5mm}
\end{figure}

This is the blow--up of the first del Pezzo surface, 
introducing an exceptional divisor corresponding to $w_5$.
The outward primitive edge vectors for the polyhedral cone are then 
easily determined to be
\be
(1,1,0), \quad (1,1,-1), \quad (1,-1,-1), \quad (1,-1,1), \quad (1,0,1)~.\ee
We resolve the cone by simply blowing up the del Pezzo surface, 
corresponding to including the interior point $w_5=(0,0)$. This 
leads to five vertices, with edges:
\bea
p_1 & : & u_1^{(1)}=(1,1,0),\ u_2^{(1)} = (0,0,-1), \ u_3^{(1)} = 
(0,-1,1)\nn \\
p_2 & : & u_1^{(2)}=(1,1,-1),\ u_2^{(2)} = (0,0,1), \ u_3^{(2)} = 
(0,-1,0)\nn \\
p_3 & : & u_1^{(3)}=(1,-1,-1),\ u_2^{(3)} = (0,0,1), \ u_3^{(3)} = 
(0,1,0)\nn \\
p_4 & : & u_1^{(4)}=(1,-1,1),\ u_2^{(4)} = (0,0,-1), \ u_3^{(4)} = 
(0,1,0)\nn \\
p_5 & : & u_1^{(5)}=(1,0,1),\ u_2^{(5)} = (0,-1,0), \ u_3^{(5)} = 
(0,1,-1)~.\eea

\begin{figure}[!th]
\vspace{5mm}
\begin{center}
\epsfig{file=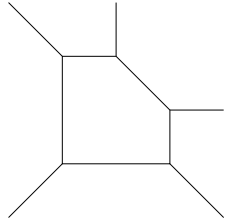,width=2.9cm,height=2.9cm}\\
\end{center}
\caption{Canonical bundle over $dP_2$.} \label{webdp2} \vspace{5mm}
\end{figure}

Rather than give the full character, we simply state the result for 
the volume:
\be
V(b) = \frac{7b_1^2+2b_1b_2+2b_1b_3-b_2^2-b_3^2+2b_2b_3}{
(b_1+b_2)(b_1+b_2-b_3)(b_1-b_2-b_3)(b_1-b_2+b_3)(b_1+b_3)}~.\ee
Setting $b_1=3$, it is straightforward to determine that the 
critical point, inside the Reeb cone, lies at 
\be
b_{*2}=b_{*3}=\frac{-57+9\sqrt{33}}{16}\ee
and that the volume at the critical point is
\be
V(b_*) = \frac{59+11\sqrt{33}}{486}~.\ee

\subsubsection{An orbifold resolution: $Y^{p,q}$ singularities}

Recall that the $Y^{p,q}$ singularities are affine toric Gorenstein singularities 
generated by four rays, with toric data $w_1=(0,0)$, $w_2=(p-q-1,p-q)$, 
$w_3=(p,p)$, $w_4=(1,0)$ \cite{toric}. This includes our earlier 
example of the complex cone over the first del Pezzo surface, although 
here we use a different basis for convenience.
\begin{figure}[!th]
\vspace{5mm}
\begin{center}
\epsfig{file=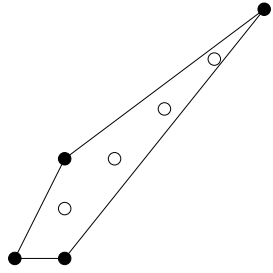,width=3.5cm,height=3.5cm}\\
\end{center}
\caption{Toric diagram for $Y^{5,3}$.} \label{y52} \vspace{5mm}
\end{figure}
The outward edge vectors for the polyhedral cone are then 
easily determined to be
\be
(0,p-q,-p+q+1), \quad (p,q,-1-q), \quad (p,-p,p-1), \quad (0,0,1)~.\ee
We now \emph{partially resolve} the cone by blowing up the Fano 
corresponding to the interior point\footnote{Note that \emph{any} 
interior point would do.} $w=(1,1)$. This 
leads to a non--compact polytope $P\subset\R^3$ with four vertices, with outward--pointing edge vectors:
\bea
p_1 & : & u_1^{(1)}=(0,p-q,-p+q+1),\ u_2^{(1)} = (0,-1,1), \ u_3^{(1)} = 
(1,-p+q+1,p-q-2)\nn \\
p_2 & : & (p-1)u_1^{(2)}=(p,q,-1-q),\ (p-1)u_2^{(2)} = (-1,p-q-1,-p+q+2), \nn \\ 
   & & (p-1)u_3^{(2)} = 
(0,-p+1,p-1)\nn \\
p_3 & : & (p-1)u_1^{(3)}=(p,-p,p-1),\ (p-1)u_2^{(3)} = (0,p-1,-p+1), \nn \\ 
   & & (p-1)u_3^{(3)} = 
(-1,1,0)\nn \\
p_4 & : & u_1^{(4)}=(0,0,1),\ u_2^{(4)} = (1,-1,0), \ u_3^{(4)} = 
(0,1,-1)~.\eea
The normalisations here ensure that we correctly get the 
corresponding weights for the torus action that enter 
the orbifold localisation formula.
Note that they are generally \emph{rational} vectors, for vertices 
$p_2$ and $p_3$. Indeed, it is straightforward to show that the three 
primitive outward--pointing edge vectors at these vertices 
span $\Z^3$ over $\mathbb{Q}$, but not over $\Z$. In both cases, $\Z^3$ modulo 
this span is isomorphic to $\Z_{p-1}$. This immediately implies 
that these vertices are $\Z_{p-1}$ orbifold singularities, and thus 
the orders of these fixed points are $d_{p_2}=d_{p_3}=p-1$. On the 
other hand, the vertices $p_1$ and $p_4$ are smooth, and thus 
$d_{p_1}=d_{p_4}=1$. 
\begin{figure}[!th]
\vspace{5mm}
\begin{center}
\epsfig{file=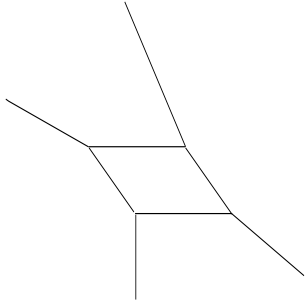,width=4cm,height=4cm}\\
\end{center}
\caption{Partially resolved polytope $P$ for $Y^{5,3}$.} \label{weby52} \vspace{5mm}
\end{figure}
We must now use the localisation formula for orbifolds, which 
includes the inverse orders $1/d_{p}$ at each vertex $p$ as a multiplicative 
factor. This is straightforward to compute:
\bea
V(b) & = & \sum_{p_{A},A=1,\ldots,4}\frac{1}{d_{p_A}}\prod_{i=1}^3 \frac{1}{(b,u_{i}^{(A)})} \nn \\
& = & \frac{p\left[p(p-q)b_1+q(p-q)b_2+q(2-p+q)b_3\right]}{b_3(pb_1-pb_2+(p-1)b_3)((p-q)b_2+(1-p+q)b_3)(pb_1+qb_2-(1+q)b_3)}\nn\eea
which is indeed the correct expression \cite{MSY}.

%Note that one could also compute the character from this partial resolution. 
%However, as we already mentioned, the localisation formula for the 
%character for orbifold resolutions is rather more involved than for completely 
%smooth resolutions. We simply chose not to give these details here as they 
%are a little technical, but 
%one can indeed write down a general formula. In the case at hand, since 
%the fixed points are isolated, the situation is not too bad. The 
%vertices $p_2$ and $p_3$ contribute to the character in a slightly 
%more complicated fashion -- namely, one must sum over $p-1$ images. 
%We leave it as an exercise for the interested reader to work out 
%the details.

%%%%%%%%%%%%%%%%%%%%%%%%%%%%%%%%%%%%%%%%%%%%%%%%%%%%%%%%%%%%%%%%%%%%%%%%%%

\subsection*{Acknowledgments}
\noindent 
We would like to thank S. Benvenuti, D. Goldfeld, P. A. Grassi, A. Hanany, 
J. F. Morales and R. Thomas for interesting discussions.
D. M. and J. F. S 
would also like to thank the Institute for Mathematical Sciences, Imperial 
College London, for hospitality during the final stages of this work. 
J. F. S. is supported by NSF grants DMS--0244464, 
DMS--0074329 and DMS--9803347. S.--T. Y. is supported in part by 
NSF grants DMS--0306600 and DMS--0074329.
 
%%%%%%%%%%%%%%%%%%%%%%%%%%%%%%%%%%%%%%%%%%%%%%%%%%%%%%%%%%%%%%%%%%%%%%%%%%

\appendix

%%%%%%%%%%%%%%%%%%%%%%%%%%%%%%%%%%%%%%%%%%%%%%%%%%%%%%%%%%%%%%%%%%%%%%%
\section{The Reeb vector field is holomorphic and Killing}
\label{reebiskilling}

In this appendix we give a proof that $\xi=J(\rdr)$ is both Killing 
and holomorphic. This fact is well--known in the literature, although it 
seems the derivation is not (however, see \cite{Boyer:1998sf}).
Thus, for completeness, we give one here.

We begin with the following simple formulae for covariant derivatives on $X$:
\bea
& & \nabla_{r\partial/\partial r} \left(\fracrdr \right)  = \fracrdr, 
 \quad \nabla_{r\partial/\partial r} Y = \nabla_Y \left(\fracrdr \right)=Y 
\nn \\ 
& & \nabla_Y Z = \nabla^L_Y Z -g_X(Y,Z)\fracrdr ~,\eea
which may easily be checked by computing the Christoffel 
symbols of the metric $g_X= \diff r^2 +r^2 g_L$.
Here $\nabla$ denotes the Levi--Civita connection on $(X,g_X)$, 
$\nabla^L$ is that on $(L,g_L)$, and $Y,Z$
are vector fields on $L$, viewed as vector fields on $X_0=\mathbb{R}_+\times L$. 
A straightforward calculation, using $\nabla J=0$, then shows that $\xi$ is in fact a
Killing vector field on $X$:
\be
g_X(\nabla_Y\xi, Z) = g_X(\nabla_Y \left[J(r\tfrac{\partial}{\partial r})\right],Z) 
= g_X(J(\nabla_Y r\tfrac{\partial}{\partial r}), Z) = 
g_X(JY,Z)\ee
where $Z$ and $Y$ are any two vector fields on $L$. The last term is 
skew in $Z$ and $Y$. Similarly,
\be
g_X(\nabla_{r\partial/\partial r}\xi, Y) = g_X(\xi,Y) ~,%= g_X (J\rdr,Y) = -g_X(\nabla_Y\xi, \rdr)
\label{skrew}
\ee
and 
\be 
g_X(\nabla_Y\xi,\rdr) = - g_X (JY,\rdr) = - g_X(Y,\xi)~,
\ee
so that  (\ref{skrew})
is also skew; while the diagonal element 
\be
g_X(\nabla_{\rdr}\xi,\rdr)=0\ee
clearly vanishes. 
Thus we conclude that $g_X(\nabla_U\xi, V)$ is skew in $U,V$ for any 
vector fields $U$ and $V$ on $X$, and hence $\xi$ is Killing. One
can similarly check that $\xi$ pushes forward to a unit 
Killing vector on $L$, 
where we identify $L=X\mid_{r=1}$.

In fact $\rdr$ and $\xi$ are both holomorphic vector fields. 
Indeed, for any vector fields $U$ and $V$, we have the general formula
\be
(\mathcal{L}_U J) V = (\nabla_U J) V + J \nabla_V U - \nabla_{JV} U\ee
relating the Lie derivative to the covariant derivative. Using this and the fact that $J$ is
covariantly constant, $\nabla_U J =0 $, one now easily sees that 
\be
\mathcal{L}_{r\partial/\partial r} J = 0~, \quad \qquad \mathcal{L}_{\xi}J=0~.\ee

\section{More on the holomorphic $(n,0)$--form}
\label{moremore}

In the main text we used the fact that $\mathcal{L}_Y\Omega=0$ is 
equivalent to $\mathcal{L}_Y\Psi=0$, where $Y$ is holomorphic, 
Killing, and commutes with $\xi$. Although intuitively clear, 
one has to do a little work to prove this. We include the details 
here for completeness.

Suppose that $\mathcal{L}_Y\Psi=0$, where $\Psi$ is the canonically defined spinor 
on the K\"ahler cone $X$ and $Y$ is a holomorphic Killing vector 
field that commutes with $\xi$. This is true of all $Y\in\mathtt{t}_s$ in 
the main text. The restriction of $\Psi$ to $L$ is the spinor $\theta$. 
Writing $\Omega=\exp(f/2)K$, we of course have
$\mathcal{L}_Y K=0$. Thus we must prove that $\mathcal{L}_Yf=0$. 
Since $i\partial\bar{\partial}f=\rho= J\, \mathrm{Ric}(g_X)$,  
and $Y$ is 
both holomorphic and Killing, we immediately have
\be\label{riba}
i\partial\bar{\partial}\mathcal{L}_Y f=0~.\ee
Recall that $f=\log\|\Omega\|^2_{g}$ is degree zero under $\rdr$ and 
basic with respect to $\xi$. Thus the same is true of $\mathcal{L}_Y f$. 
We may then interpret (\ref{riba}) as a transverse equation, or, when 
$L$ is quasi--regular, as an equation on the Fano $V$. In the 
latter case it follows immediately that $\mathcal{L}_Y f = c$, a 
constant. In the irregular case, the paper \cite{french} 
also claims that a transverse $i\partial\bar{\partial}$ lemma holds 
in general. 

We now compute
\bea\label{exactdude}
\frac{i^n}{2^n}(-1)^{n(n-1)/2}c\Omega\wedge\bar{\Omega} & = & \frac{i^n}{2^n}(-1)^{n(n-1)/2}\mathcal{L}_Y(\Omega\wedge\bar{\Omega})
= \diff\left[e^f Y\lrcorner \frac{\omega^n}{n!}\right] \nn\\
& = & -\diff \left(e^f\right)\wedge\diff y_Y\wedge\frac{\omega^{n-1}}{(n-1)!}~.\eea
Here we have used the fact that $\Omega$ is closed, and recall 
that $\diff y_Y = -Y\lrcorner \omega$ where $y_Y$ is the Hamiltonian 
function for $Y$. The right hand side of (\ref{exactdude}) is clearly 
exact. Hence we may integrate this equation over $r\leq 1$ and use 
Stokes' Theorem to deduce that
\be
\frac{i^n}{2^n}(-1)^{n(n-1)/2}c\int_{r\leq1}\Omega\wedge\bar{\Omega} 
= -\int_L e^f\diff y_Y\wedge\frac{\omega_T^{n-1}}{(n-1)!}~.\ee
Now, every term in the integrand on the right hand side of this 
equation is basic with respect to $\xi$. In particular, $\xi$ 
contracted into the integrand is zero\footnote{To see that $\dd y_Y$ is
basic, simply notice that ${\dd y_Y (\xi)}= {\cal L}_\xi y_Y =\tfrac{1}{2} 
{\cal L}_\xi (r^2 \eta (Y))=0$, the last equality following from the fact that 
$\xi$ preseves $\eta,Y$  and $r$, as discussed at various points in the text.}.
However, this means that 
the integral is itself zero. Since the integral of $\Omega\wedge
\bar{\Omega}$ is certainly non--zero, we conclude that $c=0$ and 
hence that $\mathcal{L}_Y f=0$, as desired.

Conversely, suppose that $\mathcal{L}_Y\Omega=0$. From 
equation (\ref{MA}) we immediately deduce now that $\mathcal{L}_Yf=0$ 
since $Y$ is holomorphic and Killing by assumption. Thus 
$\mathcal{L}_Y K=0$. Hence
\be\label{cute}
0 = \mathcal{L}_Y K = (\mathcal{L}_Y\bar{\Psi}^c)\gamma_{(n)}\Psi 
+ \bar{\Psi}^c\gamma_{(n)}\mathcal{L}_Y\Psi = 
2\bar{\Psi}^c\gamma_{(n)}\mathcal{L}_Y\Psi~.\ee
Consider now $\mathcal{L}_Y\Psi$. In fact this must be proportional 
to $\Psi$. An easy way to see this is to go back to the isomorphism 
(\ref{iso}). The splitting of $\Lambda^{0,*}(X)$ into differential 
forms of different degrees is realised on the space of spinors $\mathcal{V}$ 
via the Clifford action of the K\"ahler form $\omega$. The latter 
splits the bundle $\mathcal{V}$ into eigenspaces
\be
\mathcal{V}=\bigoplus_{a=0}^n \mathcal{V}_a\ee
where $\mathcal{V}_a$ is an eigenspace of $\omega\cdot$ with 
eigenvalue $i(n-2a)$. Moreover, $\dim\mathcal{V}_a = \binom{n}{a}$ and
\be
\mathcal{V}_a \cong \Lambda^{0,a}(X)~.\ee 
Recall now that $\Psi$ corresponds to a section of $\Lambda^{0,0}(X)$ 
under the isomorphism (\ref{iso}); hence $\Psi$ has eigenvalue 
$in$ under the Clifford action of $\omega$. Indeed, one can check this 
eigenvalue rather straighforwardly, without appealing to the 
isomorphism (\ref{iso}). 

We may now consider $\mathcal{L}_Y\Psi$. Since $Y$ is holomorphic and 
Killing, it preserves $\omega$. Thus the Clifford action commutes past 
the Lie derivative, and we learn that $\mathcal{L}_Y\Psi$ has the 
same eigenvalue as $\Psi$. But since this eigenbundle is one--dimensional, 
they must in fact be proportional: $\mathcal{L}_Y\Psi = F\Psi$ for 
some function $F$. Thus (\ref{cute}) says that
\be
0 = 2F\bar{\Psi}^c\gamma_{(n)}\Psi=2FK~.\ee
Since $K$ is certainly non--zero, we conclude that $F=0$, and we are 
done.

\section{Variation formulae}
\label{variations}

In this appendix we derive the first and second variation 
formulae (\ref{firstder}), (\ref{secondder}).

\subsection{First variation}

Recall that we linearise the equations for deforming the Reeb 
vector field around a given background K\"ahler cone with 
K\"ahler potential $r^2$. We set
\bea
\xi(t) & = & \xi + t Y\label{Bone}\\
r^2(t) & = & r^2(1+t\phi)~.\eea
We work to first order in $t$. Note that contracting (\ref{Bone}) with $J$ we have
\be
r(t)\frac{\partial}{\partial r(t)}=\fracrdr-tJ(Y)~.\ee
Expanding
\be
\mathcal{L}_{r(t)\partial/\partial r(t)}r^2(t)=2r^2(t)\ee
to first order in $t$ gives
\be
\mathcal{L}_{\rdr}\phi = 2\mathcal{L}_{J(Y)}\log r = -2\eta(Y)~.\ee
Recall we also require
\be
\mathcal{L}_{\xi(t)}r^2(t)=0\ee
which gives, again to first order,
\be\label{diffphixi}
\mathcal{L}_{\xi}\phi = -2\diff\log r(Y)=0~.\ee
In particular, note that when $Y=0$ we recover that $\phi$ 
should be homogeneous degree zero and basic, and 
thus gives a transverse K\"ahler deformation. Note also 
that the right hand side of (\ref{diffphixi}) is 
zero if and only if $Y$ is a holomorphic Killing vector 
field of the background. 

We may use these equations to compute the derivative of the volume 
$\vol[L]$, which we think of as $\vol[\xi]$, in the direction $Y$.
We write this as $\diff\vol[L](Y)$. Arguing much as in section \ref{EHaction}, 
we obtain
\be\label{jarvis}
\diff\vol[L](Y) = -n\int_L \phi\diff\mu + \frac{n}{2}\int_{r\leq1}\diff\diff^c(r^2\phi)\wedge
\frac{\omega^{n-1}}{(n-1)!}~.\ee
The first term arises from the variation of the domain, as in equations (\ref{domainchange}),
(\ref{domainchange2}). However, one must be careful to 
note that here $\phi=\phi(r,x)$ is a function of both $r$ and the point $x\in L$, where $L$ is 
the unperturbed link $L=X\mid_{r=1}$. We must integrate up to the hypersurface $r(t)=1$,
which one can check is, to first order in $t$, given by $r=1-(1/2)t\phi(r=1,x)$. Thus 
one should replace $\phi$ by $\phi(r=1)$ in (\ref{domainchange}).

Using Stokes' theorem on the second term on the right hand side of (\ref{jarvis}), 
the first term is cancelled, as before,
leaving 
\be
\diff\vol[L](Y) = \frac{n}{2}\int_L \diff^c\phi\wedge\frac{\omegatr^{n-1}}{(n-1)!}~.\ee
Now
\be
\xi\lrcorner\diff^c\phi = \mathcal{L}_{\rdr}\phi = -2\eta(Y)\label{B10}\ee
where in the last equality we have used the linearised equation. 
We thus have
\be
\diff\vol[L](Y)= -n\int_L\eta(Y)\diff\mu\ee
which is formula (\ref{firstder}) in the main text.

\subsection{Second variation}

We now take the second variation of the volume. We write
\be
\diff\vol[L](Y) = -n(n+1)\int_{r\leq1}(\diff^c r^2)(Y)\frac{\omega^n}{n!}\label{B12}\ee
as described in the main text. We now again deform
\bea
r^2(t) & = & r^2(1+t\psi)\nn \\
r(t)\frac{\partial}{\partial r(t)} & = & \fracrdr - tJ(Z)\eea
giving linearised equations
\bea\label{linguy}
\mathcal{L}_{\rdr}\psi & = & -2\eta(Z)\nn \\
\mathcal{L}_{\xi}\psi & = & -2\diff \log r(Z)=0~.\eea
In fact the second equation again will not be used. The derivative of 
(\ref{B12}) gives
\bea\label{boris}
\frac{1}{n(n+1)}\diff^2\vol[L](Y,Z) &=&  \int_L \psi \eta(Y)\diff\mu 
-\int_{r\leq1}(\diff^c(r^2\psi))(Y)\frac{\omega^n}{n!}\nn\\
&-&\int_{r\leq 1} 2r^2\eta(Y)
\frac{1}{4}\diff\diff^c(r^2\psi)\wedge\frac{\omega^{n-1}}{(n-1)!}~.\eea
The three terms occur from the variation in domain, the variation 
of $\diff^c(r^2)$, and the variation of the measure, respectively.
The last term in (\ref{boris}) may be integrated by parts, with respect to $\diff$, giving 
the two terms
\bea\label{johnson}
-\frac{1}{2}\int_L \eta(Y)\diff^c(r^2\psi)\wedge\frac{\omegatr^{n-1}}{(n-1)!}
-\int_{r\leq1}(Y\lrcorner\omega)\wedge\diff^c(r^2\psi)\wedge
\frac{\omega^{n-1}}{(n-1)!}~,\eea
where we have used $2Y\lrcorner \omega= - \diff (r^2 \eta(Y))$.
Expanding the first term in (\ref{johnson}), with respect to 
$\diff^c$, gives
\bea\label{bach}
-\int_L \psi\eta(Y)\diff\mu+\int_L\eta(Y)\eta(Z)\diff\mu~.\eea
To produce the form of the second term in (\ref{bach}), we have 
used the trick (\ref{B10}) of writing the linearised equation (\ref{linguy})
as $(\diff^c\psi)(\xi)=-2\eta(Z)$. Now, the first term in (\ref{bach}) 
precisely cancels the first term in (\ref{boris}). Hence we are left with
\bea
&&\frac{1}{n(n+1)}\diff^2\vol[L](Y,Z)  = \int_L\eta(Y)\eta(Z)\diff\mu \nn\\
&&\qquad\qquad -\int_{r\leq1}(\diff^c(r^2\psi))(Y)\frac{\omega^n}{n!}+
Y\lrcorner\omega\wedge\diff^c(r^2\psi)\wedge\frac{\omega^{n-1}}{(n-1)!}~.\eea
Finally, we note the identity
\bea
0 & = & Y\lrcorner\left[\diff^c(r^2\psi)\wedge\frac{\omega^n}{n!}\right]\nn\\
& = & (\diff^c(r^2\psi))(Y)\frac{\omega^n}{n!}-\diff^c(r^2\psi)
\wedge Y\lrcorner\omega\wedge\frac{\omega^{n-1}}{(n-1)!}~.\eea
Thus we have shown that 
\be
\diff^2\vol[L](Y,Z)  = n(n+1)\int_L \eta(Y)\eta(Z)\diff\mu\ee
which is equation (\ref{secondder}) in the main text.

%%%%%%%%%%%%%%%%%%%%%%%%%%%%%%%%%%%%%%%%%%%%%%%%%%%%%%%%%%%%%%%%%%%%

\end{document}